\documentclass[a4paper,11pt,notitlepage,amssymb]{article}
\pdfoutput=1
\usepackage{graphicx}

\usepackage[margin=20mm]{geometry}
\usepackage{setspace}
\usepackage[font={small}, labelfont={bf,small}]{caption}
\usepackage{etoolbox,siunitx}
\robustify\bfseries
\usepackage{cite}
\usepackage{hyperref}
\usepackage{multirow}
\usepackage{makecell}

\usepackage[dvipsnames]{xcolor}
\usepackage{latexsym}
\usepackage{cite}

\usepackage{xspace}
\usepackage{amssymb}
\usepackage{amsmath}
\usepackage{bm}
\usepackage{siunitx}
\usepackage[capitalise]{cleveref}
\usepackage{xparse}
\usepackage{booktabs}
\usepackage{setspace} 
\usepackage{pdflscape}

\DeclareCaptionLabelFormat{sfiglab}{\textbf{Supplementary Figure #2}}
\DeclareCaptionLabelFormat{stablab}{\textbf{Supplementary Table #2}}

\newcommand{\FigCausalGraphBreitblue}{Supplementary Fig.~1}
\newcommand{\FigRTNetLRFL}{Supplementary Fig.~2}
\newcommand{\FigCCDF}{Supplementary Fig.~3}
\newcommand{\FigCausalGraphNoStaff}{Supplementary Fig.~4}
\newcommand{\FigCausalGraphNoAggr}{Supplementary Fig.~5}
\newcommand{\FigOutletsAlignment}{Supplementary Fig.~6}
\newcommand{\FigRankComp}{Supplementary Fig.~7}
\newcommand{\FigLaggedCorr}{Supplementary Figs.~8,~9,~10 \&~11}

\newcommand{\TabHostnames}{Supplementary Table~1}

\newcommand{\TabJaccardBreitblue}{Supplementary Table~3}

\newcommand{\TabRTNetStatsComplete}{Supplementary Table~7}
\newcommand{\TabInflOverlap}{Supplementary Table~8}
\newcommand{\TabCampStaff}{Supplementary Table~9}
\newcommand{\TabCorrVals}{Supplementary Table~10}
\newcommand{\TabCausalEffectsNoStaff}{Supplementary Table~11}

\newcommand{\TabOfficialClients}{Supplementary Table~14}
\newcommand{\TabCausalParents}{Supplementary Table~15}

\newcommand{\TabsBreitBlue}{Supplementary Tables~2,~3,~4,~5 \&~6}

\newcommand{\TabsNotAggr}{Supplementary Tables~12 and~13}

\begin{document}

 \begin{center}
   \LARGE{\textbf{Influence of fake news in Twitter during the 2016 US
       presidential election}}
 
\vspace{1cm}

\large Alexandre Bovet$^{1,2,3}$, Hern\'an A. Makse$^{1,*}$

\vspace{0.2cm} \normalsize 
\textit{1) Levich Institute and Physics
  Department, City College of New York, New York, New York 10031, USA\\
2) ICTEAM, Universit\'e Catholique de Louvain, Avenue George Lema\^itre 4, 1348 Louvain-la-Neuve, Belgium\\
3) naXys and Department of Mathematics, Universit\'e de Namur, Rempart de la Vierge 8, 5000 Namur, Belgium.\\ 
* hmakse@ccny.cuny.edu
}

\end{center}

\begin{abstract}

The dynamics and influence of fake news on Twitter during the 2016 US
presidential election remains to be clarified. Here, we use a dataset
of 171 million tweets in the five months preceding the election day to
identify 30 million tweets, from 2.2 million users, which contain a
link to news outlets. Based on a classification of news outlets
curated by \texttt{www.opensources.co}, we find that 25\% of these
tweets spread either fake or extremely biased news. We characterize
the networks of these users to find the most influential spreaders of
fake and traditional news and use causal modelling to uncover how fake
news influenced the presidential election.  We find that, while top
influencers spreading traditional center and left leaning news largely
influence the activity of Clinton supporters, this causality is
reversed for the fake news: the activity of Trump supporters
influences the dynamics of the top fake news spreaders.

\end{abstract}

\section{Introduction}

Recent social and political events, such as the 2016 US presidential
election\cite{Allcott2017}, have been marked by a growing number of
so-called ``fake news'', i.e. fabricated information that disseminate
deceptive content, or grossly distort actual news reports, shared on
social media platforms.
While misinformation and propaganda have existed since ancient times\cite{PoliticoFakeNews},
their importance and influence
in the age of social media is still not clear.
Indeed, massive digital misinformation has
been designated as a major technological and geopolitical risk by the
2013 report of the World Economic Forum\cite{howell2013digital}.
A substantial number of studies have recently investigated the phenomena
of misinformation in online social networks such as
Facebook\cite{Bessi2015PlosOne,Bessi2015,Mocanu2015,Bessi2015PlosTwo,Bessi2016,DelVicario2016,DelVicario2017}
Twitter\cite{Shao2016,Vosoughi2018,DelVicario2017,Shao2018}, YouTube\cite{Bessi2016youtube} or
Wikipedia\cite{Kumar2016}. 
These investigations, as well as
theoretical modeling\cite{DelVicario2017SRep,Askitas2017}, suggest
that confirmation bias\cite{Klayman1987} and social influence results
in the emergence, in online social networks, 
{of user communities that share similar beliefs about specific topics,
i.e. echo chambers, where unsubstantiated claims or 
true information, aligned with these beliefs, 
are as likely to propagate virally\cite{Mocanu2015,Qiu2017}\label{rev:echo_chamber}}.
A comprehensive investigation of the spread of true and false news in 
Twitter also showed that false news is
characterized by a faster and broader diffusion than true news
mainly due to the attraction of the novelty of false news\cite{Vosoughi2018}.
A polarization in
communities is also observed in the consumption of news in
general\cite{Schmidt2017,DelVicario2017Brexit} and corresponds with
political alignment\cite{Bakshy2015}.  Recent works also revealed the
role of bots, i.e. automated accounts, in the spread of
misinformation\cite{Lee2006,Bessi2016Bots,Ferrara2016,Vosoughi2018}.  In particular, Shao
\textit{et al.} found that, during the 2016 US
presidential election on Twitter, bots were responsible for the early
promotion of misinformation, that they targeted influential users
through replies and mentions\cite{Shao2017}
{\label{rev:newrefs}and that the sharing of fact-checking articles nearly disappears in the 
core of the network, while social bots proliferate\cite{Shao2018}}.
These results have raised the question
of whether such misinformation campaigns could alter public opinion
and endanger the integrity of the presidential
election\cite{Bessi2016Bots}.\\

Here, we use a dataset of 171 million tweets sent by 11 million users
covering almost the whole activity of users regarding the two main US
presidential candidates, Hillary Clinton and Donald Trump, collected
during the five months preceding election day and used to extract
and analyze Twitter opinion trend in our previous work
\cite{Bovet2017TwitterOpinion}.
We compare the spread of news coming from websites that have been described as displaying
fake news with the spread of news coming from traditional, fact-based,
news outlets with different political orientations.
We relied upon the opinion of communications scholars (see Methods for details) who have 
classified websites as containing fake news or extremely biased news.
We investigate
the diffusion in Twitter of each type of media to understand what is
their relative importance, who are the {top news spreaders} and how they
drive the dynamics of Twitter opinion.
We find that, among the 30.7 million tweets containing an URL directing
to a news outlet website, {10\% point toward websites containing fake
news or conspiracy theory and 15\% point toward websites with extremely biased news}.
When considering only
tweets originating from non-official Twitter clients, we see a
tweeting rate for users tweeting links to websites containing news classified as fake
more than four times larger than for traditional media,
suggesting a larger role of bots in the diffusion of fake news.
We separate traditional news outlets from the 
least biased to the most biased and reconstruct the information flow
networks by following retweets tree for each type of media. 
User diffusing fake news form more connected networks with less
heterogeneous connectivity than users in traditional {center and left leaning}
news diffusion networks.
While {top news spreaders} of traditional news outlets are
journalists and public figures with verified Twitter accounts, we find
that a large number of {top fake and extremely biased news spreaders}
are unknown users or users with deleted Twitter accounts.
The presence of two clusters of
media sources and their relation with the supporters of each candidate
is revealed by the analysis of the correlation of their activity.
Finally, we explore the dynamics between the {top news spreaders} and the
supporters' activity with a {multivariate causal network reconstruction\cite{Runge2015}}.
We find two different mechanisms for the
dynamics of fake news and traditional news.
The {top spreaders} of center and left leaning news outlets, who
are mainly journalists, are the main drivers of {Twitter's activity
and in particular of Clinton supporters' activity, who represent the majority in
Twitter\cite{Bovet2017TwitterOpinion}}.
For fake news, we find that it is the activity of Trump supporters that governs
their dynamics and {top spreaders} of fake news are
merely following it.\\

\section{Results}

\subsection{News spreading in Twitter}

\label{sec:fake_news_class}
To characterize the spreading of news in Twitter we analyze all the tweets in our dataset 
that contained at least one URL (Uniform Resource Locator, i.e. web address) linking to 
a website outside of Twitter.
We first separate URL in two main categories based on the websites they link to:
websites containing misinformation and traditional, fact-based, news outlets.
We use the term traditional in the sense that news outlets in this category 
follow the traditional rules of fact-based journalism and therefore 
also include recently created news outlets (e.g. \url{vox.com}).

Classifying news outlets as spreading misinformation 
or real information is a matter of individual judgment and opinion, 
and subject to imprecision and controversy.
We include a finer classification of news outlets
spreading misinformation in two sub-categories: \textit{fake news} 
and \textit{extremely biased news}.
Fake news websites are websites that have been flagged as 
consistently spreading fabricated news or conspiracy theories by several fact-checking groups.
Extremely biased websites include more controversial websites 
that not necessarily publish fabricated information but 
distort facts and may rely on propaganda, decontextualized information, or opinions distorted as facts.
We base our classification of misinformation websites 
on a curated list of websites
which, in the judgment of a media and communication research
team headed by a researcher of Merrimack College, USA, are either fake, false, conspiratorial or misleading
(see Methods).
They classify websites by analyzing several aspects, such as
if they try to imitate existing reliable
websites, if they were flagged by fact-checking groups (e.g. \texttt{snopes.com}, \texttt{hoax-slayer.com} and 
\texttt{factcheck.org}),
or by analyzing the sources cited in articles (the full explanation of their methods
is available at \url{www.opensources.co}).
{\label{rev:fake_calssification}We discard insignificant outlets accumulating less then one percent
of the total number of tweets in their category.
We classify the remaining websites in the extremely biased category according to 
their political orientation
by manually checking the bias report of each websites on
\url{www.allsides.com} and \url{mediabiasfactcheck.com}.
Details about our classification of websites spreading misinformation
is available in the Methods section.\\

We also use a finer classification for traditional news websites based 
on their political orientation.
We identify the most important traditional news outlets by manually inspecting 
the list of top {250} URL's hostnames, representing {79\%} of all URLs, shared 
on Twitter.
We classify news outlets as right, right leaning, center, left leaning 
or left 
based on their reported bias on \url{www.allsides.com} and \url{mediabiasfactcheck.com}.
The news outlets in the right leaning, center and left leaning 
categories are more likely to follow the traditional rules of fact-based journalism.
As we move toward more biased categories, websites are more 
likely to have mixed factual reporting.
As for misinformation websites, we discard insignificant outlets by keeping only websites that accumulate more
than one percent of the total number of tweets of their respective category.
Although we do not know how many news websites
are contained in the list of less popular URLs, a threshold as small as 1\% 
allows us to capture a relatively broad sample of the
media in term of popularity.
Assuming that the decay in popularity of the websites 
in each media category is similar, our measure 
of the proportion of tweets and users in each category
should not be significantly changed if we extended 
our measure to the entire dataset of tweets with URLs.
While the detail of our classification is subject to 
some subjectivity, we find that 
our analysis reveals patterns encompassing
several media categories that form 
group with similar characteristic.
Our results are therefore robust 
to changes of classification within 
these larger group of media.\\

We report the hostnames in each categories along with the number of tweets 
with a URL pointing toward them in \TabHostnames{}.
Using this final separation in seven classes, we identify in our dataset 
(we give the top hostname as en example in parenthesis): {16} hostnames
corresponding to fake news websites (e.g. \url{thegatewaypundit.com}),
{17 hostnames for extremely biased (right) news websites (e.g. \url{breitbart.com})},
{7 hostnames for extremely biased (left) news websites (e.g. \url{dailynewsbin.com})},
{18} hostnames for left news websites (e.g. \url{huffingtonpost.com}), 
{19} hostnames for left leaning news websites (e.g. \url{nytimes.com}),
{13} hostnames for center news websites (e.g. \url{cnn.com}),
{7}  hostnames for right leaning websites (e.g. {\url{wsj.com}})
and {20 hostnames} for right websites (e.g. \url{foxnews.com}).\\

We identified {30.7} million tweets with an URLs directing to a news outlet website, 
sent by {2.3} million users.
An important point when comparing the absolute number of tweets 
and users contributing to the spread of 
different types of news is the bias introduced by the keywords selected during
the data collection.
Indeed, if we had used keywords targeting specific news outlets or hashtags concerning
specific news event, it would be impossible to perfectly control the bias toward fake and 
reliable news or representation of the political orientation of the tweet sample.
Here, we used neutral keywords in term of media representation, the names of the two main 
candidates to the presidential
election (see Methods), in order to collect a sample representative of the real 
coverage of the election on Twitter by all media sources.\\

\begin{table}[b]
{
\footnotesize
\centering
\begin{tabular}{lS[table-format = 7]
		 SS[table-format = 7]
		 SSSSS}
\toprule
{} &        $N_\textrm{t}$ &  $p_\textrm{t}$ &        $N_\textrm{u}$ &  $p_\textrm{u}$ & $N_\textrm{t}/N_\textrm{u}$ & $p_\textrm{t,n/o}$ & $p_\textrm{u,n/o}$ & $N_\textrm{t,n/o}/N_\textrm{u,n/o}$ \\
\midrule
fake news          &  2991073 & 0.10 &   204899 & 0.05 &   14.60 &      0.19 &      0.03 &               80.35 \\
extreme bias (right)     &  3969639 & 0.13 &   294175 & 0.07 &   13.49 &      0.09 &      0.03 &               36.52 \\
right news         &  4032284 & 0.13 &   416510 & 0.10 &    9.68 &      0.11 &      0.04 &               24.80 \\
right leaning news &  1006746 & 0.03 &   272347 & 0.06 &    3.70 &      0.18 &      0.06 &               11.39 \\
center news        &  6322257 & 0.21 &  1032722 & 0.24 &    6.12 &      0.20 &      0.05 &               26.68 \\
left leaning news  &  7491344 & 0.24 &  1272672 & 0.30 &    5.89 &      0.14 &      0.04 &               18.64 \\
left news          &  4353999 & 0.14 &   674744 & 0.16 &    6.45 &      0.14 &      0.05 &               16.64 \\
extreme bias (left)      &   609503 & 0.02 &    99743 & 0.02 &    6.11 &      0.06 &      0.03 &               11.46 \\
\bottomrule
\end{tabular}
}
\caption{\textbf{Tweet and user volume corresponding to each media category in Twitter}. 
Number, $N_\textrm{t}$, and proportion, $p_\textrm{t}$, of tweets with a URL
  pointing to a website belonging to one of the media
  categories.  Number, $N_\textrm{u}$, and proportion, $p_\textrm{u}$, of users having
  sent the corresponding tweets, and average number of tweets per
  user, $N_\textrm{t}/N_\textrm{u}$, for each category.  Proportion of tweets sent by
  non-official clients, $p_\textrm{t,n/o}$, proportion of users having sent
  at least one tweet from an non-official client, $p_\textrm{u,n/o}$, and
  average number of tweets per user sent from non-official clients,
  $N_\textrm{t,n/o}/N_{u,n/o}$.}
\label{tab:url_stats}
\end{table}

\begin{figure}[tb]
\centering
\includegraphics[width=0.5\linewidth]{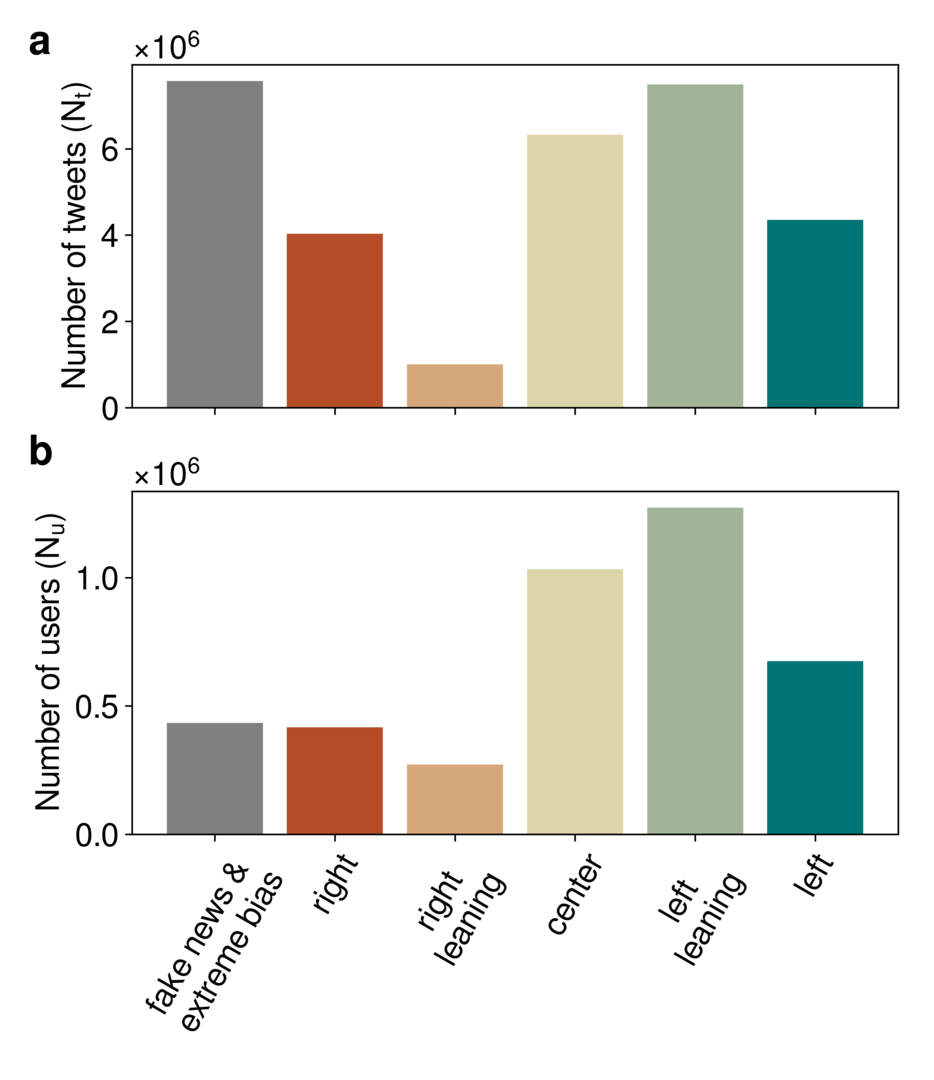}
\caption{\textbf{Importance of different types of
news outlets in Twitter}.
Number of distinct tweets (\textbf{a}) and number of distinct users having sent tweets (\textbf{b}) with a URL
  pointing to a website belonging to one of following categories:
  fake or extremely biased, right, right leaning,
  center, left or left leaning news outlets.
  While the tweet volume of fake and extremely biased news is comparable 
  to the tweet volumes of center and left volume (\textbf{a}), users posting
  fake and extremely biased news are around twice more active in average (see Table\,\ref{tab:url_stats}).
  Consequently, the share of users posting fake and extremely biased
  news (\textbf{b}) is smaller {(12\%)} than the share of tweets directing toward
  fake and extremely biased news websites {(25\%)}.
}
\label{fig:num_tweets_num_users}
\end{figure}

We see a large number of tweets linking to fake news websites and extremely 
biased news websites (Fig.~\ref{fig:num_tweets_num_users}a and Table~\ref{tab:url_stats}).
However, the majority of tweets linking to news outlets points toward
left leaning news websites closely followed by center news websites.
Tweets directing to left and left leaning news websites represent 
together 38\% of the total and tweets directing towards center news outlets
represents 21\%.
Tweets directing to fake and extremely biased news websites
represents a share of {25\%}.
When considering the number of distinct users having sent the tweets instead of the number of tweets
(Fig.~\ref{fig:num_tweets_num_users}b and Table~\ref{tab:url_stats}),
the share of left and left leaning websites increases to {43\%} and the share of center news to 
{29\%},
while the share going to fake news and extremely biased news is equal
to {12\%} (the share of users differ slightly from Table~\ref{tab:url_stats} when grouping categories as users 
may belong to several categories).
The number of tweets linking to websites
producing fake and extremely biased news is comparable with the number for center, left and left
leaning media outlets.
However, users posting links to fake news or extreme bias {(right)} websites are, 
in average, more active than users posting 
links to other news websites (Table\,\ref{tab:url_stats}).
In particular, they post around twice the number of tweets compared to users 
posting links towards center or left leaning news outlets.\\

The proportion of tweets sent by, and users using, non-official Twitter clients
(Table~\ref{tab:url_stats}) allows to evaluate the importance of automated posting in each category.
Details about our classification of official Twitter clients are available in the Methods.
We see that the two top categories are fake news and center news
with around 20\% of tweets being sent from non-official accounts.
When considering the proportion of users sending tweets from non-official clients,
the number are very similar for all categories, around 4\%,
showing that the automation of posting plays an important role across all media categories.
Indeed, non-official clients includes a broad range of clients, from 
``social bots'' to applications used to facilitate the management of professional
Twitter accounts.
A large discrepancy between sources arises when we consider the average
number of tweets per users sent from non-official clients (Table\,\ref{tab:url_stats}).
Users using non-official clients to send tweets with links directing to 
websites displaying fake news tweeted an average of {80} times during the 
collection period, which is more than twice the value for other types
of news outlets.
This high activity from non-official clients suggests an abnormal presence of bots.
The role of bots in the diffusion of fake news has already been documented\cite{Shao2017,Shao2018} 
as well as their presence in the Twitter discussions during 2016 US election\cite{Bessi2016Bots}.\\

{\label{rev:breitblue}We note that Breitbart News
is the most dominant media outlet in
term of number of tweets among the right end of the outlet 
categories with 1.8 million tweets (see \TabHostnames{}).
We examine the relation between Breitbart and the rest of 
the media outlets in Supplementary Note 1,
\TabsBreitBlue{} as well as \FigCausalGraphBreitblue{}.
Our analysis shows that removing Breitbart 
from the extreme bias category does not change our results significantly}.\\

\subsection{Networks of information flow}

To investigate the flow of information we build the retweet networks 
for each category of news websites, i.e. when a user $u$ retweets (a retweet allows a user 
to rebroadcast the tweet of an other user to his followers) the tweet of a user $v$ that contains a URL
linking to a website belonging to one of the news media category, we add a link, or edge, going
from node $v$ to node $u$ in the network.
{The direction of the links represents 
the direction of the information flow between Twitter users.}
We do not consider multiple links with the same direction between the same two users
and neither consider self-links, i.e. when a user retweet her/his own tweet.
The out-degree of a node is its number of out-going links and is equal to the number
of different users that have
retweeted at least one of her/his tweets.
Its in-degree is its number of in-going links and represents the number 
of different users she/he retweeted.\\

\begin{figure}
\centering
\includegraphics[width=0.91\linewidth]{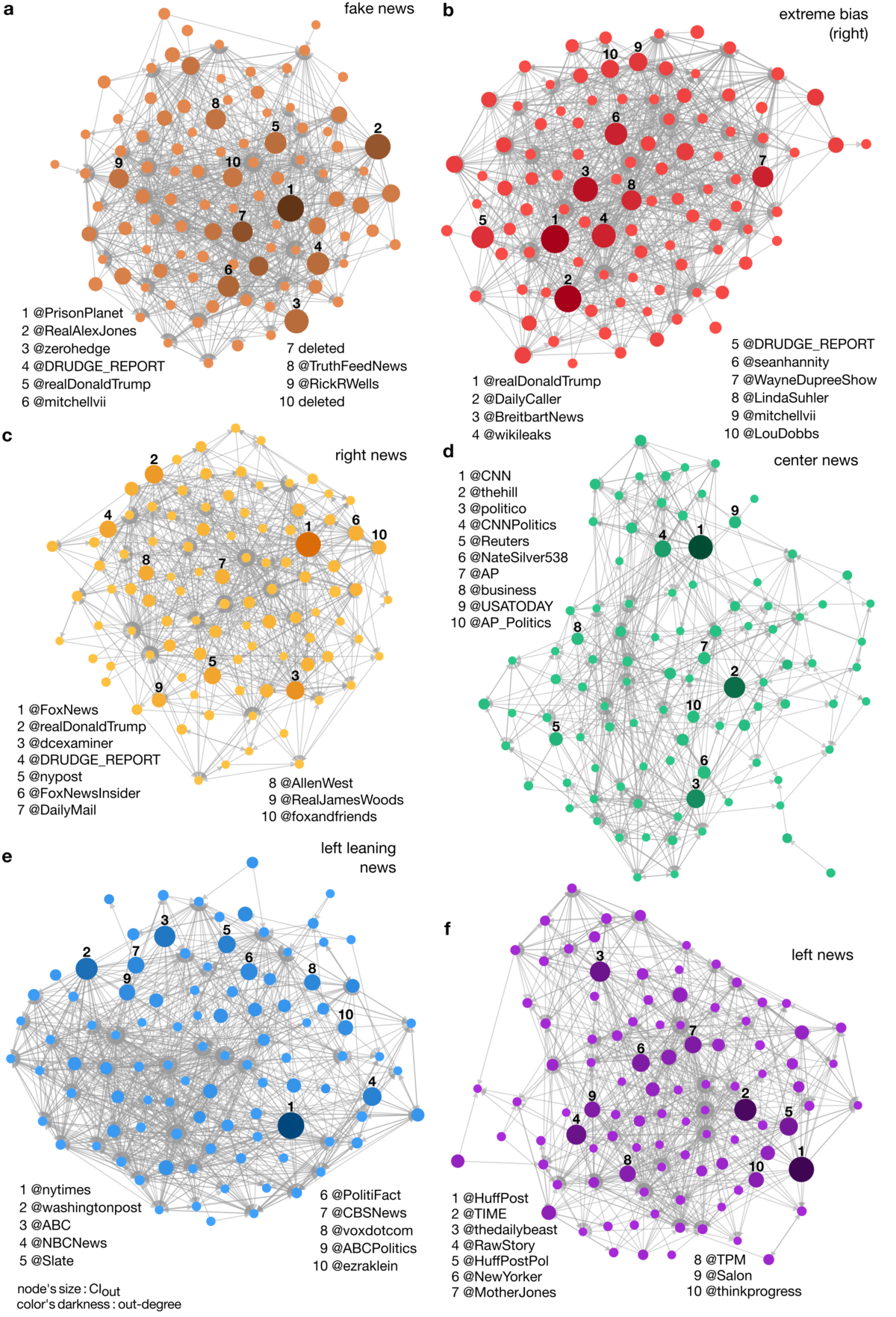}
\caption{\textbf{Retweet networks formed by the top 100 {news spreaders} of 
different media categories}.
Retweet networks for fake news (\textbf{a}), {extreme bias (right)} news (\textbf{b})
right news (\textbf{c}), center news (\textbf{d}), left leaning news (\textbf{e})
and left news (\textbf{f}) 
showing only the top
100 {news spreaders} ranked according to their collective influence.
The direction of the links represents the flow of information
between users.
The size of the nodes is proportional to their 
Collective Influence score, CI$_{\textrm{out}}$,
and the shade of the nodes' color represents their out-degree, 
{i.e. the number of different users that have
retweeted at least one of her/his tweets with a URL directing to a news outlet},
from dark (high out-degree) to light (low out-degree).
The network of fake (\textbf{a}) and {extreme bias (right)} (\textbf{b}) are characterized
by a connectivity that is larger in average and less heterogeneous
than for networks of center and left leaning news (Table\,\ref{tab:retweet_graph_stats}).
}
\label{fig:retweet_networks}
\end{figure}

\begin{table}[b!]
\centering
{
\footnotesize
\begin{tabular}{lS[table-format = 6]S[table-format = 7]S
	         S[separate-uncertainty]
	         S[separate-uncertainty,table-format=1.2(1)]
	         S[table-format = 6]S[table-format = 4]}
\toprule
{} &  {$N$ nodes} &  {$N$ edges} &  {$\left<k\right>$} & {$\sigma(k_{\textrm{out}})/\left<k\right>$} & {$\sigma(k_{\textrm{in}})/<k>$} &  {$\max(k_{\textrm{out}})$} &  {$\max(k_{\textrm{in}})$} \\
\midrule
{fake news}                          &       175605 &      1143083 &     6.51 &                         32 \pm 4 &                   2.49 \pm 0.06 &                       42468 &                       1232 \\
{extreme bias (right)}                          &       249659 &      1637927 &     6.56 &                         36 \pm 6 &                   2.73 \pm 0.03 &                       51845 &                        588 \\
{right}                              &       345644 &      1797023 &     5.20 &                        44 \pm 11 &                   2.70 \pm 0.04 &                       86454 &                        490 \\
{right leaning}                      &       216026 &       495307 &     2.29 &                        45 \pm 11 &                   1.72 \pm 0.02 &                       32653 &                        129 \\
{center}                             &       864733 &      2501037 &     2.89 &                        75 \pm 39 &                   2.69 \pm 0.06 &                      229751 &                        512 \\
{left leaning}                       &      1043436 &      3570653 &     3.42 &                        59 \pm 19 &                   3.38 \pm 0.10 &                      145047 &                        843 \\
{left}                               &       536903 &      1801658 &     3.36 &                        47 \pm 12 &                   3.50 \pm 0.08 &                       58901 &                        733 \\
{extreme bias (left)}                           &        78911 &       277483 &     3.52 &                         33 \pm 6 &                   2.49 \pm 0.08 &                       23168 &                        648 \\
\bottomrule
\end{tabular}
}
\caption{\textbf{Retweet networks characteristics for each news source categories}.
We show the number of nodes and edges (links) of the networks,
the average degree, $\left<k\right>=\left<k_{\textrm{in}}\right>=\left<k_{\textrm{out}}\right>$, (the in-/out-degree 
of a node is the number of in-going/out-going links attached to it).
{In a directed network, the average in-degree and out-degree are always equal.}
The out-degree of a node, i.e. a user, is equal to the number of different users that have
retweeted at least one of her/his tweets.
Its in-degree represents the number of different users she/he retweeted.
The ratio of the standard deviation and the average of the in- and out-degree
distribution, $\sigma(k_{\textrm{in}})/\left<k\right>$ and $\sigma(k_{\textrm{out}})/\left<k\right>$, 
measures the heterogeneity of the connectivity of each networks.
As the standard deviation of heavy-tailed degree distributions can depend on the network size,
we computed the values of $\sigma(k_{\textrm{in}})/\left<k\right>$ and
$\sigma(k_{\textrm{out}})/\left<k\right>$
by taking the average, and standard error, of \SI{1000} independent samples, 
of \SI{78911} values each, drawn from the in-
and out- degree distributions of each network.
}
\label{tab:retweet_graph_stats}
\end{table}

Figure \ref{fig:retweet_networks} shows the networks formed
by the top 100 {news spreaders} of the 6 most important 
retweet networks.
{The retweet networks for right leaning and extreme bias (left) news 
is shown in \FigRTNetLRFL{}.}
We explain in Section (\ref{sec:influencers}) and in the Methods
how the {news spreaders} are identified.
A clear difference is apparent between the networks representing the
flow of fake and {extremely biased (right)} news and the networks for left
leaning and center news (Table~\ref{tab:retweet_graph_stats} and \FigCCDF{}).
The left leaning and center news
outlets correspond to larger networks in term of number of nodes and
edges, revealing their larger reach and influence in Twitter.  However,
the retweet networks corresponding to fake and {extremely biased (right)} news
outlets are the most dense with an average degree $\left<k\right> \simeq 6.5$.
The retweet network for right news has characteristics in between
those two groups with a slightly larger size than the networks for
fake and {extremely biased (right)} news and a larger average degree than
center news.
These results show that users
spreading fake and extremely biased news, although in smaller numbers,
are not only more active in average (Table~\ref{tab:url_stats}),
{\label{rev:connected} but also connected (through retweets) to 
more users in average than users in the traditional news networks.}
Table~\ref{tab:retweet_graph_stats} also shows that the
center and left leaning networks have the most heterogeneous out-degree
distribution and the fake news retweet networks has the less
heterogeneous out-degree distribution.
{\label{rev:hetero_size}We measure the heterogeneity of the 
distribution with a bootstrapping procedure (see Table \ref{tab:retweet_graph_stats})
to ensure the independence of the measure on the networks' sizes.
Our analysis indicates that the larger networks (center, left leaning) 
differ from the smaller ones not just by their size but also by their structure.}
The heterogeneity of the degree
distribution plays an important role in spreading processes on
networks, indicating a strong hierarchical diffusion cascade
from hubs to intermediate degree, and finally to small degree classes\cite{Barthelemy2004,Vespignani2011}.
{\label{rev:weighted_nets}The characteristics of the weighted retweet networks, taking into account 
multiple interactions between users, reveal the same patterns than the unweighted networks
(\TabRTNetStatsComplete{}).}
{\label{rev:degree_CCDF}Table~\ref{tab:retweet_graph_stats} and \FigCCDF{}a reveals the
existence of users with very large out-degree ($k_{\textrm{out}} > 5\times 10^5$),
in the center and left leaning networks,
i.e. very important broadcasters of information,
which are not present in other networks.
This suggests that
different mechanisms of information diffusion could be at play in the
center and left leaning news networks, where high degree nodes may
play a more important role, than in the fake and extremely biased news
networks.\\

{\label{rev:subgraphs}We note that a difference between the largest
networks, i.e center and left leaning news, and the fake and extremely
biased networks is that the former have typically access to more broadcasting technologies,
which may be disruptive to 
understanding diffusion patterns based on network data\cite{Goel2012}.
The structural differences we observe may be explained by the fact that 
there is something different about the way that the 
people in these networks organize and share information but it
may also be the case that there are subgroups of users in the 
center and left leaning news networks that form diffusion networks with a
similar structure as the smaller fake and extremely biased news networks and
then also have a large number of other individuals added to these subgroups due to
the presence of important broadcast networks that feed their ideology or information 
needs.}\\

While inspecting specific accounts is not the goal of this study,
looking at the two accounts with the maximum $k_{\textrm{out}}$ and  $k_{\textrm{in}}$
reveals an interesting contrast between
users of both networks.
The user with the largest out-degree of the center news network
is the verified account of the Cable News Network, CNN, (\texttt{@CNN}), 
which regularly posts links towards its own website using mainly
the non-official professional client Sprinklr (\texttt{www.sprinklr.com}).
The user with the largest in-degree of the fake news network is
the user \texttt{@Patriotic\_Folks}, which, at the moment of this 
writing, seems to belong to a deceiving user, 
whose profile description contains the hashtag \textit{\#MAGA} and refer to
a website belonging to our fake news website list (\texttt{thetruthdivision.com}).
The name of the 
account is ``Annabelle Trump'' and its profile
picture is a young woman wearing cow-boy 
clothes (a reverse image search on the web reveals
that this profile image is not authentic as it comes in fact
from the catalog of a website selling western clothes). 
Most of its tweet are sent from the official Twitter Web Client,
suggesting that a real person is managing the account, and contains URLs
directing to the same fake news website.
However, having a high in-degree does not indicate that this user has 
an important influence. Indeed, its out-degree is approximately 3.5 times 
smaller than its in-degree and, as we explain in the next section, 
influence is poorly measured by local network properties such as in- or out-degree.

\subsection{Top news spreaders}
\label{sec:influencers}

\begin{figure}
\centering
\includegraphics[width=0.55\linewidth]{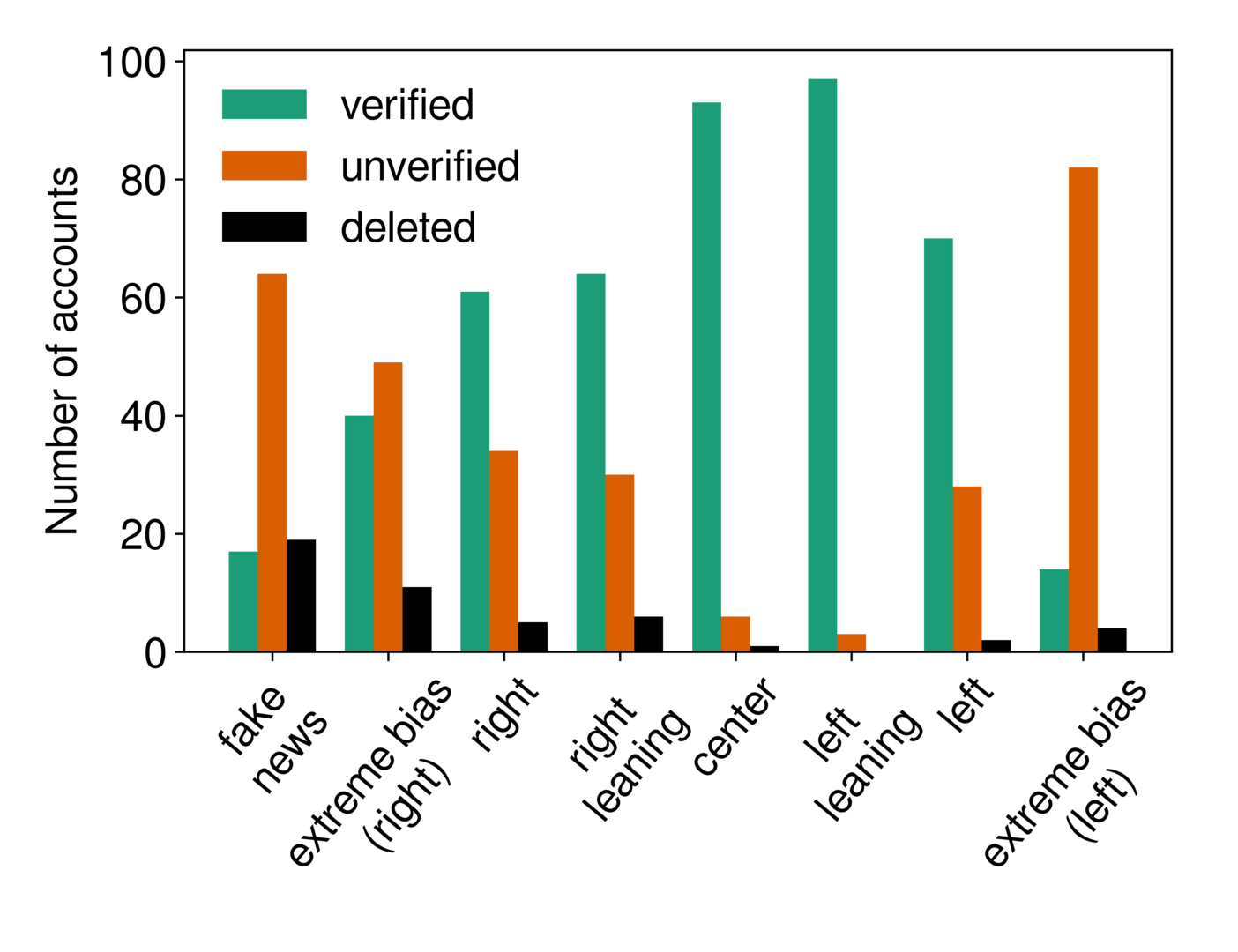}
\caption{\textbf{Types of top {news spreaders} accounts per media category}.
Proportion of verified (green), unverified (orange) and deleted (black)
accounts among the top 100 {news spreaders} in each media category.}
\label{fig:bar_plot_influencers}
\end{figure}

\begin{figure}
\centering
\includegraphics[width=\linewidth]{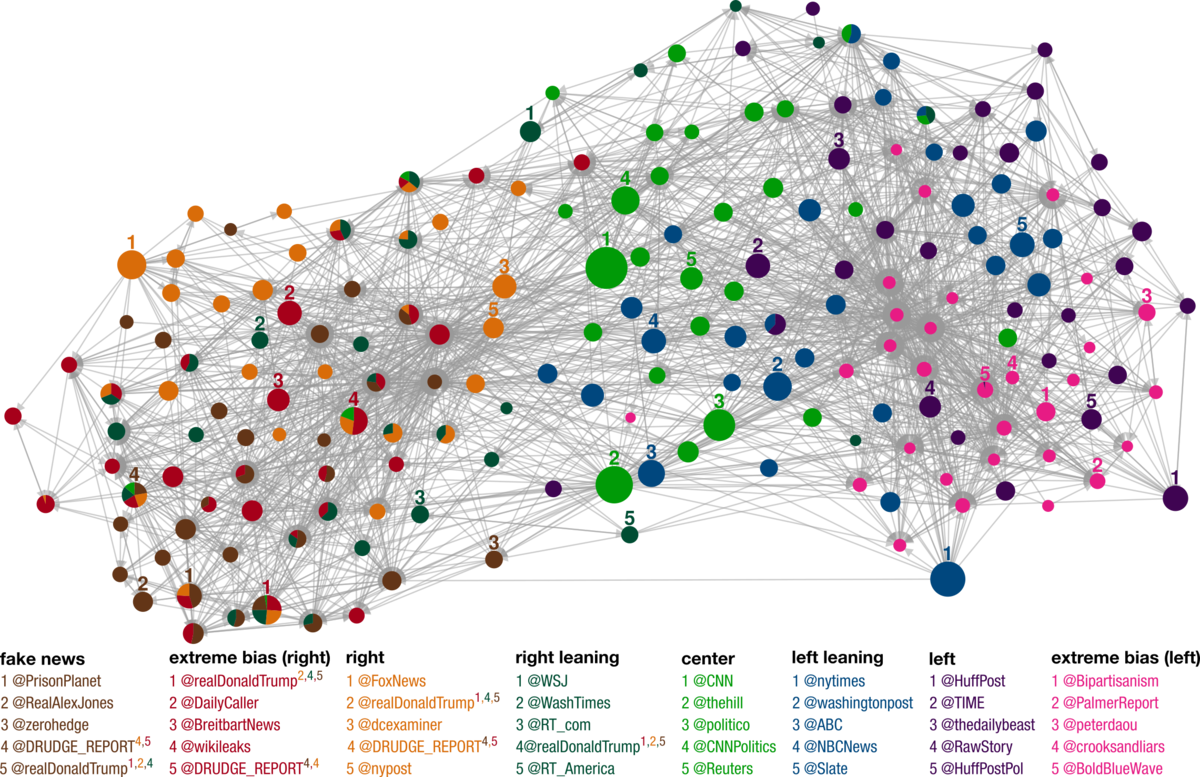}
\caption{\textbf{Retweet network formed by the top 30 influencers of 
each media category}.
The direction of the links represents the flow of information
between users.
The size of the nodes is proportional to their out-degree in the complete combined network,
i.e. the number of different users that have retweeted at least one of her/his tweets
with a URL directing to a news outlet,
and the color of the nodes indicates to which news category they belong.
Nodes that belong to several news categories are represented by pie charts
where the size of each slice is proportional to their CI$_{\textrm{out}}$ 
ranking, taking into accounts only their rank among the top 30.}
\label{fig:retweet_networks_combined}
\end{figure}

In order to uncover the most influential users of each retweet network, we use the 
Collective Influence (CI) algorithm\cite{Morone2015} which is based on the
solution of the optimal network percolation.
For a Twitter user to be highly ranked by the CI algorithm, she/he does not necessarily
need to be directly retweeted by many users, but she/he needs to be surrounded 
by highly retweeted users (see Methods for more details).\\

\begin{table}[htb!]
 \scriptsize
\centering

\begin{tabular}{rllll}
\toprule
{rank} &                 fake news                           &       extreme bias (right) news                                  &                         right news                    & right leaning news \\
       &  (7 verified, 2 deleted,                           &     (15 verified, 1 deleted,                           &              (23 verified, 0 deleted,                 &   (20 verified, 1 deleted                \\
       &                           16 unverified)             &                             9 unverified)              &                                      2 unverified)    &                           4 unverified)  \\
\midrule
1  &     {@PrisonPlanet}\,\checkmark &  {@realDonaldTrump}\,\checkmark &          {@FoxNews}\,\checkmark &              {@WSJ}\,\checkmark \\
2  &    {@RealAlexJones}\,\checkmark &      {@DailyCaller}\,\checkmark &  {@realDonaldTrump}\,\checkmark &        {@WashTimes}\,\checkmark \\
3  &                       {@zerohedge} &    {@BreitbartNews}\,\checkmark &       {@dcexaminer}\,\checkmark &           {@RT\_com}\,\checkmark \\
4  &                   {@DRUDGE\_REPORT} &        {@wikileaks}\,\checkmark &                   {@DRUDGE\_REPORT} &  {@realDonaldTrump}\,\checkmark \\
5  &  {@realDonaldTrump}\,\checkmark &                   {@DRUDGE\_REPORT} &           {@nypost}\,\checkmark &       {@RT\_America}\,\checkmark \\
6  &      {@mitchellvii}\,\checkmark &      {@seanhannity}\,\checkmark &   {@FoxNewsInsider}\,\checkmark &      {@WSJPolitics}\,\checkmark \\
7  &                         deleted                        &  {@WayneDupreeShow}\,\checkmark &        {@DailyMail}\,\checkmark &                   {@DRUDGE\_REPORT} \\
8  &                   {@TruthFeedNews} &                     {@LindaSuhler} &        {@AllenWest}\,\checkmark &   {@KellyannePolls}\,\checkmark \\
9  &                      {@RickRWells} &      {@mitchellvii}\,\checkmark &   {@RealJamesWoods}\,\checkmark &        {@TeamTrump}\,\checkmark \\
10 &                         deleted                        &         {@LouDobbs}\,\checkmark &    {@foxandfriends}\,\checkmark &         {@LouDobbs}\,\checkmark \\
11 &    {@gatewaypundit}\,\checkmark &     {@PrisonPlanet}\,\checkmark &        {@foxnation}\,\checkmark &  {@rebeccaballhaus}\,\checkmark \\
12 &                        {@infowars} &   {@DonaldJTrumpJr}\,\checkmark &         {@LouDobbs}\,\checkmark &       {@WSJopinion}\,\checkmark \\
13 &                   {@Lagartija\_Nix} &                  {@gerfingerpoken} &   {@KellyannePolls}\,\checkmark &      {@reidepstein}\,\checkmark \\
14 &   {@DonaldJTrumpJr}\,\checkmark &       {@FreeBeacon}\,\checkmark &    {@JudicialWatch}\,\checkmark &                         deleted                        \\
15 &                   {@ThePatriot143} &                 {@gerfingerpoken2} &     {@PrisonPlanet}\,\checkmark &  {@JasonMillerinDC}\,\checkmark \\
16 &                     {@V\_of\_Europe} &        {@TeamTrump}\,\checkmark &        {@wikileaks}\,\checkmark &       {@DanScavino}\,\checkmark \\
17 &                  {@KitDaniels1776} &                  {@Italians4Trump} &        {@TeamTrump}\,\checkmark &     {@PaulManafort}\,\checkmark \\
18 &                  {@Italians4Trump} &       {@benshapiro}\,\checkmark &    {@IngrahamAngle}\,\checkmark &         {@SopanDeb}\,\checkmark \\
19 &                        {@\_Makada\_} &   {@KellyannePolls}\,\checkmark &    {@marklevinshow}\,\checkmark &                      {@asamjulian} \\
20 &                    {@BigStick2013} &       {@DanScavino}\,\checkmark &        {@LifeZette}\,\checkmark &    {@JudicialWatch}\,\checkmark \\
21 &  {@conserv\_tribune}\,\checkmark &                         deleted                        &         {@theblaze}\,\checkmark &                        {@\_Makada\_} \\
22 &                     {@Miami4Trump} &                 {@JohnFromCranber} &      {@FoxBusiness}\,\checkmark &          {@mtracey}\,\checkmark \\
23 &                     {@MONAKatOILS} &                     {@true\_pundit} &  {@foxnewspolitics}\,\checkmark &                  {@Italians4Trump} \\
24 &                        {@JayS2629} &                   {@ThePatriot143} &                    {@BIZPACReview} &        {@Telegraph}\,\checkmark \\
25 &                      {@ARnews1936} &                        {@RealJack} &   {@DonaldJTrumpJr}\,\checkmark &    {@RealClearNews}\,\checkmark \\

\midrule
{rank} &           center news &                    left leaning news                                            &      left news  &                                     extreme bias (left) news \\
       &          (24 verified, 0 deleted, &        (25 verified, 0 deleted                                     &      (25 verified, 0 deleted, &                           (7 verified, 1 deleted, \\
       &           1 unverified)           &         0 unverified)                                              &      0 unverified) &                                   17 unverified)   \\
\midrule
1  &              {@CNN}\,\checkmark &         {@nytimes}\,\checkmark &       {@HuffPost}\,\checkmark &   {@Bipartisanism}\,\checkmark \\
2  &          {@thehill}\,\checkmark &  {@washingtonpost}\,\checkmark &           {@TIME}\,\checkmark &    {@PalmerReport}\,\checkmark \\
3  &         {@politico}\,\checkmark &             {@ABC}\,\checkmark &  {@thedailybeast}\,\checkmark &       {@peterdaou}\,\checkmark \\
4  &      {@CNNPolitics}\,\checkmark &         {@NBCNews}\,\checkmark &       {@RawStory}\,\checkmark &  {@crooksandliars}\,\checkmark \\
5  &          {@Reuters}\,\checkmark &           {@Slate}\,\checkmark &    {@HuffPostPol}\,\checkmark &                   {@BoldBlueWave} \\
6  &    {@NateSilver538}\,\checkmark &      {@PolitiFact}\,\checkmark &      {@NewYorker}\,\checkmark &       {@Shareblue}\,\checkmark \\
7  &               {@AP}\,\checkmark &         {@CBSNews}\,\checkmark &    {@MotherJones}\,\checkmark &                         {@Karoli} \\
8  &         {@business}\,\checkmark &       {@voxdotcom}\,\checkmark &            {@TPM}\,\checkmark &                  {@RealMuckmaker} \\
9  &         {@USATODAY}\,\checkmark &     {@ABCPolitics}\,\checkmark &          {@Salon}\,\checkmark &                   {@GinsburgJobs} \\
10 &      {@AP\_Politics}\,\checkmark &       {@ezraklein}\,\checkmark &  {@thinkprogress}\,\checkmark &                    {@AdamsFlaFan} \\
11 &  {@FiveThirtyEight}\,\checkmark &     {@nytpolitics}\,\checkmark &           {@mmfa}\,\checkmark &                       {@mcspocky} \\
12 &        {@bpolitics}\,\checkmark &        {@guardian}\,\checkmark &        {@joshtpm}\,\checkmark &    {@Shakestweetz}\,\checkmark \\
13 &       {@jaketapper}\,\checkmark &     {@NYDailyNews}\,\checkmark &          {@MSNBC}\,\checkmark &                        deleted                        \\
14 &                   {@DRUDGE\_REPORT} &         {@latimes}\,\checkmark &          {@NYMag}\,\checkmark &                        {@JSavoly} \\
15 &           {@cnnbrk}\,\checkmark &    {@BuzzFeedNews}\,\checkmark &       {@samstein}\,\checkmark &                {@OccupyDemocrats} \\
16 &  {@businessinsider}\,\checkmark &        {@Mediaite}\,\checkmark &      {@JuddLegum}\,\checkmark &                   {@ZaibatsuNews} \\
17 &            {@AC360}\,\checkmark &  {@HillaryClinton}\,\checkmark &       {@mashable}\,\checkmark &                       {@wvjoe911} \\
18 &             {@cnni}\,\checkmark &      {@nytopinion}\,\checkmark &   {@theintercept}\,\checkmark &    {@DebraMessing}\,\checkmark \\
19 &     {@brianstelter}\,\checkmark &     {@CillizzaCNN}\,\checkmark &    {@DavidCornDC}\,\checkmark &                     {@SayNoToGOP} \\
20 &   {@KellyannePolls}\,\checkmark &           {@MSNBC}\,\checkmark &       {@dailykos}\,\checkmark &                    {@coton\_luver} \\
21 &        {@wikileaks}\,\checkmark &           {@KFILE}\,\checkmark &     {@JoyAnnReid}\,\checkmark &                     {@EJLandwehr} \\
22 &         {@SopanDeb}\,\checkmark &     {@TheAtlantic}\,\checkmark &     {@nxthompson}\,\checkmark &                        {@mch7576} \\
23 &            {@KFILE}\,\checkmark &        {@SopanDeb}\,\checkmark &      {@thenation}\,\checkmark &                        {@RVAwonk} \\
24 &         {@BBCWorld}\,\checkmark &     {@Fahrenthold}\,\checkmark &      {@justinjm1}\,\checkmark &                         {@\_Carja} \\
25 &           {@NewDay}\,\checkmark &        {@BuzzFeed}\,\checkmark &    {@ariannahuff}\,\checkmark &                    {@Brasilmagic} \\
\bottomrule
\end{tabular}

\caption{\textbf{Top 25 CI {news spreaders} of the retweet networks corresponding to each media category}. 
Verified users have a checkmark (\checkmark) next to their user name.
Verifying its accounts
is a feature offered by Twitter, that ``lets people know that an
account of public interest is
authentic'' (\url{help.twitter.com/en/managing-your-account/about-twitter-verified-accounts}).
Unverified accounts do not have a checkmark and accounts marked as \textit{deleted} have been deleted, either by Twitter
or by the users themselves.
}
 \label{tab:influencers}
\end{table}

We find that top {news spreaders} of left leaning and center news are
almost uniquely verified accounts belonging to news outlets or
journalists (Table~\ref{tab:influencers}).
A very different situation for {news spreaders} of the fake
news and extremely biased news websites is revealed, where, among
verified accounts of news websites and journalists, we also find a large
number of unknown, unverified, users that are not public figures but
are important {news spreaders} in Twitter (Fig.\,\ref{fig:bar_plot_influencers} and Table~\ref{tab:influencers}).
We also find deleted accounts,
that could have been deleted either by Twitter for infringing their
rules and policies or by the users themselves, mostly in the fake and
extremely biased news spreaders.
{\label{rev:deleted_accounts}We find that, based on the timestamp of their last 
tweet in our dataset, 
24 out of the 28 accounts had tweeted after election day (November 8th, 2016)
indicating that they were deleted after the election.
Deleted accounts were extremely active, 
with a median number of tweets of 2224 (minimum: 156, 1st quartile: 1400, 
3rd quartile 6711 and maximum: 15930).
In comparison, the median number of tweets per users for our entire dataset is 2.
We also find that 21 deleted accounts used an unofficial Twitter client
(the most used one by deleted accounts is \texttt{dlvr.it}).}
The list of the right, right
leaning and left news top {spreaders} form a mix of verified and unverified
accounts.
Figure \ref{fig:retweet_networks} shows the retweet networks formed by 
the top 100 spreaders of each category and Fig.~\ref{fig:retweet_networks_combined} 
shows the combined retweet network formed by top 30 news spreaders 
of all media categories and reveals the separation of
the top news spreaders in two main clusters and the relative importance
of the top spreaders. 
{The sets of top 100 fake news, extremely biased (right), right and right leaning 
{news spreaders} have an important overlap, $>30$
(Fig.  \ref{fig:retweet_networks_combined} and \TabInflOverlap{}).}
{\label{rev:verified_accounts}Fake and extremely biased news is mostly spread by
unverified accounts which could be due to the fact that some
accounts are trying to hide their real identity but also
to the fact that audiences of the fake and extremely biased news
are more likely to listen to ``non-public'' figures
due to their distrust of the establishment.}\\

We distinguish three types of unverified accounts:
1) unverified accounts that are not necessarily misleading or deceiving, for example
@zerohedge, @DRUDGE\_REPORT or @TruthFeedNews make their affiliation to their respective news websites
clear, although their identities or the ones of their websites administrators is not always clear;
2) unverified accounts that make their motif clear in their choice of screen-name,
e.g. @Italians4Trump or @Miami4Trump, although the real identity of the persons
behind such accounts is also usually undisclosed;
3) finally, unverified accounts that seem to be real persons with profile pictures and user names,
e.g @Lagartija\_Nix, @ThePatriot143, @BigStick2013, @LindaSuhler, @gerfingerpoken or @AdamsFlaFan, but
are not public figures. 
Whether such users are authentic, social bots or fake users operated by someone else is
not clear.
However, our results show that such users are not present in the top {news spreaders} of
the center and left leaning news,
while they have a high prevalence in the fake and extremely biased categories.

{\label{rev:camp_staff}Another observation is the presence 
of members of the campaign staffs 
of each candidate in the top {news spreaders}
(see Supplementary Note 2 and \TabCampStaff{}).
We see more users linked to the 
campaign staff of Donald Trump (13), and with higher ranks in term of influence,
than to the campaign staff of Hillary Clinton (3), revealing the more
important direct role of the Trump team in the diffusion of news in Twitter.}

\subsection{News spreading dynamics}
\label{sec:news_dynamics}

To investigate the news spreading dynamics of the different media
categories on Twitter, we analyze the correlations between the time
series of tweeting rate measured for each category. The Twitter
activity time series are constructed by counting the number of tweets
with a URL directing toward a website belonging to each of the media
category at a 15 minute resolution. In addition to the activity related to
each media group, we also consider the time series of the activity of
the supporters of each presidential candidates.
We classify supporters based on the content of their tweets using a supervised
machine learning algorithm trained on a dataset
obtained from the network of hashtag co-occurrences.
The full detail of our method and the validation 
of its opinion trend with the national polling average of the
New York Times is described in ref. \cite{Bovet2017TwitterOpinion}.
We use our full dataset of 
tweets concerning the two candidates, namely 171 million tweets 
sent by 11 million distinct users during more than five months.
After removing automated tweets (see Methods),
we have a total of 157 million tweets.
This represents an average of 1.1 million tweets per day 
(standard deviation of 0.6 million) sent by an average of 
about 375,000 distinct users per day (standard deviation of 190,000).
A majority of users, 64\%, is in favor of Hillary Clinton while 28\% is in favor of Donald Trump
(8\% are unclassified as they have the same number of tweets in each camp).
However, we find that Trump supporters are, in average, 1.5 times more active 
than Clinton supporters\cite{Bovet2017TwitterOpinion}.
The supporters therefore represent the general Twitter population
commenting on the candidate of the election.\\

We removed the trend and circadian cycles present in the time series with the
widely used STL (seasonal-trend decomposition procedure based on Loess) method\cite{Cleveland1990}, which is 
a robust iterative filtering method allowing to separate a time series in seasonal (in this case, daily),
trend and remainder components (see Methods).

\begin{figure}[tb]
\centering
\includegraphics[width=0.47\linewidth]{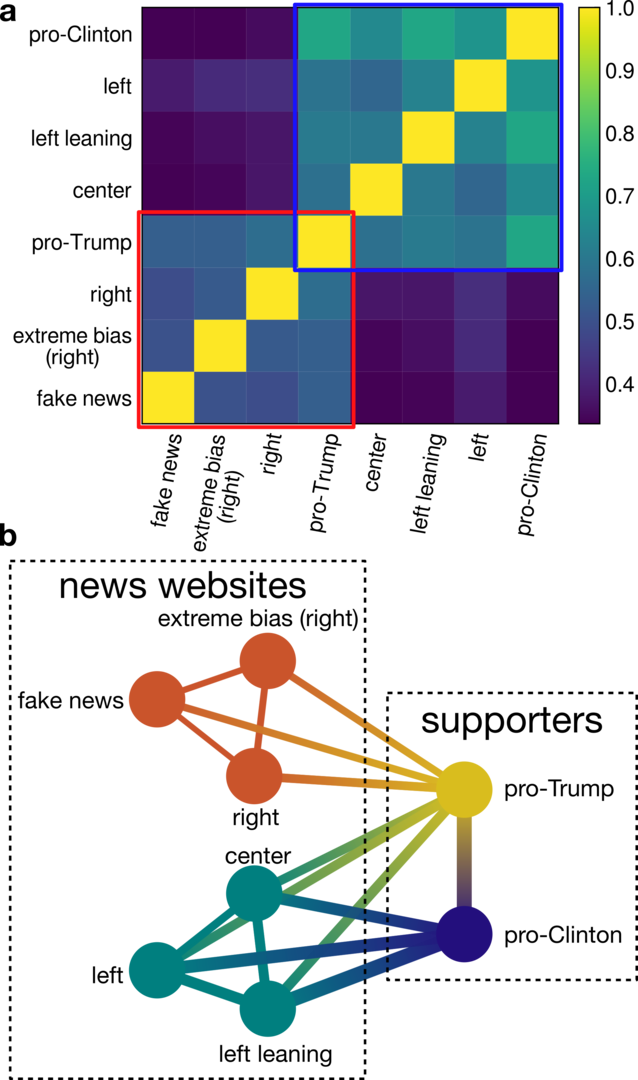}
\caption{
\textbf{Activity correlation between news outlets and supporters}.
(\textbf{a}) Pearson cross-correlation coefficients between activity time series
related to the different types of news outlets,
Trump supporters and Clinton supporters.
(\textbf{b}) Graph showing the correlation relations between 
the types of news websites and the supporters.
The edges of the graph represent correlations larger than
$r_0 = 0.49$.
Fake news, {extreme bias (right)} and right websites form a first cluster,
indicated by a red square in (\textbf{a}) and shown in orange in (\textbf{b}),
while center, left leaning and left news websites
form a second cluster, indicated by a blue square in (\textbf{a}) and shown in blue in (\textbf{b}).
The activity of Trump supporters is equally correlated with all news sources
and the activity of Clinton supporters, which represents the largest activity,
is mainly
correlated with the second media cluster and only poorly with the first one.
}
\label{fig:tweet_vol_cross_cor}
\end{figure}

The separation of the media sources in two correlated clusters
is revealed when 
using a threshold of $r_0 = 0.49$, corresponding to 
the place of the largest gap between the sorted correlation values
(Fig.~\ref{fig:tweet_vol_cross_cor}).
{The value of each cross-correlation coefficient is
reported in \TabCorrVals{}.}
The first activity cluster (indicated by a red square in Fig.~\ref{fig:tweet_vol_cross_cor}a)
comprises the fake, {extreme bias (right)} and right leaning news. 
The second activity cluster (indicated by a blue square) 
is made of the center, left and left leaning news sources.
The activities of right leaning and extremely biased (left) news are only 
poorly correlated with the other news categories or 
supporters (see \TabCorrVals{}).
We observe the following patterns between the media groups and the supporters dynamics:
the activity of Clinton supporters has a higher correlation with the second cluster 
than with the first one while
the activity of Trump supporters is equally correlated with the two clusters.
{\label{rev:corr_patterns}This indicate that Trump supporters are likely 
to react to any type of news while Clinton 
supporters mostly react to center and news on the left
and tend to ignore news coming from the right side.}

{\label{rev:polarization}These results indicate that the media 
included in the two clusters respond to 
two different news dynamics and show that 
the polarization of news observed at the structural level in 
previous works\cite{Bakshy2015,Schmidt2017,DelVicario2017Brexit}
also corresponds to a separation in dynamics.
This separation could be showing that Americans with different 
political loyalties prefer different news sources but could
also be due to the fact that supporters prefer the news that their candidate
prefers\cite{Margolin2018}.}\\

In order to investigate the causal relations between news media sources
and Twitter dynamics, {\label{rev:causal_algo}we use a multivariate causal network reconstruction 
of the links between the activity of top news {spreaders} and supporters of the presidential 
candidates based on a causal discovery algorithm\cite{Spirtes2000,Runge2012,Runge2015}.
The causal network reconstruction tests the
independence of each pair of time-series, for several time lags, 
conditioned on potential causal parents with a 
non-parametric conditional independence test\cite{Zhang2011,Strobl2017}
(see Methods).
We use the causal algorithm as a variable selection
and perform a regression of a linear model using only
the true causal link discovered.
We consider linear causal effects
for their reliable estimation and interpretability.
This permits us to compare the causal effect 
as first order approximations, estimate the uncertainties
of the model and reconstruct a causal directed weighted networks\cite{Runge2015}.
In this framework, the causal effect between a time series $X^i$ and $X^j$ 
at a time delay $\tau$, $I^{\textrm{CE}}_{i\rightarrow j} (\tau)$, is equal 
to the expected value of $X^j_t$ (in unit of standard deviation) 
if $X^i_{t-\tau}$ is perturbed by one standard deviation\cite{Runge2015}.

\label{rev:causal_suff}An assumption of causal discovery is 
causal sufficiency, i.e. the 
fact that every common cause of any two or more variables is in the system\cite{Spirtes2000}.
Here, causal sufficiency is not satisfied since Twitter's activity
is only the observed part of a larger social system
and the term ``causal'' must be understood to be meant 
relative to the system under study.}
As for the cross-correlation analysis, we use the residuals of the STL
filtering of the \SI{15}{\minute} tweet volume time series (Fig.~\ref{fig:causal_graph}a-b).

\begin{figure}[tb]
\centering
\includegraphics[width=\linewidth]{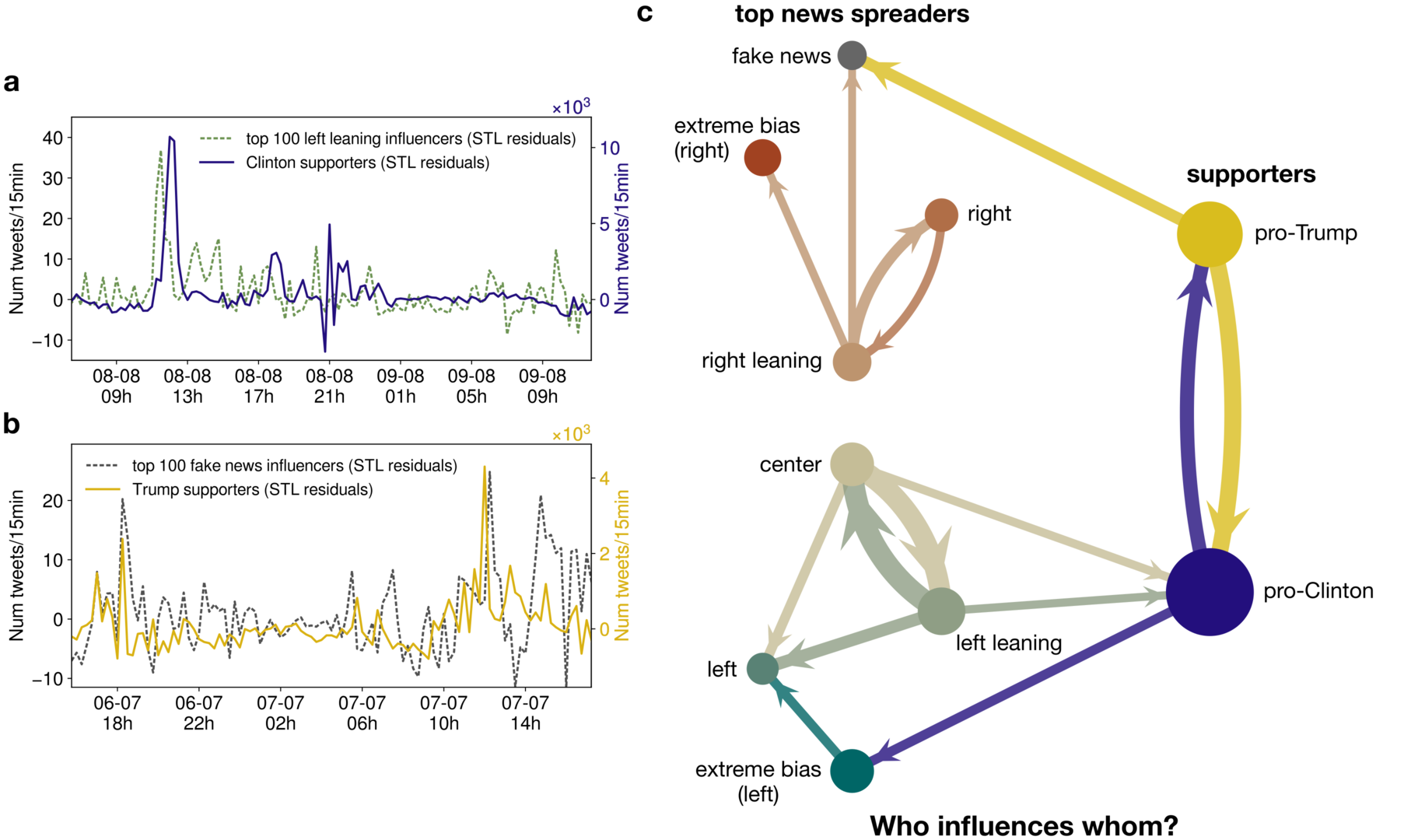}
\caption{{\textbf{Granger causal network reconstruction between top news spreaders and supporters activity}.
(\textbf{a}) Activity time series corresponding to the top 100 left leaning {news spreaders}
{(dashed)} and the Clinton supporters {(continuous, right vertical axis)}.
(\textbf{b}) Activity time series of the top 100 fake news {spreaders} (dashed) and 
the Trump supporters (continuous, right vertical axis). 
We show the residuals of the STL filtering after the removal of the seasonal (daily) 
and trend components.
A {causal effect} seems apparent from the top 100 left leaning {news spreaders}
to the Clinton supporters (\textbf{a}).
Peaks in the left leaning {news spreaders} activity {(yellow, dashed)} tend to precede peaks in 
the activity of Clinton supporters (blue).
A {causal effect} relation from the Trump supporters to 
the top 100 fake news {news spreaders} (\textbf{b}) seems also apparent.
(\textbf{c}) Graph showing the maximal causal effects between the activity of the top 100 news spreaders
of each media category (left)
and the activity of the presidential candidate supporters (right) computed over the entire five months.
Arrows indicate the direction of a the maximal causal effect ($>0.05$) between two activity time series.
The width of each arrow is proportional to the strength of the causation and
the size of each node is proportional to the auto-correlation of each time series.
The center and left leaning top news spreaders are the news spreaders that 
show the strongest causal effect on the supporters activity.
The values of the causal effects between each activity time series are
shown in Table\,\ref{tab:causal_effects}}.}
\label{fig:causal_graph}
\end{figure}

\begin{table}
{
\footnotesize
\sisetup{detect-weight=true,detect-inline-weight=math}
\begin{tabular}{l*{5}{S[separate-uncertainty,table-format=1.4(1)]}}
\toprule
\multicolumn{1}{r}{$\swarrow$} &    {pro-Clinton} &      {pro-Trump} &      {fake news} &        {extreme bias (right)} &            {right} \\
\midrule
{pro-Clinton}   &  		 0.65 \pm 0.01 &  \bfseries  0.14 \pm 0.01 &  0.029 \pm 0.007 &    0.021 \pm 0.006 &    0.002 \pm 0.006 \\
{pro-Trump}     &  \bfseries  0.11 \pm 0.02 &    0.46 \pm 0.01 &  0.009 \pm 0.006 &    0.003 \pm 0.001 &  0.0014 \pm 0.0009 \\
{fake news}     &  		0.015 \pm 0.003 &  \bfseries  0.10 \pm 0.01 &    0.14 \pm 0.01 &   \bfseries   0.05 \pm 0.01 &      0.03 \pm 0.01 \\
{extreme bias (right)}     &  		  0.02 \pm 0.01 &  0.009 \pm 0.002 &    0.03 \pm 0.01 &      0.21 \pm 0.01 &  \bfseries    0.04 \pm 0.01 \\
{right}         & 		 0.009 \pm 0.002 &  0.025 \pm 0.008 &    0.03 \pm 0.01 &      0.02 \pm 0.01 &      0.18 \pm 0.01 \\
{right leaning} & 		 0.018 \pm 0.008 & \bfseries 0.038 \pm 0.008 &    0.02 \pm 0.01 &      0.01 \pm 0.01 &   \bfseries   0.07 \pm 0.01 \\
{center}        &   \bfseries 0.04 \pm 0.01 & \bfseries 0.023 \pm 0.010 &  0.021 \pm 0.007 &  0.0020 \pm 0.0007 &    0.009 \pm 0.008 \\
{left leaning}  &   \bfseries	 0.04 \pm 0.01 & \bfseries 0.015 \pm 0.006 &  0.003 \pm 0.001 &  0.0010 \pm 0.0005 &    0.009 \pm 0.007 \\
{left}          &   		 0.03 \pm 0.01 &    0.03 \pm 0.01 &  0.010 \pm 0.008 &    0.002 \pm 0.001 &      0.01 \pm 0.01 \\
{extreme bias (left)}      &    \bfseries  0.08 \pm 0.01 &    0.03 \pm 0.02 &  0.031 \pm 0.009 &      0.03 \pm 0.01 &  0.0025 \pm 0.0008 \\

\midrule

\multicolumn{1}{r}{$\swarrow$} &    {right leaning} &         {center} &   {left leaning} &             {left} &         {extreme bias (left)} \\
\midrule
{pro-Clinton}   &    0.003 \pm 0.001 &  \bfseries 0.065 \pm 0.008 & \bfseries 0.062 \pm 0.008 &    0.017 \pm 0.009 &    0.006 \pm 0.006 \\
{pro-Trump}     &  0.0020 \pm 0.0009 &  \bfseries 0.038 \pm 0.006 &  \bfseries 0.033 \pm 0.008 &    0.020 \pm 0.007 &    0.015 \pm 0.006 \\
{fake news}     &    \bfseries  0.06 \pm 0.01 &  0.037 \pm 0.009 &  0.016 \pm 0.002 &    0.014 \pm 0.008 &    0.022 \pm 0.009 \\
{extreme bias (right)}     &   \bfseries   0.06 \pm 0.01 & \bfseries 0.039 \pm 0.009 &  0.018 \pm 0.002 &    0.026 \pm 0.009 &    0.027 \pm 0.009 \\
{right}         &   \bfseries   0.09 \pm 0.01 & \bfseries 0.044 \pm 0.009 &  0.016 \pm 0.002 &  0.0026 \pm 0.0009 &  \bfseries  0.033 \pm 0.008 \\
{right leaning} &      0.22 \pm 0.01 & \bfseries 0.042 \pm 0.009 &  0.033 \pm 0.009 &  0.0014 \pm 0.0008 &  0.0027 \pm 0.0008 \\
{center}        &    0.012 \pm 0.010 &  0.266 \pm 0.009 &  \bfseries  0.18 \pm 0.01 &    0.019 \pm 0.010 &    0.013 \pm 0.008 \\
{left leaning}  &    0.005 \pm 0.003 &  \bfseries  0.18 \pm 0.01 &  0.299 \pm 0.009 &    0.012 \pm 0.008 &    0.003 \pm 0.002 \\
{left}          &    0.015 \pm 0.008 &  \bfseries  0.08 \pm 0.01 &  \bfseries  0.10 \pm 0.01 &    0.164 \pm 0.010 &    \bfseries  0.07 \pm 0.01 \\
{extreme bias (left)}      &    0.005 \pm 0.009 &  \bfseries 0.034 \pm 0.009 &  \bfseries 0.045 \pm 0.009 &      0.03 \pm 0.01 &      0.27 \pm 0.01 \\
\bottomrule
\end{tabular}
}
\caption{{\textbf{Causal effects between the top spreaders and the candidates supporters}.
We show the value of the maximal causal effect 
$I^{\textrm{CE,max}}_{i\rightarrow j}= \max_{0<\tau\leq\tau_{\textrm{max}}} \left| I^{\textrm{CE}}_{i\rightarrow j} (\tau) \right|$
between each pair $(i,j)$ of activity time series, where $\tau_{\textrm{max}}=18\times15\si{min}=4.5\si{h}$ is the maximal time lag considered, with 
standard errors (s.d., see Methods).
The arrow indicate the direction of the causal effect.
For each activity time series, we indicate in bold the three most important 
drivers of activity (excluding themselves).
}}
\label{tab:causal_effects}
\end{table}

{\label{rev:infl_hypo}We consider only the activity of the top 100
  {news spreaders} since, by definition of CI, they are the most
  important sources of information.  Therefore, within the limitation
  of considering Twitter as a closed system, they are the most likely
  set of users to trigger the activity of the rest of the population.
  We test this hypothesis with Granger causal analysis.}

{Our causal analysis takes into account self-links,
i.e. the auto-correlation of each time series, 
and reveals that they are the strongest causal effect
for all time series. 
Since we are interested in the cross-links, we leave the 
self-links aside for the rest of the discussion.}
The center and left leaning {news spreaders} have the strongest
causation on the supporters activity, with a stronger effect
on the Clinton supporters than on the Trump supporters
(Table~\ref{tab:causal_effects} and Fig.~\ref{fig:causal_graph}c).
Since the Clinton supporters dominate Twitter activity, 
they also are the main drivers of the global activity.
The other top {news spreaders}
have only a small or negligible effect on the supporters activity.
{In particular,  extreme bias (left), left, right leaning and right news 
spreaders are more influenced by the activity of 
Clinton and Trump supporters than the opposite.}
{We also observe that Trump supporters have a significant causal 
effect on the fake news spreaders' activity and Clinton supporters
have a significant effect on extreme bias (left) spreaders' activity
(Fig.~\ref{fig:causal_graph}c).}
This suggests that they are in fact following Twitter activity rather than
driving it.
{\label{rev:inter_causal}Regarding the causal relations in-between news spreaders, 
center news spreaders are the most central driver as
they are among the top three drivers of all news spreaders (Table~\ref{tab:causal_effects}).
Strong mutual causal effects are revealed between center and left leaning spreaders.
Right leaning top spreaders are driving 
the activity of the right, extremely biased (right) and fake news spreaders.
The two supporter groups have also strong mutual 
causal effects.}

These results reveal two very different 
dynamics of news diffusion for traditional, center and left leaning, news
and misinformation.
{Center and left leaning news spreaders are the most influential and 
are driving the supporters activity.
On the other hand, the dynamics of fake news spreaders seems to
be governed by the ensemble of Trump supporters.}

{\label{rev:causal_suff_int}The interpretation of the discovered causal effects
must be understood within the limitation that we
do not measure the diffusion of news outside
of Twitter.
Indeed, the fact that center and left leaning news {spreaders}
have a causal effect on the Clinton supporters could be explained
by the fact that they simply are the first to be ``activated''
by some news appearing, for example on television, 
while the supporters take more time to be ``activated''
by the same news.
However, we have other indications that the {news spreaders} 
are directly causing at least part of the supporters' activity,
namely that the {top news spreaders} are precisely the most
important source of news retweets. 
Moreover, if the external driver is an other media outside of Twitter 
and that the center/left leaning {news spreaders}, who are
almost all journalists, are the first to be activated, 
it is very likely that the media channel outside of 
Twitter is related to the journalists.
In this case, even if the causation is indirect, we still
identify the correct driver through
the affiliation of the journalists.
More importantly, while we
observe a strong causal effect between center/left
leaning {news spreaders} and the supporters, 
we do not observe a significant causal effect
between other {news spreaders} and the supporters.
This indicates that, even if the causal driver 
could be outside of Twitter, the diffusion mechanisms 
of traditional and fake news are very likely 
different.}

We investigate the influence of the presence of staff
members of candidates' teams in Supplementary Note 2,
\FigCausalGraphNoStaff{} and \TabCausalEffectsNoStaff{}.
We observe no significant changes in the causal relations
after having removed all users linked to the campaigns.
We also repeated our analysis after having removed news 
aggregators from our dataset (see Supplementary Note 3,
\FigCausalGraphNoAggr{}, \TabsNotAggr{}) and found that
news aggregators are not responsible for the observed
differences in dynamics.

\section{Discussion}

Using a dataset of tweets collected during the five months preceding the 2016 
presidential elections, we investigated the spread of content classified as fake news 
and compared its importance and influence with traditional, fact-based, media.
We find that fake {news represents 10\%} and {extremely biased news 15\%} of the 
tweets linking to a news outlet media. 
However, taking into account the difference in user activity
decreases the share of fake and extremely biased news to {12\%}.
Although we find approximately the same ratio of users using  
automated Twitter clients in each media category, we find 
that automated accounts diffusing fake news are much more
active than the automated accounts diffusing other types 
of news.
This results confirms the role of bots in the diffusion of fake news,
that has been shown using a different method of bot detection\cite{Shao2017},
and shows that automated accounts also play a role, although smaller,
in the diffusion of traditional news.\\

We analyzed the structure of the information diffusion
network of each category of news and found that 
fake and {extremely biased (right)} news diffusion networks are 
more densely connected,{\label{rev:conn} i.e. in average users retweet more people and are more retweeted},  
and have less heterogeneous 
connectivity distributions than traditional, center and 
left leaning, news diffusion networks.
The heterogeneity of the degree
distribution is known to play an important role in 
in spreading processes on
networks\cite{Barthelemy2004,Vespignani2011}.
Spreading in networks with heterogeneous connectivity
usually follows a hierarchical dynamics in which 
the information propagates from higher-degree to lower-degree classes\cite{Vespignani2011}.\\

We discovered the {top news spreaders} of each type of news by computing
their collective influence\cite{Morone2015} and found 
very different profiles of fake and extremely biased news {top spreaders}
compared to traditional news {spreaders}.
While traditional news {spreaders} are mostly journalists with 
verified Twitter accounts, fake and extremely biased news {top spreaders}
include unverified accounts with deceiving profiles.\\

Analyzing the Twitter activity dynamics of the news diffusion corresponding
to each media category, we reveal the existence
of two main clusters of media in term of activity correlation
which is consistent with the findings of previous works\cite{Bessi2015PlosOne,Bessi2015,Mocanu2015,Bessi2015PlosTwo,Bessi2016,DelVicario2016}
that revealed the separation in polarized communities
of online social media news consumers.
We also show that right news media outlets
are clustered together with fake news.
Finally, a causality analysis between the {top news spreaders}
activity and the activity of presidential candidate supporters
revealed that {top news spreaders} of center and left leaning news outlets
are the ones driving Twitter activity while {top news spreaders} of fake
news are in fact following Twitter activity, 
particularly Trump supporters activity.\\

{\label{rev:media_coverage}Our analysis focuses on news concerning the candidate 
of the presidential election published from the most popular news outlets
and therefore its results cannot be directly generalized to the entire Twitter population.
Nevertheless, our investigation provides new insights into the dynamics of news
diffusion in Twitter.}
Namely, our results suggests that fake and
extremely biased news are governed by a different diffusion mechanisms
than traditional center and left leaning news.  Center and left
leaning news diffusion is driven by a small number of influential
users, mainly journalists, and follow a diffusion cascade in a network
with heterogeneous degree distribution which is typical of diffusion
in social networks\cite{Vespignani2011}, while the diffusion of fake
and extremely biased news seems to not be controlled by a small set of
influencers but rather to take place in {more connected clusters and
to be the result of a collective behavior.}

\section*{Methods}

\subsection*{Twitter data collection and processing}

We collected tweets continuously using the Twitter Search API from
June 1st, 2016 to November 8th, 2016.
We gather a total of  171 million
tweets in the English language, mentioning the two top
candidates from the Republican Party (Donald J. Trump) and Democratic
Party (Hillary Clinton) by using two different queries with the
following keywords: \textit{hillary OR clinton OR hillaryclinton} and
\textit{trump OR realdonaldtrump OR donaldtrump}.\\

We extracted the URLs from tweets by using the 
\texttt{expanded\_url} field attached to each tweet containing at least one URL.
A large number of URL were redirecting links using URL shortening services
(e.g. \texttt{bit.ly}, \texttt{dlvr.it} or \texttt{ift.tt}).
News websites sometimes also uses shortened versions of their 
hostnames (e.g. \texttt{cnn.it}, \texttt{nyti.ms}, \texttt{hill.cm} or \texttt{politi.co}).
We programmatically resolved shortened URLs, using the Python Requests library, 
in order to find their final destination URL and extracted the hostname
of each final URL in our dataset.\\

To identify tweets that may originate from bots, we extract the name of 
the Twitter client used to post each tweet from their \textit{source} field
and kept only tweets originating from 
an official twitter client.
Third-party clients represents a
variety of applications, form applications mainly used by professional
for automating some tasks (e.g. \url{www.sprinklr.com} or
\url{dlvrit.com}) to manually programmed bots, and are used to post
$\leq 8\%$ of the total number of tweets.
When a programmatic access to Twitter is gained 
through its API to send tweets, the value of the \textit{source} field
of automated tweets corresponds to the name, which must be unique, given 
to the ``App'' during the creation of access tokens.
\TabOfficialClients{} shows the
clients we consider as official and the corresponding number of tweets
with URLs originating from each client.
The number of tweets with a URL originating from official clients represents 82\% 
of the total number of tweets with a URL.
This simple method allows to identify tweets that have not been 
automated and scales very easily to large datasets contrary
to more sophisticated methods\cite{Varol2017}.
{\label{rev:botometer}Indeed, Botometer is not well 
suited for historical data as it requires 
several tweets per users (up to 200) and results of 
a Twitter search of tweets (up to 100) mentioning each users,
which we cannot do retroactively.
We compared our method with the results of Botometer 
(see Methods section of ref. \cite{Bovet2017TwitterOpinion})
and found that our method has a good accuracy
but suffer from a relatively high number of false positive
compared to Botometer.}
Advanced bots might not be detected by our method, 
but this is also a problem for more advanced methods that relies 
on a training set of known bots\cite{Varol2017}.
We remove all tweets sent from non-official clients when 
computing the activity of supporters but we keep
them when building the retweet networks, as we 
want to include automated accounts that play
a role in the diffusion of news.

\subsection*{News outlets classification}

Among the 55 million tweets with URLs linking outside of Twitter,
we identified tweets directing to websites containing fake news by matching
the URLs' hostname with a curated list of websites, which, in the judgment of 
a media and communication research team headed by Melissa Zimdars of Merrimack College, USA,
are either fake, false, conspiratorial or misleading.
The list, freely available at \url{www.opensources.co}, 
classifies websites in several categories, such as ``Fake News'',
``Satire'' or ``Junk Science''. 
For our study, we construct two non-overlapping set of websites: \textit{fake news} websites 
and \textit{extremely biased} websites.
The set of fake news website is constructed by joining the hostnames listed under the
categories ``Fake News'' and ``Conspiracy Theory'' by \url{www.opensources.co}.
The following definitions of these {two} categories are given on \url{www.opensources.co}
\begin{itemize}
 \item ``Fake News'': sources that entirely fabricate information, 
 disseminate deceptive content, or grossly distort actual news reports,
 \item ``Conspiracy Theory'': sources that are well-known promoters of kooky conspiracy theories.
\end{itemize}

The set of extremely biased websites contains hostnames appearing in the 
category ``Extreme Bias'' (defined as sources that come from a particular point 
of view and may rely on propaganda, decontextualized information, and opinions distorted as facts
by \url{www.opensources.co})
but not in any of the categories used to construct the set of fake news.
Hostnames in each categories along with the number of tweets 
with a URL pointing toward them are reported in \TabHostnames{}.
We discard insignificant outlets accumulating less then one percent
of the total number of tweets in their category.\\

{\label{rev:extreme_class}Websites classified in the extremely biased (right)
category, respectively extremely biased (left) category,
have a ranking between \textit{right} bias and \textit{extreme right} bias,
respectively \textit{left} and \textit{extreme left}, on \url{mediabiasfactcheck.com}.
The bias ranking on \url{www.allsides.com} of these same websites
is \textit{right}, respectively \textit{left}, (corresponding to 
the most biased categories of \url{www.allsides.com}).
The website \url{mediabiasfactcheck.com} also reports a 
level of factual reporting for each websites and we find that 
all the websites classified in the extremely bias category have a
level of factual reporting which is \textit{mixed} or worse.
We also find that all the websites remaining in the fake news category have a bias 
between \textit{right} and \textit{extreme right} on \url{mediabiasfactcheck.com}.}
The website \url{www.allsides.com} rates media bias using a 
combination of several methods such as blind surveys,
community feedback and independent research 
(see \url{www.allsides.com/media-bias/media-bias-rating-methods} 
for a detailed explanation of the media bias rating methodology used by AllSides),
and \url{mediabiasfactcheck.com} scores media bias by evaluating wording, sourcing 
and story choices as well as political endorsement (see \url{mediabiasfactcheck.com/methodology} for
an explanation of Media Bias Fact Check methodology).}\\

{\label{rev:establishment}A potential issue with the methodology of OpenSources
is the blurring of the assessment of “bias,” which has to do with news content,
with the assessment of “establishment”, which has to do with news form.
Specifically,
their steps 4-6 indicate they count thinks like use of the 
Associated Press style guide and the production quality of 
the website. These criteria thus conflate adherence to establishment
norms --which are likely to be correlated with things like budgets 
for professional design, fact-checking, editorial oversight
etc.-- with lack of bias. That is, if two media sites present
the same news, but one does it in a less established format, 
it may be considered “extremely biased.”
For this reason, we manually reassessed the bias of 
each website in the extreme bias categories 
on \url{mediabiasfactcheck.com} and \url{allsides.com} 
to validate their bias, as these two websites 
do not list the rejection of the establishment as
a criteria for their bias assessment.
However, even if we do not use the criteria
of adherence or rejection of the establishment in our
classification, websites in the extreme bias (right) and
extreme bias (left) categories are more likely
to not adhere with the establishment as this
variable seems to be highly correlated with political bias.}\\

In order to validate our classification, we compare it 
to the domain-level ideological alignment scores of news outlets obtained
by Bakshy \textit{et al.} \cite{Bakshy2015}
which is based on the average self declared ideological alignment of Facebook users sharing
URLs directing to news outlets.
We find a $R^2 = 0.9$ for the linear regression between the ideological alignment
found by Bakshy \textit{et al.} and our classification where we mapped our categories between -3 and 3
(see \FigOutletsAlignment{}).
{\label{rev:top_urls}Supplementary Data file \texttt{SuppData\_top\_urls\_per\_category.csv}
contains the top 10 URLs of each media category along with 
notes about their classification on fact checking websites (when available),
links to the fact checking websites and additional information.
We observe that the classification of the most popular URLs 
is well aligned with the label assigned to their domains.}\\

{\label{rev:domain_lev}
We investigate the influence and importance of news 
at the domain level and not at the article level.
Since a website classified as fake may contain
factual articles and vice versa,
domain-level classification 
implies a level of imprecision.
However, it allows us to reveal the integrated 
effect of news outlets over more than five months
and to measure the relative importance of 
each type of news by classifying all URLs
directing to important news outlets.
Moreover, classifying domains instead of 
URL (or article) allows to consider the extended 
effect of each type of news.
Indeed, when a Twitter user follows a URL to a 
news article containing factual information 
on a website publishing mostly fake news, 
she/he will be exposed to the other articles
containing fake news on the websites.
Therefore, this particular fact-based news 
ultimately increases the potential 
influence of fake news.}

\subsection{Collective influence algorithm in directed networks}

We use the Collective Influence (CI) algorithm\cite{Morone2015} applied 
to directed networks to find the most influential nodes of the 
information retweet networks.
The Collective Influence algorithm is based on the solution of
the optimal percolation of random networks which consists
of identifying the minimal set of nodes, the \textit{super-spreaders},
whose removal would
dismember the network in many disconnected and non-extensive components.
The fragmentation of the network is measured by the size
of the largest connected component, called the giant component of the network.
The CI algorithm considers influence as an emergent collective property,
not as a local property such as the node's degree, and has been shown
to be able to identify super-spreaders of information in social networks\cite{Morone2016,Teng2016}.
Here, we consider a directed version of the algorithm where we target
the \textit{super-sources} of information.

The procedure is as follow\cite{Morone2016}: 
we first compute the value of $\textrm{CI}_{\ell,\textrm{out}}(i)$
for all nodes $i = 1, ..., N$ as 

\begin{equation} 
 \textrm{CI}_{\ell,\textrm{out}}(i) = \left(k_{\textrm{out}}(i) - 1 \right)
 \sum_{\substack{j\in \partial B_\textrm{out}(i,\ell) \\ k_{\textrm{out}}(j) > 0}} 
 \left(k_{\textrm{out}}(j)  - 1\right),
\end{equation}

where $\ell$ is the radius of the ball  around each node we consider, here we use $\ell=2$,
$k_{\textrm{out}}(i)$ is the out-degree of node $i$ and $\partial B_\textrm{out}(i,\ell)$ is the
set of nodes situated at a distance $\ell$ from node $i$ computed by following outgoing paths
from $i$.
The node with the largest $\textrm{CI}_{\ell,\textrm{out}}$ value is then removed from the
network and the value of $\textrm{CI}_{\ell,\textrm{out}}$ of nodes whose value is changed
by this removal is recomputed. 
This procedure is repeated until the size of the weakly connected largest component becomes 
negligible.
The order of removal of the nodes corresponds to the final ranking of the network {top news spreaders} shown in
Table~\ref{tab:influencers}.\\

{\label{rev:rank_comp}A comparison of the ranking obtained by the CI algorithm with 
rankings obtained by considering out-degree (high degree centrality) 
and Katz centrality\cite{Katz1953} (\FigRankComp{}) shows that high degree (HD) and Katz rankings of the top 100
CI {spreaders} fall mostly within the top 100 ranks of these two other measures
with only a small number of top CI {spreaders} having a poor HD or Katz ranking.
Note that the CI algorithm is especially good at identifying 
influential nodes that are locally weakly connected but are
influent on a larger scale\cite{Morone2015}.}

\subsection{Time series processing}

We find that a 15 minute
resolution offers a sufficiently detailed sampling of Twitter
activity.
Indeed, a representative time scale of Twitter activity is
given by the characteristic retweet delay time, i.e. the typical time
between an original tweet and its retweet.  We find that the median
time of the retweet delay distribution in our dataset is
\SI{1}{\hour}\,\SI{57}{\minute} and the distribution has a log-normal
shape (first quartile at \SI{20}{\minute} and third quartile at
\SI{9}{\hour}\,\SI{11}{\minute}).  We tested the consistency of our
results using a resolution of \SI{5}{\minute} and \SI{1}{\hour} and did
not see significant changes.\\

In order to perform the cross-correlation and causality analysis
of the activity time series, we processed the time 
series to remove the trend and circadian activity cycles and to deal with 
missing data points.
For each missing data points, we remove the entire day corresponding
to the missing observation in order to keep the period of 
the circadian activity consistent over the entire time series.
This is necessary to apply filtering technique to remove the 
periodic component of the time series.
When removing an entire day, we consider that the day starts and ends
at 4 am, corresponding to the time of the day with lowest Twitter activity.
We removed a total of 24 days, representing 15\% of our observation period.
We then applied a STL (seasonal-trend decomposition procedure based on Loess)\cite{Cleveland1990}
procedure to extract the trend, seasonal and remainder components of each
activity time series.
We only consider the remainder components for the cross-correlation and causality
analysis.
We set the seasonal period of the STL filter equal to the number of observations per day, $n_p = 96$,
and the seasonal smoothing period to $n_s = 95$,
such that the seasonal component is smooth and the remainder component retains
the higher frequency signal containing the activity of interest.
Varying the value of the smoothing period to $n_s = 47$ does not change significantly the results.

{\label{rev:causal_method}
\subsection{Causal analysis}

The STL procedure removes the trend and circadian pattern in the time
series, resulting in stationary time series (the stationarity of each 
time series is confirmed by an augmented Dickey--Fuller test\cite{MacKinnon1994}).
{\label{rev:standardized_average}Before performing the causal analysis, we
also standardized each time series in order to 
remove any influence of the difference in absolute values of time series.
The causal analysis is performed using the 
entire time period (more than 5 months) and therefore 
reveals causal effects that are observed ``in average'' 
over the entire time period.}\\

In order to infer the causal relations between the activity of the {top news spreaders} and 
the supporters, we use a multivariate causal discovery algorithm based on the 
PC algorithm\cite{Spirtes2000} and further adapted for multivariate time series
by Runge \textit{et al.}\cite{Runge2012,Runge2015,Runge2017}.
Considering an ensemble of stochastic processes $\mathbf{X}$ the algorithm proceeds as follow.
First, for every time series $Y\in \mathbf{X}$ the sets of preliminary parents is
constructed by testing their independence at a range of time lags:
$\mathcal{P}_{Y_t} = \{X_{t-\tau} | 0 < \tau \leq \tau_{\textrm{max}}, Y_{t} \not\perp\! X_{t-\tau} \}$.
As this set also contains indirect links, they are then 
removed by testing if the dependence between $Y_t$ and each $X_\tau \in \mathcal{P}_{Y_t}$
vanishes when it is conditioned on an incrementally increased set of conditions $\mathcal{P}^{n,i}_{Y_t}\subseteq\mathcal{P}_{Y_t}$,
where $n$ is the cardinality of $\mathcal{P}^{n,i}_{Y_t}$ and $i$ is the index
iterating over the number of combinations of picking $n$ conditions from $\mathcal{P}_{Y_t}$.
The combinations of parents having the 
strongest dependence in the previous step are selected first\cite{Runge2015,Runge2017}.

The main free parameters are the maximum time lag $\tau_{\textrm{max}}$
and the significance level of the independence test used during the 
first step to build the set of preliminary parents which we set to $\alpha_{\textrm{PC}}=0.1$.
We set the value of the maximum time lag to $\tau_{\textrm{max}} = 18$ time steps (i.e. 270 min)
as it is the lag after which the lagged cross-correlations between each time series
falls below 0.1 in absolute value
(see \FigLaggedCorr{}).
We set the maximum number of tested combinations of the conditioning
set to 3 and we do not limit the size of the conditioning set.

We test the conditional independence 
of time series with the non-parametric RCoT test\cite{Strobl2017}.
This test uses random Fourier features to 
approximate the kernel-based conditional
independence test KCIT\cite{Zhang2011} and is at least as
accurate as KCIT while having a run time
that scales linearly with sample size\cite{Strobl2017}.
This point is crucial for our case given the size 
of our dataset (13\,152 time points $\times$ 10 time 
series $\times$ 18 time lags).
We set the number of Fourier features to $n_f = 400$.

\label{rev:fdr}
We select the significant final causal links by applying 
a Benjamini-Hochberg False Discovery Rate correction\cite{Benjamini1995}
to the $p$-values of the conditional independence tests
with a threshold level of 0.05.
FDR corrections allow to control the expected proportion of false positive.
The final causal links, i.e. parents of each time series, are 
reported in \TabCausalParents{}.
\\

Following the procedure of refs. \cite{Eichler2010,Runge2015},
We then regress a linear model:

\begin{equation}
\mathbf{X}_t = \sum_{\tau =1}^{\tau_{\textrm{max}}} \bm{\Phi}(\tau)\mathbf{X}_{t-\tau} + \varepsilon_t,
\label{eq:lin_reg}
\end{equation}

where all time series are standardized and 
only coefficients corresponding to true causal links are 
estimated while all the other ones are kept equal to zero,
i.e. $\Phi_{ij}(\tau) \neq 0 $ only for $X^i_{t-\tau} \rightarrow X^j_t$.
The causal effect between a time series $X^i$ and $X^j$ 
at a time delay $\tau$ can be computed from the regressed coefficients 
as:

\begin{equation}
I^{\textrm{CE}}_{i\rightarrow j} (\tau) = \bm{\Psi}_{ij}(\tau),
\end{equation}

where $\bm{\Psi}(\tau)$ is computed from the relation 
$\bm{\Psi}(\tau)=\sum_{s=1}^{\tau}\bm{\Phi}(s)\bm{\Psi}(\tau-s)$
, with $\bm{\Phi}(0) = \mathbf{I}$.
Here, $\bm{\Psi}_{ij}(\tau)$ gives the sum over the products of path coefficients 
along all causal paths up to a 
time lag $\tau$.
The causal effect $I^{\textrm{CE}}_{i\rightarrow j} (\tau)$ represents the expected value 
of $X^j_t$ (in unit of standard deviation) 
if $X^i_{t-\tau}$ is perturbed by one standard deviation\cite{Runge2015}.

To reconstruct the causal network, we are interested in the 
aggregated effects and therefore use the lag with maximum effect:

\begin{equation}
I^{\textrm{CE,max}}_{i\rightarrow j} = \max_{0<\tau\leq\tau_{\textrm{max}}} \left| I^{\textrm{CE}}_{i\rightarrow j} (\tau) \right|.
\end{equation}

We estimate the standard errors of each causal effects 
with a residual-based bootstrap procedure (similarly to ref. \cite{Runge2015}).
We employ 200 bootstrap surrogates time series generated 
by running model (\ref{eq:lin_reg}) with a joint random sample
$\varepsilon^*_t$ (with replacement) of the original multivariate
residual time series $\varepsilon_t$ and compute the standard deviation
of the $I^{\textrm{CE,max}}_{i\rightarrow j}$ values.
}

\subsection{Code availability}
The analysis and plotting scripts
allowing to reproduce the results of this paper are available 
at \url{https://github.com/alexbovet/information_diffusion}.
The Python module used for the network analysis (graph-tool)
is available at \url{https://graph-tool.skewed.de}.
The causal discovery algorithm software (TIGRAMITE)
is available at \url{https://jakobrunge.github.io/tigramite}.
The code for the conditional independence test (RCIT and RCoT software) 
is available at \url{https://github.com/ericstrobl/RCIT}.
The code for the LOESS processing is available at 
\url{https://github.com/jcrotinger/pyloess}.

\subsection{Data availability}

The raw Twitter data cannot be directly shared as it could infringe
the Twitter Developer Terms. However, we are sharing the tweet IDs of
the data we collected which would allow anyone to download the tweets
we used for this study directly from Twitter. The full datasets
analyzed in this study are available under the limits of Twitter's
Developer Terms at \url{http://kcore-analytics.com}.  The
classification of news as 'fake' news or 'extremely biased' news is a
matter of opinion, rather than a statement of fact. This opinion
originated in publically available datasets from fact-checking
organizations (i.e. www.opensources.co). The conclusions contained in
this article should not be interpreted as representing those of the
authors.

\section*{Acknowledgements}

\noindent
A. Bovet thanks the Swiss National Science Foundation (SNSF)
and the Flagship European Research Area Network (FLAG-ERA)
Joint Transnational Call  “FuturICT  2.0” for the
financial support provided and R. Lambiotte for helpful
comments.\\

\section*{Author contributions statement}

H. A. M. and A. B. conceived the project.  A. B. performed the analysis
and prepared figures.
A. B. and H. A. M. wrote the manuscript.\\

\section*{Additional information}

\textbf{Competing interests:} H. A. M. has shares in KCore Analytics, LLC. A. B. declares no competing interests.

\pagebreak

\setcounter{equation}{0}
\setcounter{figure}{0}
\setcounter{table}{0}
\makeatletter

\captionsetup[figure]{labelformat=sfiglab}
\captionsetup[table]{labelformat=stablab}

 \begin{center}
   \LARGE{\textbf{Influence of fake news in Twitter during the 2016 US
       presidential election\\
       Supplementary Information}}
 
\vspace{1cm}

\large Alexandre Bovet$^{1,2,3}$, Hern\'an A. Makse$^3$

\vspace{0.2cm} \normalsize 
\textit{1) ICTEAM, Universit\'e Catholique de Louvain, Avenue George Lema\^itre 4, 1348 Louvain-la-Neuve, Belgium\\
2) naXys and Department of Mathematics, Universit\'e de Namur, Rempart de la Vierge 8, 5000 Namur, Belgium.\\
3) Levich Institute and Physics
  Department, City College of New York, New York, New York 10031, USA\\
* hmakse@lev.ccny.cuny.edu
}
  
\end{center}

\section*{Supplementary Note 1}

Breitbart News (extreme bias (right))
is the most dominant media outlet in
term of number of tweets among the right end of the outlet 
categories with 1.8 million tweets (see \TabHostnames{}).
Breitbart is closely aligned with the Trump campaign as 
Steve Bannon, who co-founded Breitbart, eventually joined Trump’s campaign
as its chief executive.
We also consider separately the websites \texttt{shareblue.com} and \texttt{bluenationreview.com}
in \TabsBreitBlue{} and \FigCausalGraphBreitblue{}
as they were purchased by David Brock, a political operative of the Hillary Clinton campaign
(\url{https://en.wikipedia.org/wiki/David_Brock}).
We examine the relation between \texttt{breitbart.com}, \texttt{shareblue.com} and \texttt{bluenationreview.com}  and the rest of 
with the extremely biased outlets in \TabsBreitBlue{} as well as \FigCausalGraphBreitblue{}.
For this analysis, outlets in the extreme bias (right) news category are split in two sub-categories: 
Breitbart and the rest of extreme bias (right) news (extreme bias (right) \textbackslash breitbart).
Extreme bias (left) news are also split in two sub-categories: Shareblue + Bluenationreview (SB+BNR)
and the rest of extreme bias (left) news (extreme bias (left) \textbackslash (SB+BNR)).
Our analysis reveals that, although Breitbart represents the 
largest tweet share of the extreme bias (right) category, the majority (66\%)
of users sharing links directing toward Breitbart also share 
links toward other websites of the extreme bias (right) category 
(\TabJaccardBreitblue{}).
We also find similar characteristics in term of average activity,
retweet network structure, activity correlation and causal 
relations between Breitbart and the rest of the extreme bias (right) category.
Removing Breitbart 
from the extreme bias category and treating it as a separated category 
does not change our results significantly.
Concerning  \texttt{shareblue.com} and \texttt{bluenationreview.com},
we find that they form a minority group of the extreme bias (left) category with a strong overlap (69\%)
of users with the rest of the extreme bias (left) category and that our results are not changed significantly
when we consider them as a separated category.

\section*{Supplementary Note 2}

We observe the presence 
of several member of the campaign staffs 
of each candidate in the top {news spreaders}.
We report the ranking in each news categories of campaign staffers
among the top 100 news spreaders in 
\TabCampStaff{}.
We see more users linked to the 
campaign staff of Donald Trump (13) than to the campaign staff of 
Hillary Clinton (3).
We also see that Trump staffers have higher ranks in term of influence 
and cover a broader spectrum 
of media categories (fake news (3), extreme bias (right) (9), right (9), right leaning
(8), center (8) and left leaning (1)) than Clinton staffers 
(center (1), left leaning (2), left (1) and extreme bias (left) (1)).
This reveals that the Trump team played an important direct role 
in the diffusion of news in Twitter.

Although members of the 
Trump team are prevalent in the top spreaders of 
fake, extremely biased (right), right and right leaning news,
the causal analysis reveals that they are not 
driving the activity of Trump and Clinton supporters 
which is more importantly influenced by 
the top center and left leaning spreaders, consisting 
mainly of journalists.
To verify the importance of users linked to the candidates' teams,
we repeated the causal analysis after having removed all
users linked to the campaigns.
We report these results in \FigCausalGraphNoStaff{} and \TabCausalEffectsNoStaff{}.
We observe no significant changes in the causal relations between 
the different groups as the relations are still 
dominated by center and left leaning top spreaders.

\section*{Supplementary Note 3}

A possible distinction between the diffusion 
mechanisms of different news outlets could
be due to the fact that some websites aggregates news
from other websites instead of producing
news. We find four websites that, at least partly, 
aggregates news: \texttt{zerohedge.com} (fake news), \texttt{wnd.com} (extreme bias (right)), 
\texttt{realclearpolitics.com} (right leaning) and \texttt{truepundit.com} (extreme bias (right)).
To understand if the presence of news aggregators in categories
other than the center and left leaning could explain 
the difference in dynamics that we observe, 
we repeated our analysis of the dynamics after 
having removed the news aggregators from our dataset.
We report the results in \TabsNotAggr{}
and \FigCausalGraphNoAggr{}.
We observe no significant changes in the activity correlations
and and that without the news aggregators,
the top fake news, extreme bias (right) and right leaning spreaders have a smaller
causal effect on the other groups, while the 
left leaning and center influencers stay the dominant ones.
This shows that news aggregators are not responsible for the 
differences on dynamics that we observe.

\clearpage

\begin{table}
\centering
\scriptsize
 \begin{tabular}{llS[table-format = 6]
                  lS[table-format = 6]
                  lS[table-format = 6]}
\toprule
  {} &
      \multicolumn{2}{c}{fake news} &
      \multicolumn{2}{c}{extreme bias (right) news} &
      \multicolumn{2}{c}{right news} \\
  {} &            {hostnames} &  {$N$} &                       {hostnames} &  {$N$} &                   {hostnames} &  {$N$} \\
\midrule
1  &     thegatewaypundit.com &  761756 &          breitbart.com &      1854920 &                   foxnews.com &  1122732 \\
2  &            truthfeed.com &  554955 &        dailycaller.com &       759504 &               dailymail.co.uk &   474846 \\
3  &             infowars.com &  478872 &    americanthinker.com &       179696 &        washingtonexaminer.com &   462769 \\
4  &      therealstrategy.com &  241354 &                wnd.com &       141336 &                    nypost.com &   441648 \\
5  &  conservativetribune.com &  212273 &         freebeacon.com &       129077 &              bizpacreview.com &   170770 \\
6  &            zerohedge.com &  186706 &      newsninja2012.com &       127251 &            nationalreview.com &   164036 \\
7  &             rickwells.us &   78736 &            hannity.com &       114221 &                 lifezette.com &   139257 \\
8  &              departed.co &   72773 &            newsmax.com &        94882 &                  redstate.com &   105912 \\
9  &  thepoliticalinsider.com &   66426 &       endingthefed.com &        88376 &                allenbwest.com &   104857 \\
10 &        therightscoop.com &   63852 &         truepundit.com &        84967 &  theconservativetreehouse.com &   102515 \\
11 &             teaparty.org &   48757 &  westernjournalism.com &        77717 &                  townhall.com &   102408 \\
12 &       usapoliticsnow.com &   46252 &          dailywire.com &        67893 &                 investors.com &   102295 \\
13 &           clashdaily.com &   45970 &        newsbusters.org &        60147 &                  theblaze.com &    99029 \\
14 &  thefederalistpapers.org &   45831 &     ilovemyfreedom.org &        54772 &         theamericanmirror.com &    91538 \\
15 &          redflagnews.com &   45423 &    100percentfedup.com &        54596 &                       ijr.com &    71558 \\
16 &     thetruthdivision.com &   44486 &            pjmedia.com &        46542 &             judicialwatch.org &    70543 \\
17 &                      { } &     { } &       weaselzippers.us &        45199 &             thefederalist.com &    55835 \\
18 &                      { } &     { } &                    { } &          { } &                    hotair.com &    55431 \\
19 &                      { } &     { } &                    { } &          { } &        conservativereview.com &    54307 \\
20 &                      { } &     { } &                    { } &          { } &            weeklystandard.com &    50707 \\
\bottomrule
\end{tabular}

\vspace{0.5cm}

 \begin{tabular}{llS[table-format = 6]
                  lS[table-format = 6]
                  lS[table-format = 6]}
\toprule
  {} &
      \multicolumn{2}{c}{right leaning news} &
      \multicolumn{2}{c}{center news} &
      \multicolumn{2}{c}{left leaning news} \\
  {} &            {hostnames} &  {$N$} &                       {hostnames} &  {$N$} &                   {hostnames} &  {$N$} \\
\midrule
1  &                wsj.com &        310416 &              cnn.com &   2291736 &            nytimes.com &      1811627 \\
2  &    washingtontimes.com &        208061 &          thehill.com &   1200123 &     washingtonpost.com &      1640088 \\
3  &                 rt.com &        157474 &         politico.com &   1173717 &            nbcnews.com &       512056 \\
4  &  realclearpolitics.com &        128417 &         usatoday.com &    326198 &         abcnews.go.com &       467533 \\
5  &        telegraph.co.uk &         82118 &          reuters.com &    283962 &        theguardian.com &       439580 \\
6  &             forbes.com &         64186 &        bloomberg.com &    266662 &                vox.com &       369789 \\
7  &            fortune.com &         57644 &  businessinsider.com &    239423 &              slate.com &       279438 \\
8  &                    { } &           { } &           apnews.com &    198140 &           buzzfeed.com &       278642 \\
9  &                    { } &           { } &         observer.com &    128043 &            cbsnews.com &       232889 \\
10 &                    { } &           { } &  fivethirtyeight.com &    124268 &         politifact.com &       198095 \\
11 &                    { } &           { } &              bbc.com &    118176 &            latimes.com &       190994 \\
12 &                    { } &           { } &          ibtimes.com &     72424 &        nydailynews.com &       188769 \\
13 &                    { } &           { } &            bbc.co.uk &     71941 &        theatlantic.com &       177637 \\
14 &                    { } &           { } &                  { } &       { } &           mediaite.com &       152877 \\
15 &                    { } &           { } &                  { } &       { } &           newsweek.com &       149490 \\
16 &                    { } &           { } &                  { } &       { } &                npr.org &       142143 \\
17 &                    { } &           { } &                  { } &       { } &      independent.co.uk &       127689 \\
18 &                    { } &           { } &                  { } &       { } &                 cnb.cx &        87094 \\
19 &                    { } &           { } &                  { } &       { } &  hollywoodreporter.com &        84997 \\
\bottomrule
\end{tabular}

\vspace{0.5cm}

 \begin{tabular}{llS[table-format = 6]
                  lS[table-format = 6]}
\toprule
  {} &
      \multicolumn{2}{c}{left news} &
      \multicolumn{2}{c}{extreme bias (left) news} \\
  {} &            {hostnames} &  {$N$} &                       {hostnames} &  {$N$} \\
\midrule
1  &     huffingtonpost.com & 1057518 &      dailynewsbin.com &      189257 \\
2  &      thedailybeast.com &  378931 &  bipartisanreport.com &      119857 \\
3  &           dailykos.com &  324351 &  bluenationreview.com &       75455 \\
4  &           rawstory.com &  297256 &    crooksandliars.com &       73615 \\
5  &       politicususa.com &  293419 &   occupydemocrats.com &       73143 \\
6  &               time.com &  252468 &         shareblue.com &       50880 \\
7  &        motherjones.com &  210280 &           usuncut.com &       27653 \\
8  &  talkingpointsmemo.com &  199346 &                   { } &         { } \\
9  &              msnbc.com &  177090 &                   { } &         { } \\
10 &           mashable.com &  173129 &                   { } &         { } \\
11 &              salon.com &  172807 &                   { } &         { } \\
12 &      thinkprogress.org &  172144 &                   { } &         { } \\
13 &          newyorker.com &  171102 &                   { } &         { } \\
14 &       mediamatters.org &  152160 &                   { } &         { } \\
15 &              nymag.com &  121636 &                   { } &         { } \\
16 &       theintercept.com &  109591 &                   { } &         { } \\
17 &          thenation.com &   54661 &                   { } &         { } \\
18 &             people.com &   47942 &                   { } &         { } \\
\bottomrule

\end{tabular}

\caption{\textbf{Hostnames in each media category}.
We also show the number ($N$) of tweets 
with a URL pointing toward each hostname. 
Tweets with several URLs are counted multiple times.}
\label{tab:hostnames}
\end{table}

\begin{table}
\footnotesize
\centering
\begin{tabular}{lS[table-format = 7]
		 S[table-format = 1.2,round-mode=places, round-precision=2]
		 S[table-format = 7]
		 S[table-format = 1.2,round-mode=places, round-precision=2]
		 S[table-format = 2.2,round-mode=places, round-precision=2]
		 S[table-format = 1.2,round-mode=places, round-precision=2]
		 S[table-format = 1.2,round-mode=places, round-precision=2]
		 S[table-format = 2.2, round-mode=places, round-precision=2]}
\toprule
{} &        $N_t$ &  $p_t$ &        $N_u$ &  $p_u$ & $N_t/N_u$ & $p_{t,n/o}$ & $p_{u,n/o}$ & $N_{t,n/o}/N_{u,n/o}$ \\
\midrule
extreme bias (right) news     &  3969639 & 0.13 &   294175 & 0.07 &   13.49 &      0.09 &      0.03 &               36.52 \\
breitbart     	   &  1849871 & 0.060 &   163707 & 0.037 &  11.300 &     0.090 &     0.036 &              28.202 \\
extreme bias (right) \textbackslash breitbart&  2119876 & 0.069 &   238517 & 0.054 &   8.888 &     0.099 &     0.033 &              26.945 \\[0.1cm]
extreme bias (left) news      &   609503 & 0.02 &    99743 & 0.02 &    6.11 &      0.06 &      0.03 &               11.46 \\
SB+BNR             &   126191 & 0.004 &    28888 & 0.007 &   4.368 &     0.036 &     0.031 &               5.113 \\
extreme bias (left) \textbackslash (SB+BNR)  &   483325 & 0.016 &    90367 & 0.021 &   5.348 &     0.069 &     0.033 &              11.374 \\
\bottomrule
\end{tabular}
\caption{\textbf{Tweet and user volume corresponding to extremely biased news in Twitter}. 
Number, $N_t$, and proportion, $p_t$, of tweets with a URL
  pointing to a website belonging to one of media
  categories.  Number, $N_u$, and proportion, $p_u$, of users having
  sent the corresponding tweets, and average number of tweets per
  user, $N_t/N_u$, for each category.  Proportion of tweets sent by
  non-official clients, $p_{t,n/o}$, proportion of users having sent
  at least one tweet from an non-official client, $p_{u,n/o}$, and
  average number of tweets per user sent from non-official clients,
  $N_{t,n/o}/N_{u,n/o}$.
  The average number of tweets per users and the proportion
of tweets sent from unofficial clients are very similar 
for each sub-categories.}
\label{tab:url_stats_breitblue}
\end{table}

\begin{table}
 \footnotesize
\centering
\begin{tabular}{l*{6}{S[table-format = 1.2]}}
\toprule
{} &  {extreme bias        } &  {breitbart} &  {extreme bias (right)                         } &  {extreme bias       } &  {SB+BNR} &  {extreme bias (left)                        } \\
{} &  {             (right)} &  {         } &  {                     \textbackslash breitbart} &  {             (left)} &  {      } &  {                    \textbackslash (SB+BNR)} \\
\midrule
{extreme bias (right)}           &         1.00 &         0.56 &                   0.81 &        0.06 &      0.03 &                 0.06 \\
{breitbart}           &         0.56 &         1.00 &                   0.37 &        0.06 &      0.02 &                 0.06 \\
\makecell[cl]{extreme bias (right) \\ \textbackslash breitbart} &         0.81 &         0.37 &                   1.00 &        0.06 &      0.02 &                 0.06 \\[0.1cm]
{extreme bias (left)}            &         0.06 &         0.06 &                   0.06 &        1.00 &      0.29 &                 0.91 \\
{SB+BNR}              &         0.03 &         0.02 &                   0.02 &        0.29 &      1.00 &                 0.20 \\
\makecell[cl]{extreme bias (left) \\ \textbackslash (SB+BNR)}   &         0.06 &         0.06 &                   0.06 &        0.91 &      0.20 &                 1.00 \\
\bottomrule
\end{tabular}
\caption{\textbf{Jaccard indices between the sets of users in the extremely biased news categories}.
Jaccard indices between the sets of users tweeting URLs directing to extreme bias (right) news outlets, breitbart.com, extreme bias (right) minus breitbart.com 
({extreme bias (right) \textbackslash breitbart}),
extreme bias (left) news outlets,  shareblue.com and bluenationreview.com ({SB+BNR}), 
extreme bias (left) minus shareblue.com and bluenationreview.com ({extreme bias (left) \textbackslash (SB+BNR)}).
The Jaccard index between two sets $A$ and $B$ is computed as $J=A\cap B / A\cup B$.
Although breitbart represents the 
largest tweet share of the extreme bias (right) category, the majority (66\%)
of users sharing links directing toward breitbart also share 
links toward other websites of the extreme bias (right) category.
Shareblue and bluenationreview form a minority group of the extreme bias (left) category 
with a strong overlap (69\%) of users
with the rest of the extreme bias (left) category.}
\label{tab:jaccard_breitblue}
\end{table}

\begin{table}[b!]
\footnotesize
\centering
\begin{tabular}{lS[table-format = 6]S[table-format = 7]S
	         S[separate-uncertainty]
	         S[separate-uncertainty,table-format=1.2(1)]
	         S[table-format = 6]S[table-format = 4]}
\toprule
{} &  {$N$ nodes} &  {$N$ edges} &  {$<k>$} & {$\sigma(k_{\textrm{out}})/<k>$} & {$\sigma(k_{\textrm{in}})/<k>$} &  {$\max(k_{\textrm{out}})$} &  {$\max(k_{\textrm{in}})$} \\
\midrule
{extreme bias (right)}                          &       249659 &      1637927 &     6.56 &                         36 \pm 6 &                   2.73 \pm 0.03 &                       51845 &                        588 \\
{breitbart}                          &       141924 &       795504 &     5.61 &                         31 \pm 6 &                   2.33 \pm 0.02 &                       41039 &                        376 \\
{extreme bias (right) \textbackslash breitbart} &       201563 &       940161 &     4.66 &                         43 \pm 8 &                   2.28 \pm 0.03 &                       51845 &                        562 \\[0.1cm]
{extreme bias (left)}                           &        78911 &       277483 &     3.52 &                         33 \pm 6 &                   2.49 \pm 0.08 &                       23168 &                        648 \\
{SB+BNR}                             &        25956 &        59515 &     2.29 &                         45 \pm 6 &                   1.34 \pm 0.01 &                       15544 &                         65 \\
{extreme bias (left) \textbackslash (SB+BNR)}   &        70405 &       223532 &     3.17 &                         31 \pm 8 &                     2.4 \pm 0.1 &                       23168 &                        648 \\
\bottomrule
\end{tabular}
\caption{\textbf{Retweet networks characteristics for extremely biased news categories}.
We show the number of nodes and edges (links) of the networks,
the average degree, $\left<k\right>=\left<k_{\textrm{in}}\right>=\left<k_{\textrm{out}}\right>$, (the in-/out-degree 
of a node is the number of in-going/out-going links attached to it).
The out-degree of a node, i.e. a user, is equal to the number of different users that have
retweeted at least one of her/his tweets.
Its in-degree represents the number of different users she/he retweeted.
The ratio of the standard deviation and the average of the in- and out-degree
distribution, $\sigma(k_{\textrm{in}})/\left<k\right>$ and $\sigma(k_{\textrm{out}})/\left<k\right>$, 
measures the heterogeneity of the connectivity of each networks.
As the standard deviation of heavy-tailed degree distributions can depend on the network size,
we computed the values of $\sigma(k_{\textrm{in}})/\left<k\right>$ and
$\sigma(k_{\textrm{out}})/\left<k\right>$ with a bootstrap procedure.
The average degree and the heterogeneity of the degree distributions 
are similar for each sub-categories.}
\label{tab:retweet_graph_stats_breitblue}
\end{table}

\begin{landscape}
\begin{table}
  \footnotesize
\centering
\begin{tabular}{l*{12}{S[table-format = 1.2]}}
\toprule
{} &  {fake news} &  {breitbart} &  {extreme bias (right)                         } &  {right} &  {right        } &  {pro-Trump} &  {center} &  {left        } &  {left} &  {SB+BNR} &  {extreme bias (left)                        } &  {pro-Clinton} \\
{} &  {         } &  {         } &  {                     \textbackslash breitbart} &  {     } &  {      leaning} &  {         } &  {      } &  {     leaning} &  {    } &  {      } &  {                    \textbackslash (SB+BNR)} &  {           } \\
\midrule
fake news                                      &       1.00 &       0.40 &                        0.44 &   0.49 &           0.41 &       0.54 &    0.34 &          0.34 &  0.39 &                          0.13 &                                            0.30 &         0.34 \\
breitbart                                      &       0.40 &       1.00 &                        0.36 &   0.35 &           0.28 &       0.40 &    0.27 &          0.30 &  0.32 &                          0.11 &                                            0.28 &         0.29 \\
{extreme bias (right) \textbackslash breitbart}                     &       0.44 &       0.36 &                        1.00 &   0.49 &           0.29 &       0.47 &    0.28 &          0.28 &  0.35 &                          0.11 &                                            0.26 &         0.27 \\
right                                          &       0.49 &       0.35 &                        0.49 &   1.00 &           0.37 &       0.57 &    0.37 &          0.37 &  0.42 &                          0.12 &                                            0.33 &         0.36 \\
right leaning                                  &       0.41 &       0.28 &                        0.29 &   0.37 &           1.00 &       0.42 &    0.36 &          0.32 &  0.35 &                          0.15 &                                            0.23 &         0.36 \\
pro-Trump                                      &       0.54 &       0.40 &                        0.47 &   0.57 &           0.42 &       1.00 &    0.58 &          0.61 &  0.59 &                          0.18 &                                            0.39 &         0.73 \\
center                                         &       0.34 &       0.27 &                        0.28 &   0.37 &           0.36 &       0.58 &    1.00 &          0.60 &  0.55 &                          0.20 &                                            0.30 &         0.65 \\
left leaning                                   &       0.34 &       0.30 &                        0.28 &   0.37 &           0.32 &       0.61 &    0.60 &          1.00 &  0.63 &                          0.23 &                                            0.36 &         0.73 \\
left                                           &       0.39 &       0.32 &                        0.35 &   0.42 &           0.35 &       0.59 &    0.55 &          0.63 &  1.00 &                          0.20 &                                            0.38 &         0.68 \\
{SB+BNR}                  &       0.13 &       0.11 &                        0.11 &   0.12 &           0.15 &       0.18 &    0.20 &          0.23 &  0.20 &                          1.00 &                                            0.15 &         0.20 \\
{extreme bias (left) \textbackslash (SB+BNR)}  &       0.30 &       0.28 &                        0.26 &   0.33 &           0.23 &       0.39 &    0.30 &          0.36 &  0.38 &                          0.15 &                                            1.00 &         0.35 \\
pro-Clinton                                    &       0.34 &       0.29 &                        0.27 &   0.36 &           0.36 &       0.73 &    0.65 &          0.73 &  0.68 &                          0.20 &                                            0.35 &         1.00 \\
\bottomrule
\end{tabular}
\caption{\textbf{Pearson correlation coefficient between the activity corresponding to different media categories}.
The correlation profile of breitbart and
extreme bias (left) minus breitbart are very similar. 
Extreme bias (left) minus breitbart has a slightly higher correlation with the 
right new and with the pro-Trump
supporters than breitbart alone.
SB+BNR has a relatively different correlation
profile than extreme bias (left) minus SB+BNR, as it is poorly correlated
with all of other categories.
}
\label{tab:corr_breitblue}
\end{table}
\end{landscape}

\begin{table}
\scriptsize

\begin{tabular}{l*{6}{S[separate-uncertainty,table-format=1.4(1)]}}
\toprule
\multicolumn{1}{r}{$\swarrow$} &    {pro-Clinton} &      {pro-Trump} &      {fake news} &        {breitbart} & {extreme bias                                 } &            {right} \\
                               &    {           } &      {         } &      {         } &        {         } & { (right)  \textbackslash breitbart} &            {        }\\
\midrule
{pro-Clinton}                        &    0.65 \pm 0.01 &    0.14 \pm 0.01 &  0.007 \pm 0.008 &  0.0004 \pm 0.0003 &                    0.0008 \pm 0.0010 &  0.005 \pm 0.007 \\
{pro-Trump}                          &    0.13 \pm 0.02 &    0.45 \pm 0.01 &  0.004 \pm 0.005 &    0.000 \pm 0.001 &                    0.0003 \pm 0.0004 &  0.002 \pm 0.005 \\
{fake news}                          &  0.021 \pm 0.004 &    0.10 \pm 0.01 &    0.15 \pm 0.01 &      0.02 \pm 0.01 &                        0.06 \pm 0.01 &    0.01 \pm 0.01 \\
{breitbart}                          &    0.05 \pm 0.01 &    0.06 \pm 0.01 &    0.03 \pm 0.01 &      0.20 \pm 0.02 &                        0.05 \pm 0.01 &    0.02 \pm 0.01 \\
\makecell[cl]{extreme bias \\ (right) \textbackslash breitbart} &  0.015 \pm 0.009 &  0.005 \pm 0.002 &    0.01 \pm 0.01 &      0.04 \pm 0.01 &                        0.23 \pm 0.01 &    0.05 \pm 0.01 \\
{right}                              &  0.019 \pm 0.008 &  0.027 \pm 0.009 &    0.03 \pm 0.01 &      0.03 \pm 0.02 &                        0.04 \pm 0.01 &    0.17 \pm 0.01 \\
{right leaning}                      &  0.016 \pm 0.008 &  0.020 \pm 0.009 &    0.02 \pm 0.01 &      0.01 \pm 0.02 &                        0.03 \pm 0.01 &    0.06 \pm 0.01 \\
{center}                             &    0.03 \pm 0.01 &  0.011 \pm 0.006 &  0.022 \pm 0.007 &  0.0017 \pm 0.0007 &                    0.0024 \pm 0.0008 &  0.011 \pm 0.007 \\
{left leaning}                       &    0.04 \pm 0.01 &  0.007 \pm 0.003 &  0.004 \pm 0.002 &  0.0024 \pm 0.0008 &                    0.0023 \pm 0.0008 &  0.011 \pm 0.007 \\
{left}                               &    0.03 \pm 0.01 &    0.03 \pm 0.01 &  0.010 \pm 0.008 &  0.0024 \pm 0.0010 &                    0.0031 \pm 0.0009 &  0.009 \pm 0.008 \\
{(SB+BNR)}                           &    0.09 \pm 0.02 &  0.012 \pm 0.002 &  0.023 \pm 0.008 &    0.025 \pm 0.008 &                        0.03 \pm 0.01 &  0.003 \pm 0.001 \\
\makecell[cl]{extreme bias \\ (left) \textbackslash (SB+BNR)}   &    0.09 \pm 0.02 &    0.03 \pm 0.01 &    0.02 \pm 0.01 &      0.04 \pm 0.01 &                      0.027 \pm 0.008 &  0.003 \pm 0.001 \\
\bottomrule
\end{tabular}

\vspace{0.5cm}

\begin{tabular}{l*{6}{S[separate-uncertainty,table-format=1.4(1)]}}
\toprule
\multicolumn{1}{r}{$\swarrow$} &    {right leaning} &         {center} &   {left leaning} &             {left} &         {(SB+BNR)} & {extreme bias                         } \\
                               &    {             } &        {        }&    {           } &          {        }&      {        } & { (left)        \textbackslash (SB+BNR)}  \\
\midrule
{pro-Clinton}                        &    0.001 \pm 0.001 &  0.046 \pm 0.007 &  0.063 \pm 0.008 &      0.04 \pm 0.01 &  0.037 \pm 0.009 &                    0.016 \pm 0.006 \\
{pro-Trump}                          &  0.0005 \pm 0.0007 &  0.037 \pm 0.008 &  0.034 \pm 0.007 &    0.020 \pm 0.007 &  0.008 \pm 0.003 &                    0.008 \pm 0.005 \\
{fake news}                          &      0.06 \pm 0.01 &  0.026 \pm 0.010 &  0.015 \pm 0.003 &    0.013 \pm 0.009 &    0.01 \pm 0.01 &                      0.02 \pm 0.01 \\
{breitbart}                          &      0.04 \pm 0.01 &  0.042 \pm 0.009 &  0.019 \pm 0.002 &    0.004 \pm 0.001 &  0.003 \pm 0.001 &                    0.042 \pm 0.009 \\
\makecell[cl]{extreme bias \\ (right) \textbackslash breitbart} &      0.06 \pm 0.01 &  0.045 \pm 0.009 &  0.030 \pm 0.010 &    0.029 \pm 0.009 &    0.03 \pm 0.01 &                    0.010 \pm 0.009 \\
{right}                              &      0.09 \pm 0.01 &  0.043 \pm 0.009 &  0.017 \pm 0.003 &  0.0034 \pm 0.0010 &  0.035 \pm 0.008 &                    0.002 \pm 0.001 \\
{right leaning}                      &      0.22 \pm 0.01 &  0.044 \pm 0.009 &  0.034 \pm 0.009 &  0.0036 \pm 0.0009 &  0.026 \pm 0.008 &                  0.0029 \pm 0.0008 \\
{center}                             &    0.009 \pm 0.009 &  0.266 \pm 0.009 &    0.18 \pm 0.01 &    0.032 \pm 0.009 &  0.014 \pm 0.002 &                    0.030 \pm 0.009 \\
{left leaning}                       &    0.003 \pm 0.002 &    0.17 \pm 0.01 &  0.291 \pm 0.009 &    0.039 \pm 0.010 &  0.043 \pm 0.009 &                    0.028 \pm 0.008 \\
{left}                               &      0.02 \pm 0.01 &    0.08 \pm 0.01 &    0.10 \pm 0.01 &      0.16 \pm 0.01 &    0.02 \pm 0.01 &                      0.06 \pm 0.01 \\
{(SB+BNR)}                           &    0.003 \pm 0.001 &  0.028 \pm 0.009 &  0.045 \pm 0.010 &      0.03 \pm 0.01 &    0.21 \pm 0.01 &                      0.06 \pm 0.01 \\
\makecell[cl]{extreme bias \\ (left) \textbackslash (SB+BNR)}   &      0.02 \pm 0.01 &  0.026 \pm 0.010 &  0.045 \pm 0.008 &    0.034 \pm 0.010 &  0.022 \pm 0.009 &                      0.25 \pm 0.01 \\
\bottomrule
\end{tabular}

\caption{\textbf{Maximum causal effect with Breitbart and SB+BNR separated}.
Maximum causal effect values ($\pm$ s.d.) between the activity of the top 100 spreaders 
of each media category and the candidate supporters
when considering Breitbart and shareblue+bluenationreview as 
separated from extreme bias (right) and extreme bias (left), respectively.}
\label{tab:causal_effects_breitblue}
\end{table}

\begin{table}[b!]
\footnotesize
\centering
\begin{tabular}{lS[table-format = 6]S[table-format = 7]S
	         S[separate-uncertainty]
	         S[separate-uncertainty,table-format=1.2(1)]
	         S[table-format = 6]S[table-format = 4]}
\toprule
{} &  {$N$ nodes} &  {$N$ edges} &  {$<k>$} & {$\sigma(k_{\textrm{out}})/<k>$} & {$\sigma(k_{\textrm{in}})/<k>$} &  {$\max(k_{\textrm{out}})$} &  {$\max(k_{\textrm{in}})$} \\
\midrule
{fake news}                          &       175605 &      1854439 &    10.56 &                         47 \pm 7 &                   3.18 \pm 0.06 &                      104840 &                       1861 \\
{extreme bias (right)}                          &       249659 &      2699930 &    10.81 &                        56 \pm 12 &                   3.55 \pm 0.06 &                      172769 &                       1712 \\
{right}                              &       345644 &      2799298 &     8.10 &                        63 \pm 20 &                   3.57 \pm 0.08 &                      243101 &                       1998 \\
{right leaning}                      &       216026 &       611563 &     2.83 &                        55 \pm 14 &                   2.33 \pm 0.08 &                       53248 &                        468 \\
{center}                             &       864733 &      4140477 &     4.79 &                        94 \pm 55 &                     4.7 \pm 0.6 &                      680126 &                       5703 \\
{left leaning}                       &      1043436 &      4965956 &     4.76 &                        75 \pm 27 &                     4.9 \pm 0.3 &                      279049 &                       2547 \\
{left}                               &       536903 &      2707064 &     5.04 &                        65 \pm 17 &                     5.0 \pm 0.2 &                      119444 &                       1830 \\
{extreme bias (left)}                           &        78911 &       426452 &     5.40 &                         52 \pm 9 &                   3.27 \pm 0.08 &                       50415 &                       1003 \\
\bottomrule
\end{tabular}
\caption{\textbf{Weighted retweet networks characteristics}.
We show the number of nodes and edges (links) of the networks,
the average degree, $\left<k\right>=\left<k_{\textrm{in}}\right>=\left<k_{\textrm{out}}\right>$, (the in-/out-degree 
of a node is the number of in-going/out-going links attached to it).
Here, the weight of a link represents the number of retweets from a user to another.
In a directed network, the average in-degree and out-degree are always equal.
The out-degree of a node, i.e. a user, is equal to the number of times other users have
retweeted her/his tweets.
Its in-degree represents the number of times she/he retweeted other users.
The ratio of the standard deviation and the average of the in- and out-degree
distribution, $\sigma(k_{\textrm{in}})/\left<k\right>$ and $\sigma(k_{\textrm{out}})/\left<k\right>$, 
measures the heterogeneity of the connectivity of each networks.
As the standard deviation of heavy-tailed degree distributions can depend on the network size,
we computed the values of $\sigma(k_{\textrm{in}})/\left<k\right>$ and
$\sigma(k_{\textrm{out}})/\left<k\right>$ with a bootstrap procedure.
}
\label{tab:retweet_graph_stats_complete}
\end{table}

\begin{landscape}
\begin{table}
\scriptsize
\centering
\begin{tabular}{l*{15}{S[table-format = 2]}}
\toprule
{} &  {fake}      &  {fake}            &  {breitbart} &  {extreme             } &  {extreme bias        }            &  {extreme bias                                 }  &  {right} &  {right}         &  {right        }            &  {center} &  {left}         &  {left} &  {extreme            } & {extreme bias                               }  &  {SB+} \\
{} &  {news  }    &       {(no aggr.)} &              &      {bias   }          &    {(right           }             &  { (right)                          }             &          &        {leaning} &                {leaning   } &           &       {leaning} &         &      {bias  }          & { (left)                         }             &    {BNR}     \\
{} &  {      }    &                    &              &      {(right)}          &    {        no aggr.)}             &  {          \textbackslash breitbart}             &          &                  &                {(no aggr.)} &           &                 &         &      {(left)}          & {         \textbackslash (SB+BNR)}             &              \\
\midrule
fake news                &   100 &             96 &         44 &         40 &                  40 &             37 &     31 &          24 &                   20 &      10 &          3 &     0 &         0 &              0 &          0 \\
fake (no aggr.)       &    96 &            100 &         45 &         41 &                  41 &             38 &     30 &          23 &                   20 &      10 &          3 &     0 &         0 &              0 &          0 \\
breitbart           &    44 &             45 &        100 &         73 &                  76 &             46 &     40 &          33 &                   27 &      15 &          3 &     0 &         0 &              0 &          0 \\
extreme bias (right)           &    40 &             41 &         73 &        100 &                  96 &             72 &     43 &          35 &                   29 &      16 &          3 &     0 &         0 &              0 &          0 \\
\makecell[cl]{extreme bias \\ (right no aggr.)}  &    40 &             41 &         76 &         96 &                 100 &             70 &     44 &          36 &                   30 &      17 &          3 &     0 &         0 &              0 &          0 \\
\makecell[cl]{extreme bias \\ (right) \textbackslash breitbart }      &    37 &             38 &         46 &         72 &                  70 &            100 &     39 &          30 &                   28 &      16 &          3 &     0 &         0 &              0 &          0 \\
right               &    31 &             30 &         40 &         43 &                  44 &             39 &    100 &          36 &                   31 &      19 &          3 &     0 &         0 &              0 &          0 \\
right leaning          &    24 &             23 &         33 &         35 &                  36 &             30 &     36 &         100 &                   82 &      22 &          4 &     2 &         0 &              0 &          0 \\
right leaning (no aggr.) &    20 &             20 &         27 &         29 &                  30 &             28 &     31 &          82 &                  100 &      23 &          5 &     3 &         1 &              0 &          1 \\
center              &    10 &             10 &         15 &         16 &                  17 &             16 &     19 &          22 &                   23 &     100 &         18 &     9 &         1 &              0 &          2 \\
left leaning           &     3 &              3 &          3 &          3 &                   3 &              3 &      3 &           4 &                    5 &      18 &        100 &    14 &         1 &              0 &          2 \\
left                &     0 &              0 &          0 &          0 &                   0 &              0 &      0 &           2 &                    3 &       9 &         14 &   100 &        16 &             14 &         13 \\
extreme bias (left)            &     0 &              0 &          0 &          0 &                   0 &              0 &      0 &           0 &                    1 &       1 &          1 &    16 &       100 &             81 &         42 \\
\makecell[cl]{extreme bias \\ (left) \textbackslash (SB+BNR)}      &     0 &              0 &          0 &          0 &                   0 &              0 &      0 &           0 &                    0 &       0 &          0 &    14 &        81 &            100 &         26 \\
SB+BNR           &     0 &              0 &          0 &          0 &                   0 &              0 &      0 &           0 &                    1 &       2 &          2 &    13 &        42 &             26 &        100 \\
\bottomrule
\end{tabular}
\caption{\textbf{Intersection between sets of the top 100 news spreaders from each media category}.
We observe that the set of top 100 influencers 
does not change greatly when removing the 
news aggregators. 
The sets of top 100 fake news and fake news without aggregators influencers
have 96 influencers in common. 
Their are also 96 influencers in common in the top 100 sets of extreme bias (right) and 
extreme bias (right) without aggregators. 
The right leaning and right leaning without aggregators top 100 
influencers see the largest change, but still have 82 influencers in common.}
\label{tab:infl_overlap}
\end{table}
\end{landscape}

\begin{table}
 \begin{tabular}{lllllllll}
\toprule
{} & fake news & extreme bias         & right & lean       & center & lean      & left & extreme bias        \\
{} & fake      &              (right) &       &      right &        &      left &      &              (left) \\
\midrule
@realDonaldTrump (T) &    5 &         1 &     2 &          4 &     28 &        53 &      &          \\
@DonaldJTrumpJr  (T)&   14 &        12 &    25 &         62 &     84 &           &      &          \\
@DanScavino      (T)&   73 &        20 &    36 &         16 &     76 &           &      &          \\
@BreitbartNews   (T)&      &         3 &       &            &        &           &      &          \\
@EricTrump       (T)&      &        45 &    31 &            &        &           &      &          \\
@TeamTrump       (T)&      &        16 &    17 &          9 &     34 &           &      &          \\
@PaulManafort    (T)&      &        59 &    45 &         17 &     82 &           &      &          \\
@KellyannePolls  (T)&      &        19 &    13 &          8 &     20 &           &      &          \\
@JasonMillerinDC (T)&      &        60 &    26 &         15 &     43 &           &      &          \\
@seanspicer      (T)&      &           &    80 &         38 &     83 &           &      &          \\
@RealBenCarson   (T)&      &           &       &            &        &           &      &          \\
@BreitbartXM     (T)&      &        65 &       &            &        &           &      &          \\
@BreitbartTech   (T)&      &        87 &       &            &        &           &      &          \\[0.2cm]
@HillaryClinton  (C)&      &           &       &            &     67 &        17 &   51 &          \\
@JesseLehrich    (C)&      &           &       &            &        &        85 &      &          \\
@Shareblue       (C)&      &           &       &            &        &           &      &        6 \\
\bottomrule
\end{tabular}
\caption{\textbf{Collective influence ranking of Twitter users linked to the campaign staffs}.
Influence ranking of users in the campaign staffs of Donald Trump (T) and Hillary Clinton (C)
among the top 100 news spreaders of each media category.
Based on \url{http://www.p2016.org/trump/trumporggen.html} and \url{http://www.p2016.org/clinton/clintonorggen.html}.
We consider accounts related to Breitbart.com to be linked to the Trump team because of Steve Bannon who co-founded Breitbart and 
was chief executive of Donald Trump's presidential campaign (\url{https://en.wikipedia.org/wiki/Steve_Bannon}).
We consider @Shareblue to be linked to Clinton team because of David Brock, a political operative of the Hillary Clinton campaign
who purchased Shareblue (\url{https://en.wikipedia.org/wiki/David_Brock}).
}
\label{tab:camp-staff}
\end{table}

\begin{landscape}
\begin{table}
  \footnotesize
\centering
\begin{tabular}{l*{10}{S[table-format = 1.2]}}
\toprule
{} &  {fake news} &  {extreme bias (right)} &  {right} &  {right leaning} &  {center} &  {left leaning} &  {left} &  {extreme bias (left)} &  {pro-Trump} &  {pro-Clinton} \\
\midrule
fake news     &       1.00 &       0.50 &   0.49 &           0.41 &    0.34 &          0.34 &  0.39 &      0.30 &       0.54 &         0.34 \\
extreme bias (right)     &       0.50 &       1.00 &   0.52 &           0.34 &    0.34 &          0.36 &  0.41 &      0.32 &       0.53 &         0.34 \\
right         &       0.49 &       0.52 &   1.00 &           0.37 &    0.37 &          0.37 &  0.42 &      0.32 &       0.57 &         0.36 \\
right leaning &       0.41 &       0.34 &   0.37 &           1.00 &    0.36 &          0.32 &  0.35 &      0.26 &       0.42 &         0.36 \\
center        &       0.34 &       0.34 &   0.37 &           0.36 &    1.00 &          0.60 &  0.55 &      0.34 &       0.58 &         0.65 \\
left leaning  &       0.34 &       0.36 &   0.37 &           0.32 &    0.60 &          1.00 &  0.63 &      0.40 &       0.61 &         0.73 \\
left          &       0.39 &       0.41 &   0.42 &           0.35 &    0.55 &          0.63 &  1.00 &      0.41 &       0.59 &         0.68 \\
extreme bias (left)      &       0.30 &       0.32 &   0.32 &           0.26 &    0.34 &          0.40 &  0.41 &      1.00 &       0.41 &         0.38 \\
pro-Trump     &       0.54 &       0.53 &   0.57 &           0.42 &    0.58 &          0.61 &  0.59 &      0.41 &       1.00 &         0.73 \\
pro-Clinton   &       0.34 &       0.34 &   0.36 &           0.36 &    0.65 &          0.73 &  0.68 &      0.38 &       0.73 &         1.00 \\
\bottomrule
\end{tabular}
\caption{\textbf{Pearson correlation coefficient between the activity corresponding to each media categories}.}
\label{tab:corr_vals}
\end{table}
\end{landscape}

\begin{table}
\footnotesize

\begin{tabular}{l*{5}{S[separate-uncertainty,table-format=1.4(1)]}}
\toprule
\multicolumn{1}{r}{$\swarrow$} &    {pro-Clinton} &      {pro-Trump} &      {fake news} &        {extreme bias (right)} &            {right} \\
                               &    {           } &      {         } &      { no staff} &        {no staff } &            {no staff}\\
\midrule
{pro-Clinton}   &    0.65 \pm 0.01 &    0.14 \pm 0.01 &  0.003 \pm 0.006 &    0.010 \pm 0.008 &  0.011 \pm 0.008 \\
{pro-Trump}     &    0.12 \pm 0.02 &    0.45 \pm 0.01 &  0.001 \pm 0.002 &  0.0010 \pm 0.0006 &  0.002 \pm 0.001 \\
{fake news}     &  0.018 \pm 0.003 &    0.10 \pm 0.01 &    0.15 \pm 0.01 &      0.04 \pm 0.02 &    0.03 \pm 0.01 \\
{extreme bias (right)}     &    0.03 \pm 0.01 &  0.008 \pm 0.003 &    0.03 \pm 0.01 &      0.18 \pm 0.02 &  0.047 \pm 0.010 \\
{right}         &  0.009 \pm 0.002 &  0.006 \pm 0.002 &    0.03 \pm 0.01 &      0.04 \pm 0.01 &    0.20 \pm 0.01 \\
{right leaning} &  0.019 \pm 0.008 &  0.040 \pm 0.008 &    0.02 \pm 0.01 &      0.02 \pm 0.01 &    0.08 \pm 0.01 \\
{center}        &    0.03 \pm 0.01 &  0.013 \pm 0.009 &  0.012 \pm 0.008 &  0.0010 \pm 0.0006 &  0.007 \pm 0.010 \\
{left leaning}  &    0.04 \pm 0.01 &  0.008 \pm 0.005 &  0.002 \pm 0.002 &  0.0006 \pm 0.0005 &  0.008 \pm 0.008 \\
{left}          &    0.07 \pm 0.02 &    0.02 \pm 0.01 &  0.019 \pm 0.009 &  0.0026 \pm 0.0009 &  0.003 \pm 0.002 \\
{extreme bias (left)}      &    0.07 \pm 0.02 &    0.02 \pm 0.01 &    0.03 \pm 0.01 &      0.03 \pm 0.01 &  0.004 \pm 0.001 \\
\bottomrule
\end{tabular}

\vspace{0.5cm}

\begin{tabular}{l*{5}{S[separate-uncertainty,table-format=1.4(1)]}}
\toprule
\multicolumn{1}{r}{$\swarrow$} &    {right leaning} &         {center} &   {left leaning} &             {left} &         {extreme bias (left)} \\
                               &    { no staff    } &        {no staff}&   {no staff    } &          {no staff}&      {no staff   } \\
\midrule
{pro-Clinton}   &  0.0009 \pm 0.0010 &  0.054 \pm 0.008 &  0.071 \pm 0.008 &    0.013 \pm 0.009 &    0.017 \pm 0.008 \\
{pro-Trump}     &  0.0005 \pm 0.0005 &  0.016 \pm 0.003 &  0.036 \pm 0.007 &    0.027 \pm 0.008 &    0.013 \pm 0.006 \\
{fake news}     &      0.06 \pm 0.01 &  0.026 \pm 0.009 &  0.014 \pm 0.002 &    0.005 \pm 0.001 &    0.029 \pm 0.008 \\
{extreme bias (right)}     &      0.06 \pm 0.01 &  0.039 \pm 0.009 &  0.019 \pm 0.003 &    0.030 \pm 0.009 &    0.033 \pm 0.008 \\
{right}         &      0.09 \pm 0.01 &  0.043 \pm 0.009 &  0.018 \pm 0.003 &    0.031 \pm 0.009 &    0.033 \pm 0.008 \\
{right leaning} &      0.23 \pm 0.01 &  0.036 \pm 0.009 &  0.031 \pm 0.010 &  0.0032 \pm 0.0010 &  0.0025 \pm 0.0007 \\
{center}        &    0.004 \pm 0.009 &  0.261 \pm 0.010 &    0.17 \pm 0.01 &    0.018 \pm 0.008 &    0.005 \pm 0.009 \\
{left leaning}  &    0.002 \pm 0.002 &  0.138 \pm 0.010 &  0.313 \pm 0.009 &    0.015 \pm 0.008 &    0.001 \pm 0.001 \\
{left}          &    0.016 \pm 0.008 &    0.07 \pm 0.01 &    0.10 \pm 0.01 &      0.16 \pm 0.01 &      0.06 \pm 0.01 \\
{extreme bias (left)}      &      0.02 \pm 0.01 &  0.023 \pm 0.003 &  0.051 \pm 0.008 &      0.03 \pm 0.01 &      0.26 \pm 0.01 \\
\bottomrule
\end{tabular}

\caption{\textbf{Maximum causal effects without campaign staffers}.
Maximum causal effect values ($\pm$ s.d.) between the activity of the top 100 spreaders 
of each media category, where member of the staff of each candidate campaign 
(see Supplementary Table \ref{tab:camp-staff})
are removed, and the activity of the presidential candidate supporters.}
\label{tab:causal_effects_no_staff}
\end{table}

\begin{landscape}
\begin{table}
  \footnotesize
\centering
\begin{tabular}{l*{10}{S[table-format = 1.2]}}
\toprule
{} &  {fake news                 } &  {extreme bias (right)                 } &  {right} &  {right leaning                 } &  {center} &  {left        } &  {left} &  {extreme bias (left)} &  {pro-Trump} &  {pro-Clinton}  \\
{} &  {          (no aggregators)} &  {          (no aggregators)} &  {     } &  {              (no aggregators)} &  {      } &  {     leaning} &  {    } &  {        } &  {         } &  {           }  \\
\midrule
{fake (no aggregators)}          &                     1.00 &                          0.49 &     0.48 &                              0.39 &      0.33 &            0.34 &    0.38 &        0.29 &         0.53 &           0.33 \\
{extreme bias (right) (no aggregators)}     &                     0.49 &                          1.00 &     0.52 &                              0.32 &      0.34 &            0.35 &    0.41 &        0.31 &         0.52 &           0.33 \\
{right}                          &                     0.48 &                          0.52 &     1.00 &                              0.35 &      0.37 &            0.37 &    0.42 &        0.32 &         0.57 &           0.36 \\
{right leaning (no aggregators)} &                     0.39 &                          0.32 &     0.35 &                              1.00 &      0.35 &            0.31 &    0.34 &        0.25 &         0.41 &           0.35 \\
{center}                         &                     0.33 &                          0.34 &     0.37 &                              0.35 &      1.00 &            0.60 &    0.55 &        0.34 &         0.58 &           0.65 \\
{left leaning}                   &                     0.34 &                          0.35 &     0.37 &                              0.31 &      0.60 &            1.00 &    0.63 &        0.40 &         0.61 &           0.73 \\
{left}                           &                     0.38 &                          0.41 &     0.42 &                              0.34 &      0.55 &            0.63 &    1.00 &        0.41 &         0.59 &           0.68 \\
{extreme bias (left)}                       &                     0.29 &                          0.31 &     0.32 &                              0.25 &      0.34 &            0.40 &    0.41 &        1.00 &         0.41 &           0.38 \\
{pro-Trump}                      &                     0.53 &                          0.52 &     0.57 &                              0.41 &      0.58 &            0.61 &    0.59 &        0.41 &         1.00 &           0.73 \\
{pro-Clinton}                    &                     0.33 &                          0.33 &     0.36 &                              0.35 &      0.65 &            0.73 &    0.68 &        0.38 &         0.73 &           1.00 \\
\bottomrule
\end{tabular}
\caption{\textbf{Pearson correlation coefficient between the activity corresponding to each media categories without 
the news aggregators}.
We observe no significant changes in the correlation coefficients
between the analysis with (Tab. \ref{tab:corr_vals}) and without news aggregators.
The maximum difference in correlation (0.02) is between 
the right leaning and extreme bias (right).}
\label{tab:corr_vals_no_aggr}
\end{table}
\end{landscape}

\begin{table}
\footnotesize

\begin{tabular}{l*{5}{S[separate-uncertainty,table-format=1.4(1)]}}
\toprule
\multicolumn{1}{r}{$\swarrow$} &    {pro-Clinton} &      {pro-Trump} &      {fake news} &        {extreme bias} &            {right} \\
                               &    {           } &      {         } &    { (no aggr.)} &   {(right, no aggr.)} &            {}\\
\midrule
{pro-Clinton}                    &    0.65 \pm 0.01 &    0.14 \pm 0.01 &              0.015 \pm 0.005 &            0.0014 \pm 0.0005 &    0.003 \pm 0.003 \\
{pro-Trump}                      &    0.13 \pm 0.02 &    0.45 \pm 0.01 &              0.010 \pm 0.006 &            0.0011 \pm 0.0005 &  0.0010 \pm 0.0005 \\
{fake news (no aggr.)}     &    0.05 \pm 0.01 &    0.10 \pm 0.01 &                0.14 \pm 0.01 &                0.09 \pm 0.01 &      0.02 \pm 0.01 \\
{extreme bias (right) (no aggr.)}     &    0.03 \pm 0.01 &  0.005 \pm 0.003 &                0.03 \pm 0.01 &                0.20 \pm 0.01 &      0.05 \pm 0.01 \\
{right}                          &  0.023 \pm 0.008 &    0.04 \pm 0.01 &                0.02 \pm 0.01 &                0.03 \pm 0.01 &      0.19 \pm 0.01 \\
{right leaning (no aggr.)} &  0.006 \pm 0.002 &  0.002 \pm 0.002 &                0.02 \pm 0.01 &              0.012 \pm 0.010 &      0.05 \pm 0.01 \\
{center}                         &    0.04 \pm 0.01 &  0.026 \pm 0.010 &              0.012 \pm 0.007 &            0.0012 \pm 0.0007 &    0.015 \pm 0.008 \\
{left leaning}                   &    0.04 \pm 0.01 &  0.016 \pm 0.005 &              0.003 \pm 0.001 &            0.0006 \pm 0.0004 &    0.011 \pm 0.007 \\
{left}                           &    0.06 \pm 0.01 &    0.02 \pm 0.01 &              0.013 \pm 0.008 &              0.009 \pm 0.009 &    0.012 \pm 0.009 \\
{extreme bias (left)}                       &    0.09 \pm 0.02 &  0.012 \pm 0.009 &                0.04 \pm 0.01 &                0.02 \pm 0.01 &      0.00 \pm 0.01 \\
\bottomrule
\end{tabular}

\vspace{0.5cm}

\begin{tabular}{l*{5}{S[separate-uncertainty,table-format=1.4(1)]}}
\toprule
\multicolumn{1}{r}{$\swarrow$} &    {right leaning} &         {center} &   {left leaning} &             {left} &         {extreme bias       } \\
                               &    {(no aggr.)} &               &                  &                    &              {(left)}           \\
\midrule
{pro-Clinton}                    &                  0.006 \pm 0.006 &  0.046 \pm 0.007 &  0.065 \pm 0.008 &  0.022 \pm 0.009 &    0.006 \pm 0.006 \\
{pro-Trump}                      &                  0.002 \pm 0.001 &  0.032 \pm 0.007 &  0.015 \pm 0.003 &  0.025 \pm 0.007 &    0.013 \pm 0.006 \\
{fake news (no aggr.)}     &                    0.05 \pm 0.01 &  0.039 \pm 0.009 &  0.013 \pm 0.002 &  0.025 \pm 0.008 &    0.028 \pm 0.009 \\
{extreme bias (right) (no aggr.)}     &                    0.06 \pm 0.01 &  0.041 \pm 0.009 &  0.017 \pm 0.002 &  0.004 \pm 0.001 &    0.030 \pm 0.010 \\
{right}                          &                    0.08 \pm 0.01 &  0.042 \pm 0.009 &  0.018 \pm 0.003 &  0.028 \pm 0.009 &    0.034 \pm 0.009 \\
{right leaning (no aggr.)} &                    0.29 \pm 0.01 &  0.036 \pm 0.010 &    0.03 \pm 0.01 &  0.016 \pm 0.008 &  0.0022 \pm 0.0009 \\
{center}                         &                  0.005 \pm 0.007 &  0.267 \pm 0.009 &    0.18 \pm 0.01 &  0.020 \pm 0.009 &    0.021 \pm 0.008 \\
{left leaning}                   &                  0.002 \pm 0.002 &    0.18 \pm 0.01 &  0.300 \pm 0.009 &  0.013 \pm 0.008 &    0.013 \pm 0.007 \\
{left}                           &                  0.021 \pm 0.008 &    0.07 \pm 0.01 &    0.10 \pm 0.01 &  0.162 \pm 0.010 &      0.07 \pm 0.01 \\
{extreme bias (left)}                       &                  0.010 \pm 0.010 &  0.024 \pm 0.004 &    0.05 \pm 0.01 &    0.03 \pm 0.01 &      0.26 \pm 0.01 \\
\bottomrule
\end{tabular}

\caption{\textbf{Maximum causal effect without news aggregators}.
Maximum causal effect values ($\pm$ s.d.) between the activity of the top 100 spreaders 
of each media category, where news aggregators websites have removed,
and the activity of the presidential candidate supporters.
We see that our conclusions stay valid even without the news aggregators,
namely the domination of center and left leaning influencers in term 
of causal effects.
We observe a small decrease in the intensity of the causal effect of center influencers toward
Clinton supporters (0.065 to 0.046), but the effect is still the second 
most important after the left leaning influencers.
We also observe a small increase of the causal effect of Clinton supporters on the 
fake news top spreaders.
Without the news aggregators,
the top fake news, extreme bias (right) and right leaning spreaders have a smaller
causal effect on the other groups.}
\label{tab:causal_effects_not_aggr}
\end{table}

\begin{table}
\footnotesize
\centering
\begin{tabular}{lS[table-format = 7]}
\toprule
client name &    {number of tweets with a URL} \\
\midrule
          Twitter for iPhone &                  14215411 \\
          Twitter Web Client &                  13045560 \\
         Twitter for Android &                  10192781 \\
            Twitter for iPad &                   3355197 \\
                    Facebook &                   1254619 \\
                   TweetDeck &                   1079637 \\
             Mobile Web (M5) &                    951749 \\
                  Mobile Web &                    452812 \\
                      Google &                    410514 \\
         Twitter for Windows &                    200088 \\
   Twitter for Windows Phone &                    170529 \\
             Mobile Web (M2) &                    161682 \\
      Twitter for BlackBerry &                     93937 \\
                         iOS &                     72334 \\
 Twitter for Android Tablets &                     56007 \\
             Twitter for Mac &                     43993 \\
                        OS X &                     40642 \\
     Twitter for BlackBerry® &                     25140 \\
\bottomrule
\end{tabular}
\caption{\textbf{List of Twitter official clients}. We also 
display the number of tweets containing a URL and originating from
each official client.
The number of tweets with a URL originating from official clients represent 82\% 
of the total number of tweets with a URL.}
\label{tab:official_clients}
\end{table}

\begin{table}
\centering
\scriptsize
\begin{align*}
\toprule
\mathcal{P}_{0}  = & (0_{t-1}, 1_{t-1}, 8_{t-1}, 7_{t-1}, 6_{t-1}, 9_{t-1}, 0_{t-2}, 1_{t-2}, 8_{t-2}, 6_{t-2}, 7_{t-2}, 0_{t-3}, 1_{t-3}, 8_{t-3}, 3_{t-3}, 2_{t-3}, 0_{t-4}, 1_{t-4}, 8_{t-4},\\
		     & 0_{t-5}, 1_{t-5}, 8_{t-5}, 0_{t-6}, 1_{t-6}, 8_{t-6}, 0_{t-7}, 1_{t-7}, 8_{t-7}, 0_{t-8}, 1_{t-8}, 8_{t-8}, 4_{t-8}, 0_{t-9}, 1_{t-9}, 8_{t-9}, 6_{t-9}, 4_{t-9}, 0_{t-10},\\
		     & 1_{t-10}, 6_{t-10}, 0_{t-11}, 1_{t-11}, 8_{t-11}, 0_{t-12}, 9_{t-12}, 1_{t-12}, 6_{t-12}, 0_{t-13}, 8_{t-13}, 1_{t-13}, 0_{t-14}, 1_{t-14}, 0_{t-15}, 0_{t-16}, 0_{t-17},\\
		     & 1_{t-17}, 0_{t-18}, 1_{t-18})\\ 
\mathcal{P}_{1}  = & (1_{t-1}, 0_{t-1}, 9_{t-1}, 1_{t-2}, 0_{t-2}, 2_{t-2}, 6_{t-2}, 1_{t-3}, 0_{t-3}, 2_{t-3}, 8_{t-3}, 1_{t-4}, 0_{t-4}, 1_{t-5}, 0_{t-5}, 2_{t-5}, 8_{t-5}, 1_{t-6}, 0_{t-6},\\
		& 1_{t-7}, 0_{t-7}, 1_{t-8}, 0_{t-8}, 1_{t-9}, 0_{t-9}, 1_{t-10}, 0_{t-10}, 1_{t-11}, 8_{t-11}, 0_{t-11}, 1_{t-12}, 7_{t-12}, 0_{t-12}, 1_{t-13}, 8_{t-13}, 0_{t-13}, 1_{t-14}, \\
		& 2_{t-14}, 1_{t-15}, 0_{t-15}, 0_{t-16}, 7_{t-16}, 6_{t-17}, 1_{t-17}, 0_{t-17}, 0_{t-18}, 1_{t-18})\\ 
\mathcal{P}_{2}  = & (2_{t-1}, 5_{t-1}, 9_{t-1}, 3_{t-1}, 4_{t-1}, 6_{t-1}, 1_{t-1}, 2_{t-2}, 3_{t-2}, 9_{t-2}, 6_{t-2}, 1_{t-2}, 5_{t-2}, 2_{t-3}, 4_{t-3}, 1_{t-3}, 6_{t-3}, 3_{t-3}, 2_{t-4}, \\
		& 8_{t-4}, 1_{t-4}, 6_{t-5}, 5_{t-5}, 2_{t-5}, 2_{t-6}, 1_{t-6}, 5_{t-7}, 2_{t-8}, 8_{t-9}, 1_{t-9}, 6_{t-11}, 6_{t-13}, 2_{t-13})\\ 
\mathcal{P}_{3}  = & (3_{t-1}, 5_{t-1}, 2_{t-1}, 4_{t-1}, 9_{t-1}, 6_{t-1}, 0_{t-1}, 3_{t-2}, 6_{t-2}, 5_{t-2}, 9_{t-2}, 2_{t-2}, 4_{t-2}, 4_{t-3}, 4_{t-4}, 0_{t-4}, 5_{t-5}, 6_{t-5}, 8_{t-6}, \\
	        & 0_{t-7}, 4_{t-7}, 6_{t-11}, 6_{t-13}, 3_{t-17}, 5_{t-18})\\ 
\mathcal{P}_{4}  = & (4_{t-1}, 5_{t-1}, 2_{t-1}, 3_{t-1}, 6_{t-1}, 9_{t-1}, 7_{t-1}, 4_{t-2}, 5_{t-2}, 3_{t-2}, 2_{t-2}, 4_{t-3}, 2_{t-3}, 4_{t-4}, 6_{t-5}, 4_{t-5}, 5_{t-5}, 1_{t-5}, 3_{t-6}, \\
		& 1_{t-8}, 2_{t-13}, 3_{t-17})\\ 
\mathcal{P}_{5}  = & (5_{t-1}, 4_{t-1}, 2_{t-1}, 3_{t-1}, 7_{t-1}, 6_{t-1}, 5_{t-2}, 6_{t-2}, 2_{t-2}, 1_{t-2}, 3_{t-2}, 4_{t-3}, 5_{t-4}, 7_{t-4}, 6_{t-5}, 5_{t-5}, 2_{t-5}, 4_{t-6}, 0_{t-18})\\ 
\mathcal{P}_{6}  = & (6_{t-1}, 7_{t-1}, 8_{t-1}, 0_{t-1}, 5_{t-1}, 1_{t-1}, 6_{t-2}, 7_{t-2}, 8_{t-2}, 9_{t-2}, 2_{t-2}, 6_{t-3}, 7_{t-3}, 8_{t-3}, 2_{t-3}, 1_{t-3}, 6_{t-4}, 7_{t-4}, 8_{t-4}, \\
	        & 5_{t-4}, 6_{t-5}, 8_{t-5}, 4_{t-5}, 7_{t-5}, 6_{t-6}, 7_{t-6}, 8_{t-6}, 9_{t-6}, 5_{t-6}, 2_{t-6}, 6_{t-7}, 7_{t-7}, 7_{t-8}, 6_{t-8}, 2_{t-9}, 6_{t-10}, 8_{t-18})\\ 
\mathcal{P}_{7}  = & (7_{t-1}, 6_{t-1}, 8_{t-1}, 0_{t-1}, 7_{t-2}, 6_{t-2}, 8_{t-2}, 6_{t-3}, 7_{t-3}, 8_{t-3}, 7_{t-4}, 6_{t-4}, 8_{t-4}, 7_{t-5}, 6_{t-5}, 4_{t-5}, 8_{t-5}, 8_{t-6}, 6_{t-6}, \\
		& 7_{t-6}, 6_{t-7}, 6_{t-8}, 6_{t-9}, 6_{t-10}, 7_{t-11}, 6_{t-17})\\ 
\mathcal{P}_{8}  = & (8_{t-1}, 9_{t-1}, 7_{t-1}, 6_{t-1}, 8_{t-2}, 6_{t-2}, 7_{t-2}, 4_{t-2}, 5_{t-2}, 8_{t-3}, 6_{t-3}, 7_{t-3}, 9_{t-3}, 2_{t-3}, 0_{t-3}, 1_{t-3}, 7_{t-4}, 8_{t-4}, 6_{t-4}, \\
		& 9_{t-4}, 0_{t-4}, 7_{t-5}, 8_{t-5}, 2_{t-5}, 8_{t-6}, 5_{t-6}, 7_{t-6}, 6_{t-6}, 0_{t-6}, 7_{t-7}, 8_{t-7}, 9_{t-7}, 1_{t-7}, 6_{t-8}, 8_{t-8}, 8_{t-9}, 7_{t-11}, 6_{t-13})\\ 
\mathcal{P}_{9}  = & (9_{t-1}, 8_{t-1}, 7_{t-1}, 5_{t-1}, 1_{t-1}, 0_{t-1}, 9_{t-2}, 8_{t-2}, 2_{t-2}, 6_{t-2}, 1_{t-2}, 9_{t-3}, 8_{t-3}, 2_{t-3}, 9_{t-4}, 8_{t-4}, 0_{t-4}, 2_{t-5}, 8_{t-5}, \\
		& 3_{t-5}, 5_{t-5}, 3_{t-6}, 9_{t-6}, 0_{t-6}, 8_{t-7}, 7_{t-7}, 0_{t-7}, 9_{t-8}, 6_{t-8}, 3_{t-8}, 0_{t-8}, 7_{t-11}, 1_{t-11}, 9_{t-12}, 3_{t-13}, 9_{t-13}, 2_{t-13}, 6_{t-14},\\
		& 0_{t-14})\\ 
\bottomrule
\end{align*}

\caption{\textbf{Parents $\mathcal{P}$ for each time series estimated with the causal discovery algorithm}.
0 stands for pro-Clinton, 1 for pro-Trump, 2 for top fake news spreaders, 3 for top extreme bias (right) spreaders, 
4 for top right spreaders, 5 for top right leaning spreaders, 6 for top center spreaders, 
7 for top left leaning spreaders, 8 for top left spreaders and 9 for top extreme bias (left) spreaders.}
\label{tab:causal_parents}
\end{table}


\begin{figure}[tb]
\centering
\includegraphics[width=0.7\linewidth]{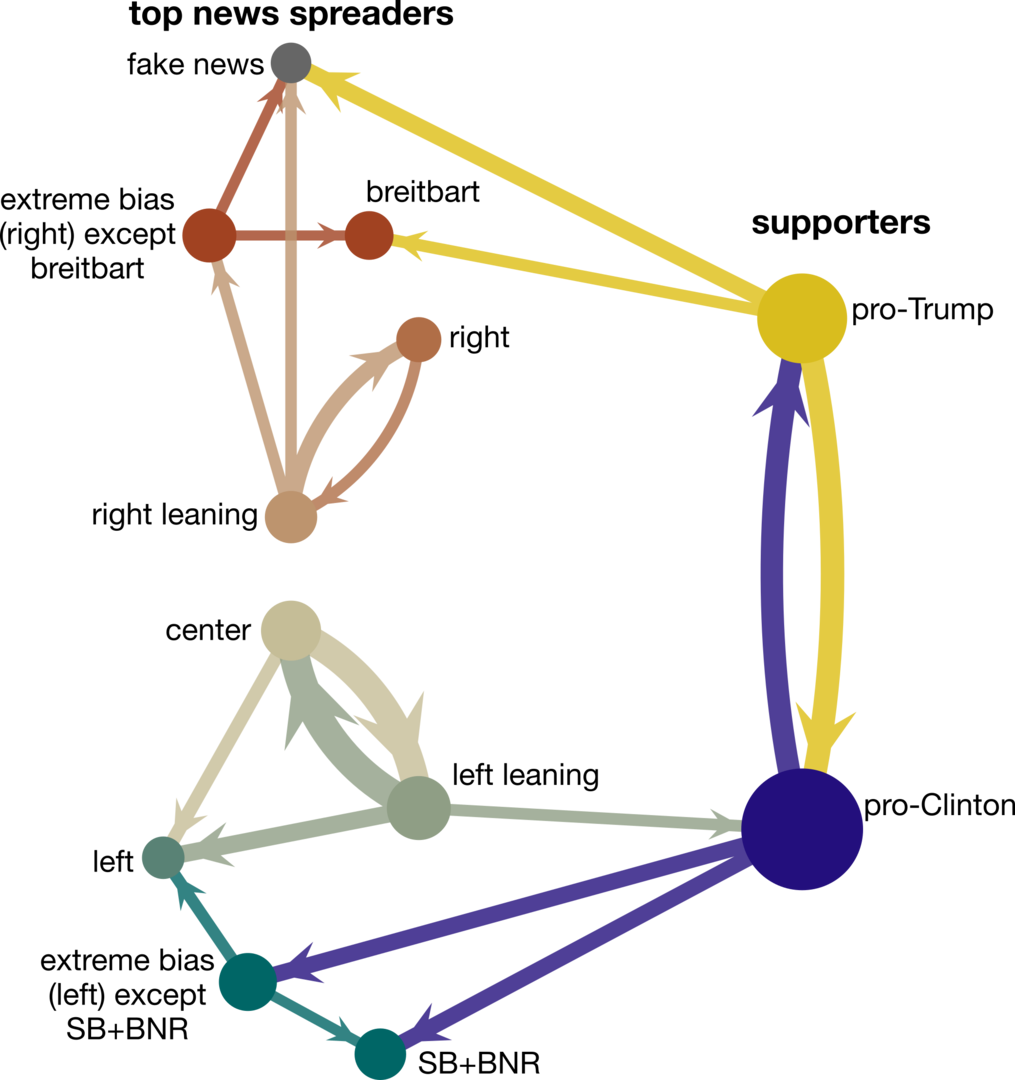}
\caption{\textbf{Causal graph obtained when considering breitbart and shareblue+bluenationreview (SB+BNR) 
as separated from extreme bias (right) and extreme bias (left), respectively}.
We only show causal effects larger than 0.05.
}
\label{fig:causal_graph_breitblue}
\end{figure}

\begin{figure}
\centering
\includegraphics[width=0.9\linewidth]{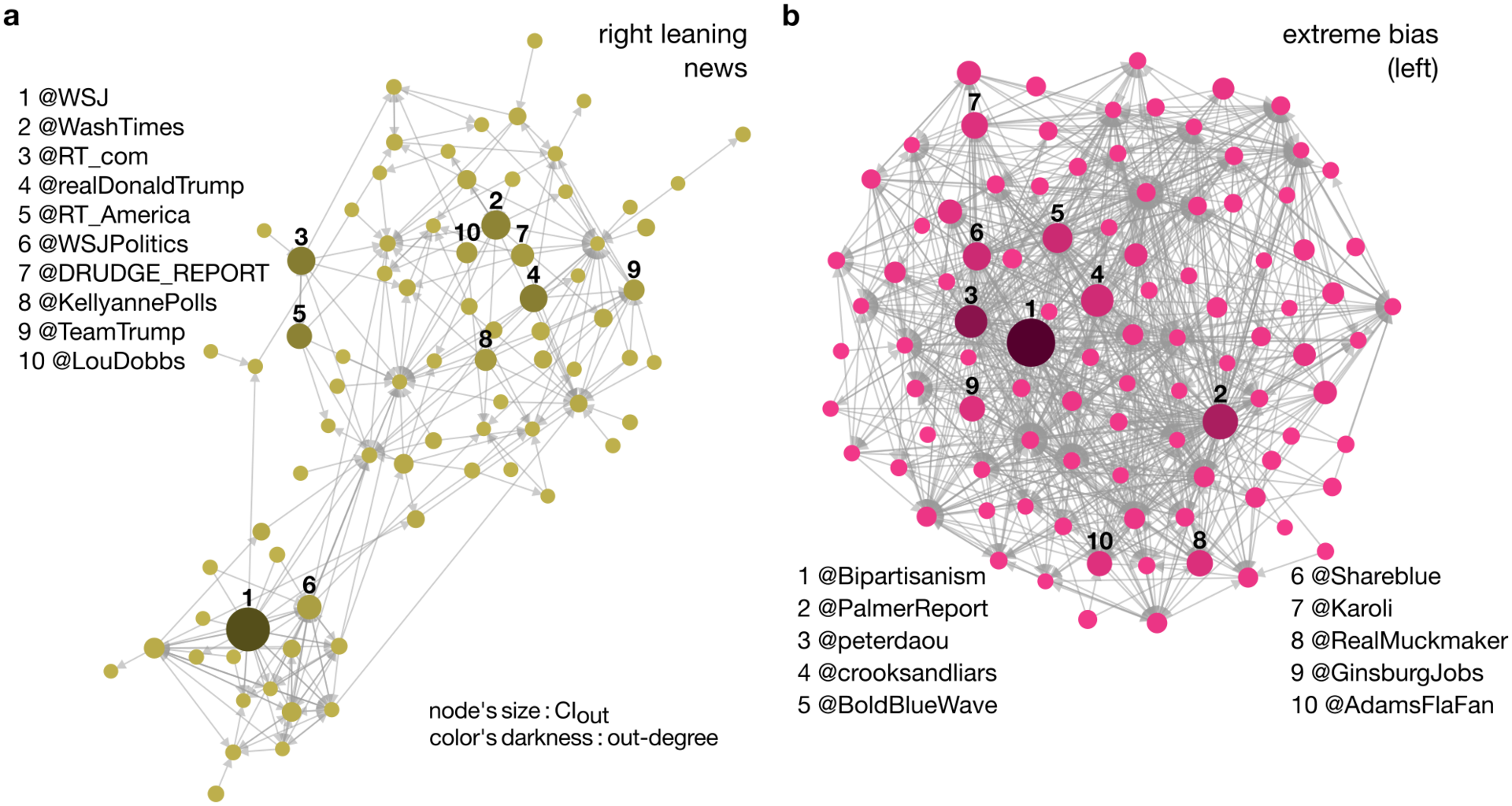}
\caption{\textbf{Retweet networks formed by the top 100 influencers of 
right leaning (a) and extreme bias (left) news (b)}.
The direction of the links represents the flow of information
between users.
The size of the nodes is proportional to their CI$_{\textrm{out}}$ 
values
and the shade of the nodes' color represents their out-degree 
from dark (high out-degree) to light (low out-degree).}
\label{fig:retweet_networks_lrfl}
\end{figure}

\begin{figure}
\centering
\includegraphics[width=\linewidth]{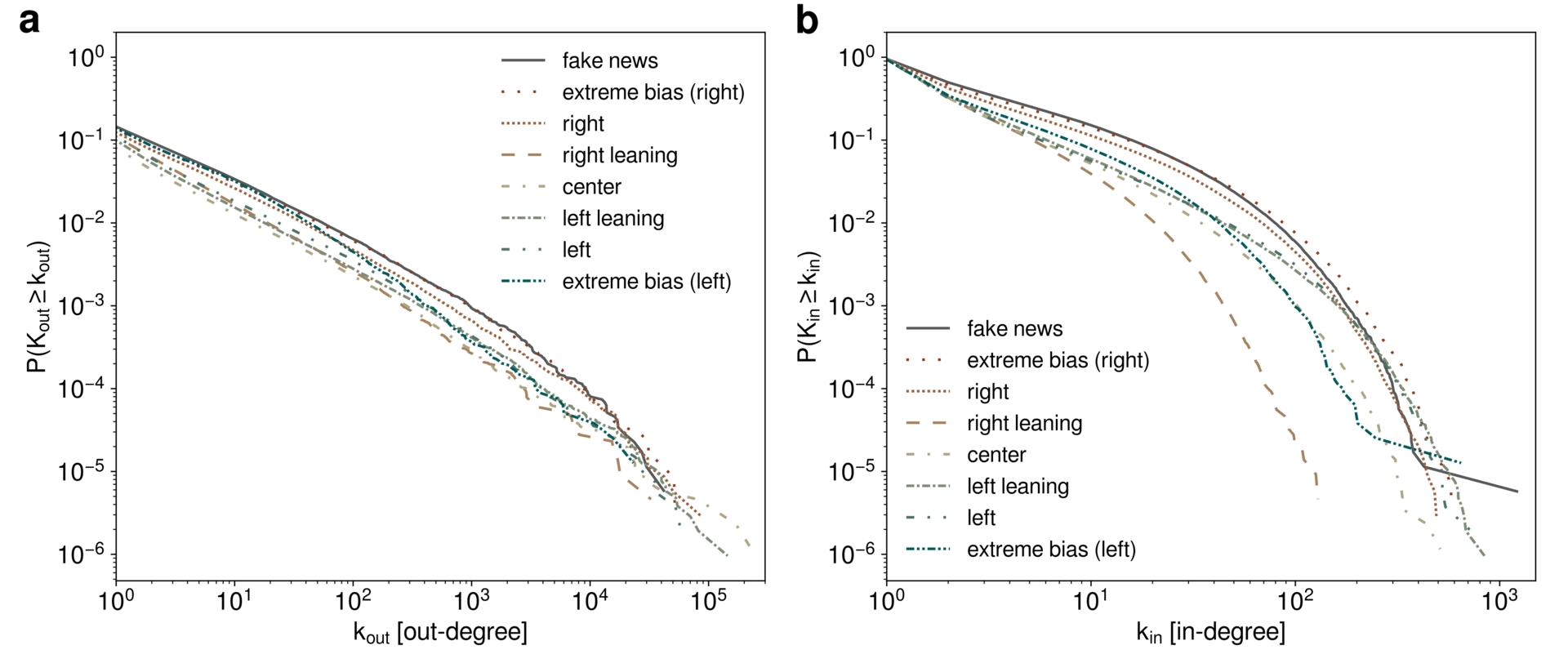}
\caption{\textbf{Empirical complementary cumulative distribution function (CCDF) of the 
out-degree (\textbf{a}) and in-degree (\textbf{b}) of the retweet networks for each media category.}
The CCDF, $P(K\geq k)$, gives the probability  that the in- (or out-) degree of a node is 
greater of equal to $k$.
{The out-degree of a node, i.e. a user, is equal to the number of different users that have
retweeted at least one of her/his tweets with a URL directing to a news outlet.
Its in-degree represents the number of different users she/he retweeted.}
The CCDF of the fake, 
extremely biased (right) and right 
networks are characterized by less steep slopes on the log-log plots
than the other distributions,
resulting in a larger average degree, thus 
indicating a wider diversity of attention from the audience of these news,
i.e. they typically retweet more people and are retweeted by more people,
than the audience of more traditional news.}
\label{fig:degree_CCDF}
\end{figure}

\begin{figure}[tb]
\centering
\includegraphics[width=0.7\linewidth]{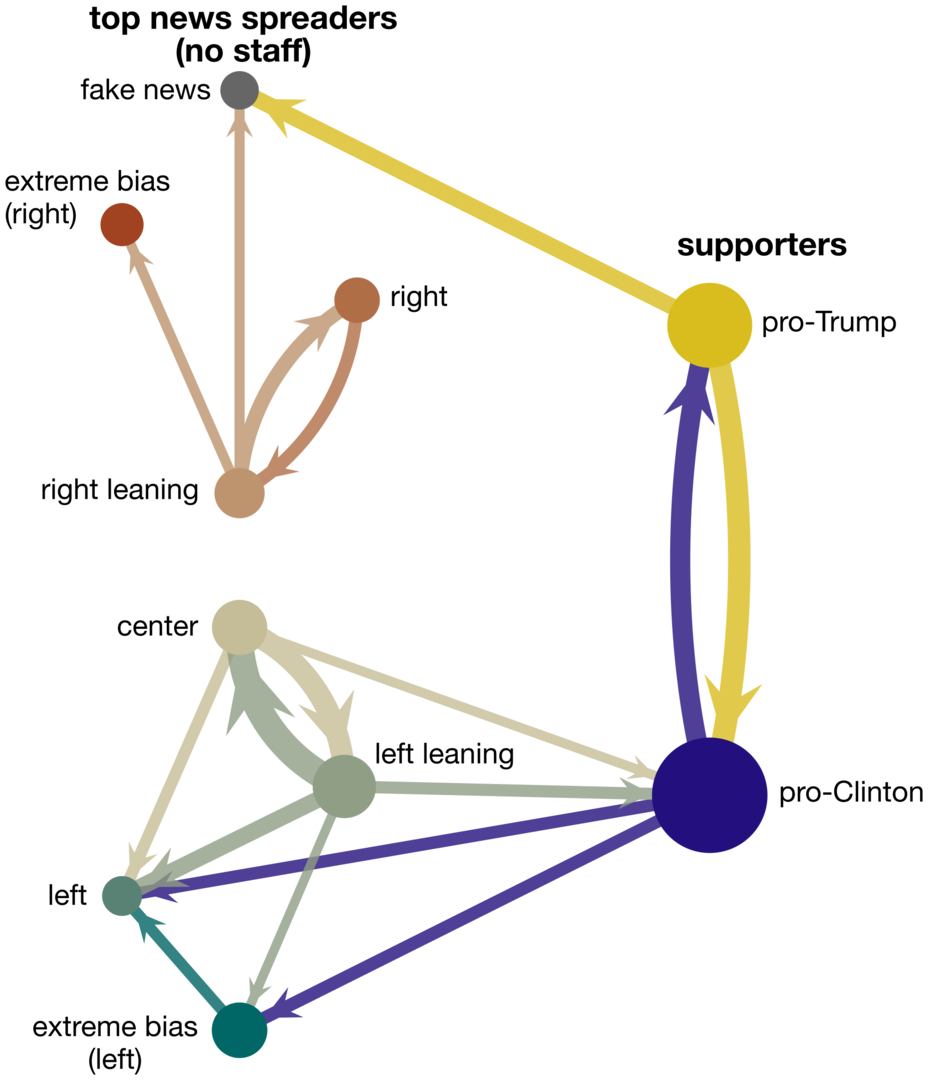}
\caption{\textbf{Causal graph obtained after removing all users linked to the campaign 
staff of each candidate from the influencers}.
We only show causal effects larger than 0.05.}
\label{fig:causal_graph_no_staff}
\end{figure}

\begin{figure}[tb]
\centering
\includegraphics[width=0.7\linewidth]{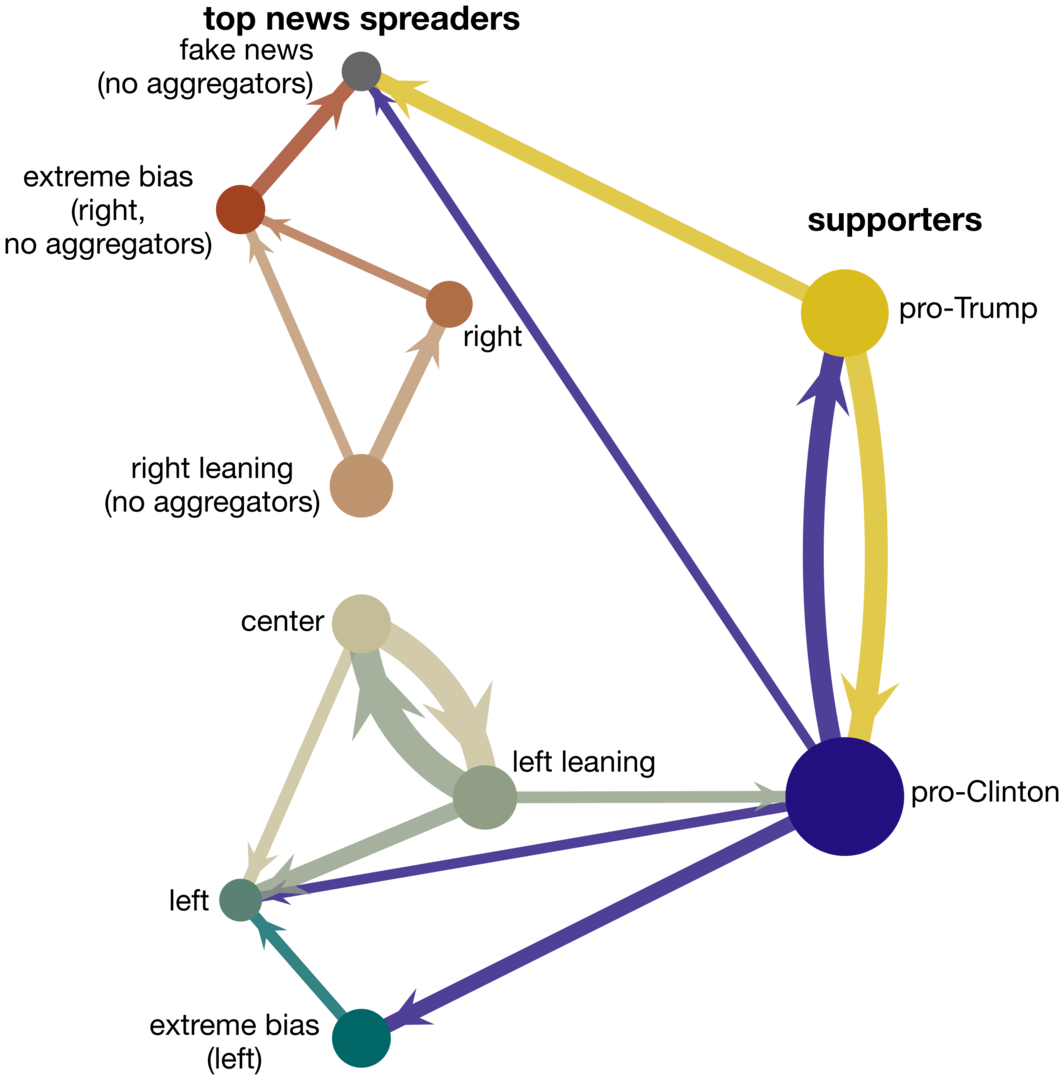}
\caption{\textbf{Causal graph obtained after removing news aggregators websites}.
We only show causal effects larger than 0.05.}
\label{fig:causal_graph_not_aggr}
\end{figure}

\begin{figure}[tb]
\centering
\includegraphics[width=0.5\linewidth]{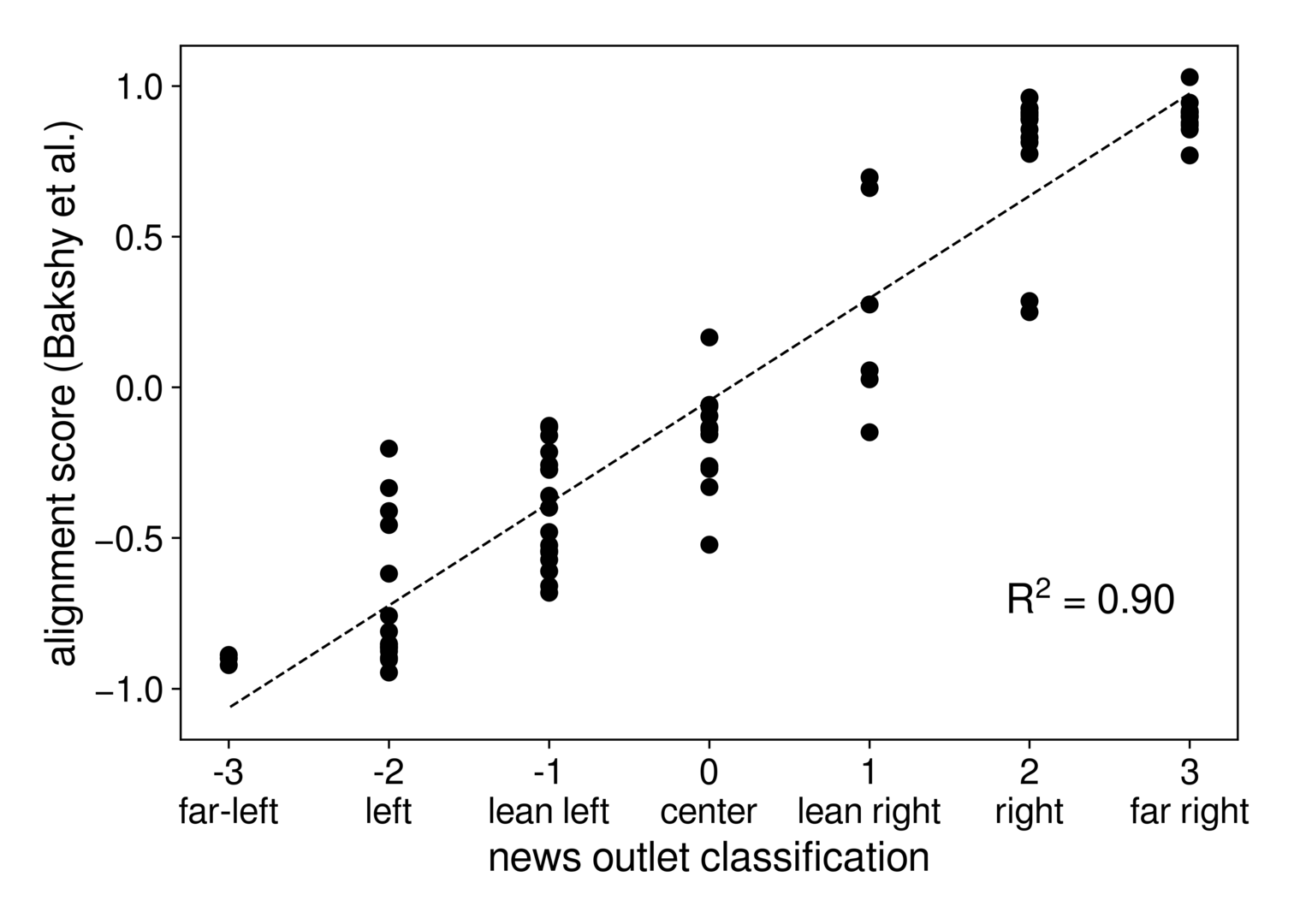}
\caption{\textbf{Comparison of the news outlet political alignment we obtained with the results of \cite{Bakshy2015}.}
}
\label{fig:outlets_alignment}
\end{figure}

\begin{figure}[tb]
\centering
\includegraphics[width=0.9\linewidth]{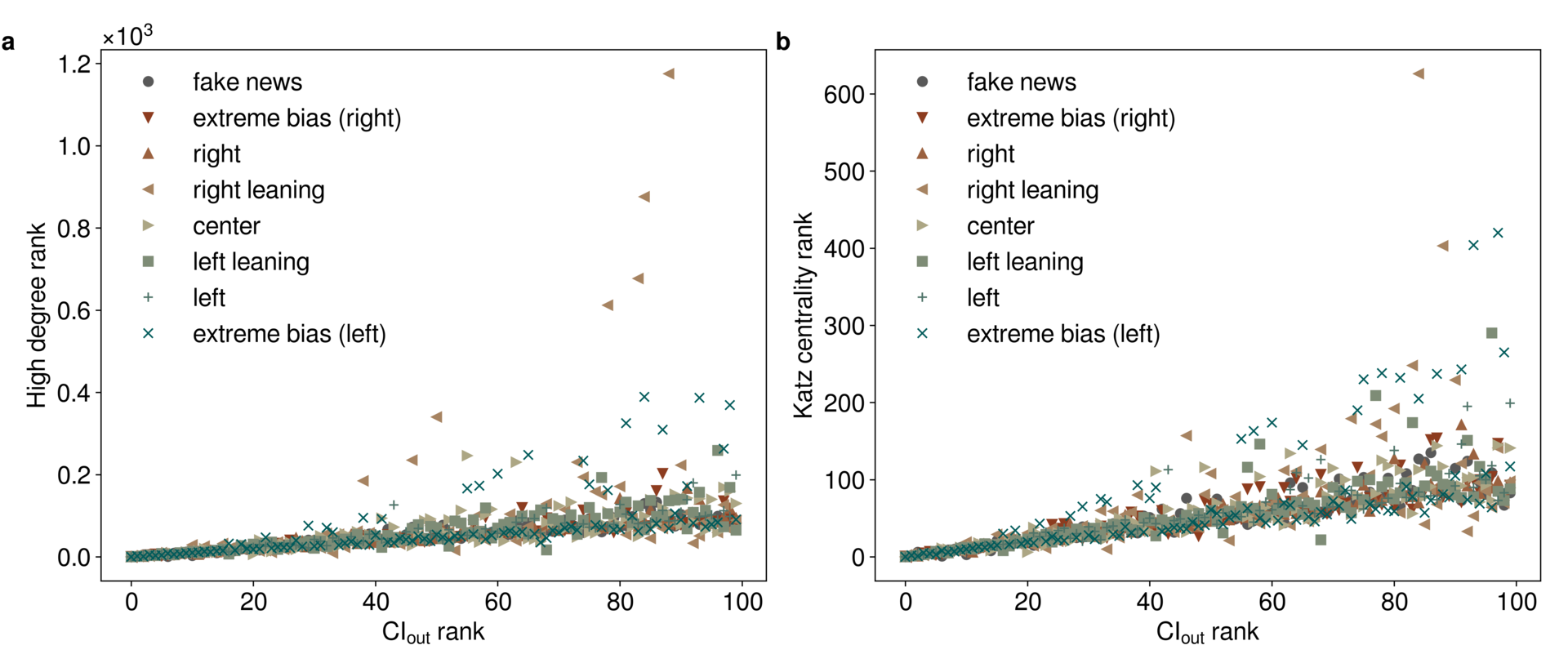}
\caption{\textbf{Comparison of Collective Influence super-spreader ranking (CI$_{\textrm{out}}$) with
High degree ranking (a) and Katz centrality ranking (b)}.}
\label{fig:rank_comparison}
\end{figure}

\begin{figure}[tb]
\centering
\includegraphics[width=\linewidth]{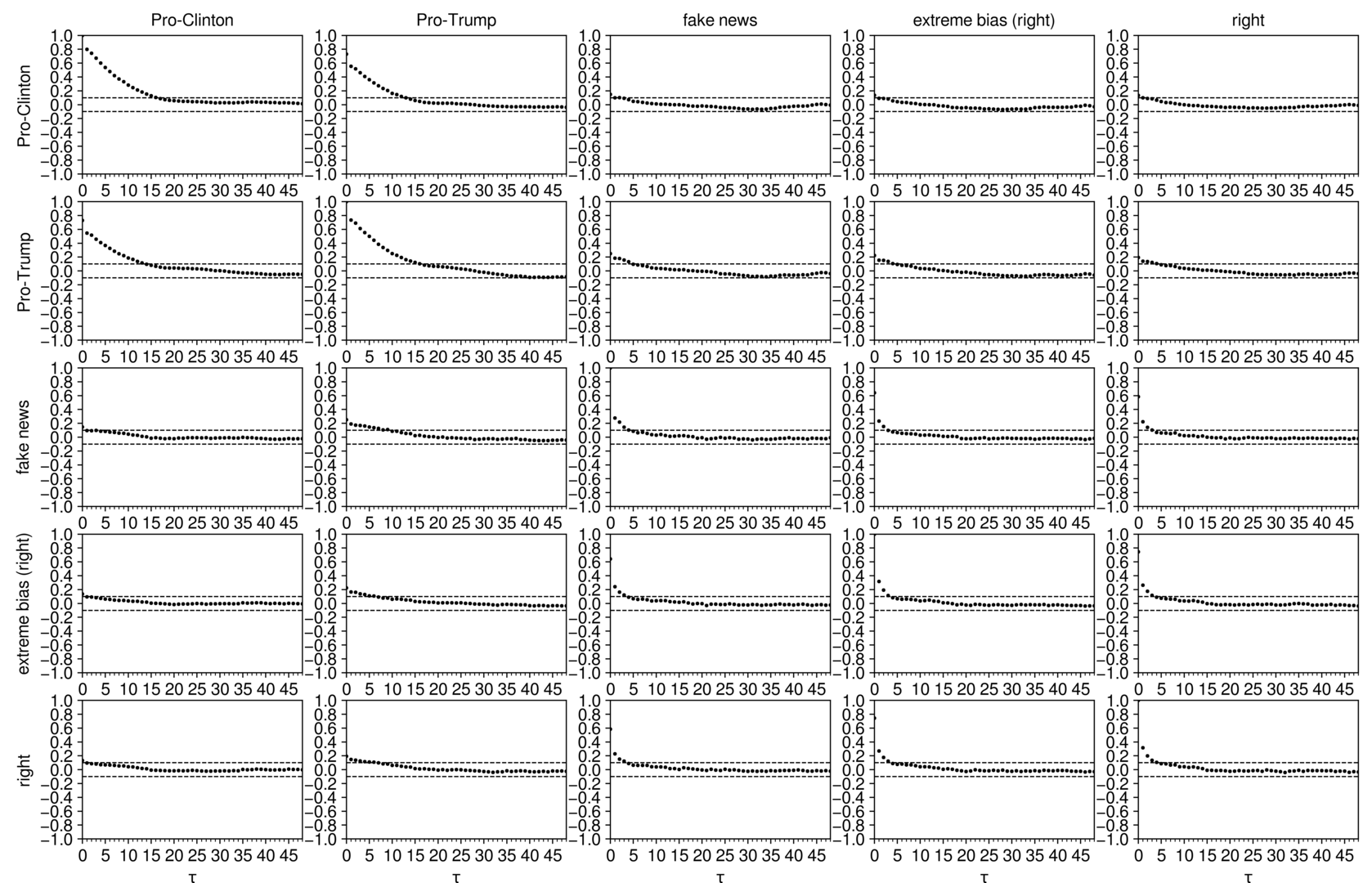}
\caption{\textbf{Pairwise lagged correlations between the activity time series of top 100 influencers 
of fake, extreme bias (right) and right news as well as and Trump and Clinton supporters}.
The time lag, $\tau$, is expressed in data time points corresponding to 15\,min interval.
The horizontal dashed line represents a correlation value of 0.1 and -0.1.}
\label{fig:lagged_corr1}
\end{figure}

\begin{figure}[tb]
\centering
\includegraphics[width=\linewidth]{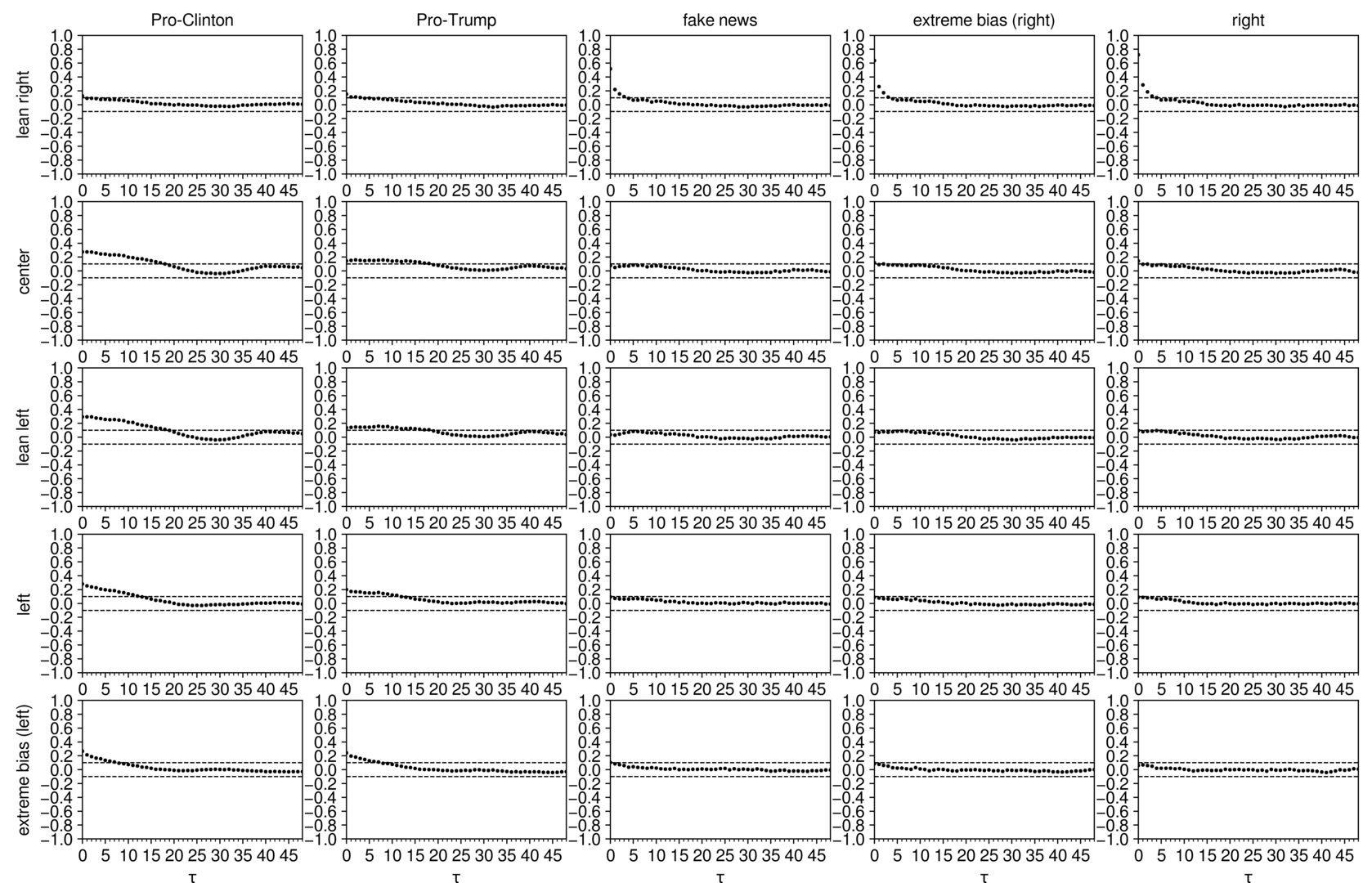}
\caption{\textbf{Pairwise lagged correlations between the activity time series of top 100 influencers 
of fake, extreme bias (right), right leaning, center, left leaning, left and extreme bias (left) news as well as and Trump and Clinton supporters}.
The time lag, $\tau$, is expressed in data time points corresponding to 15\,min interval.
The horizontal dashed line represents a correlation value of 0.1 and -0.1.}
\label{fig:lagged_corr2}
\end{figure}

\begin{figure}[tb]
\centering
\includegraphics[width=\linewidth]{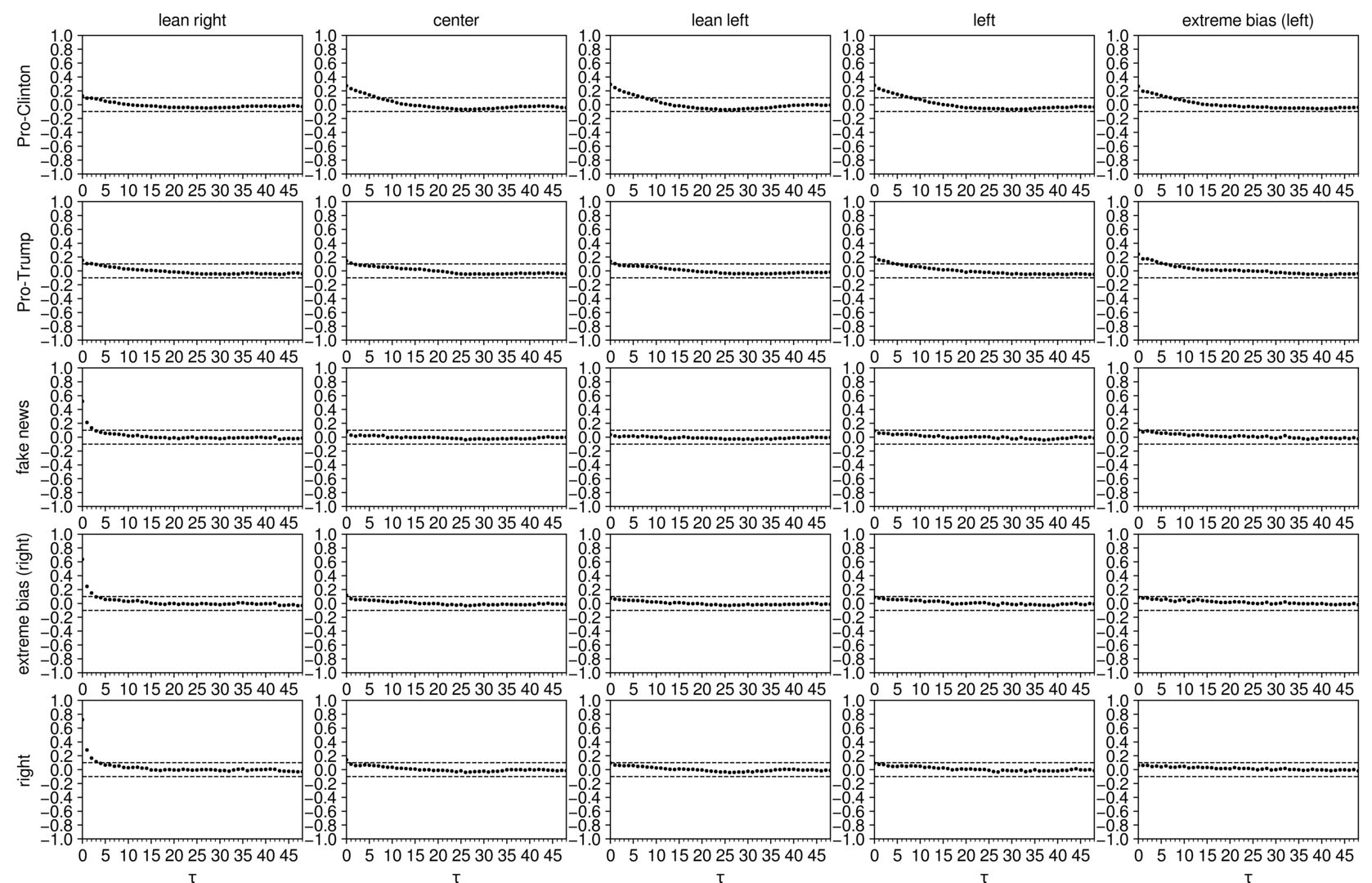}
\caption{\textbf{Pairwise lagged correlations between the activity time series of top 100 influencers 
of fake, extreme bias (right), right leaning, center, left leaning, left and extreme bias (left) news as well as and Trump and Clinton supporters}.
The time lag, $\tau$, is expressed in data time points corresponding to 15\,min interval.
The horizontal dashed line represents a correlation value of 0.1 and -0.1.}
\label{fig:lagged_corr3}
\end{figure}

\begin{figure}[tb]
\centering
\includegraphics[width=\linewidth]{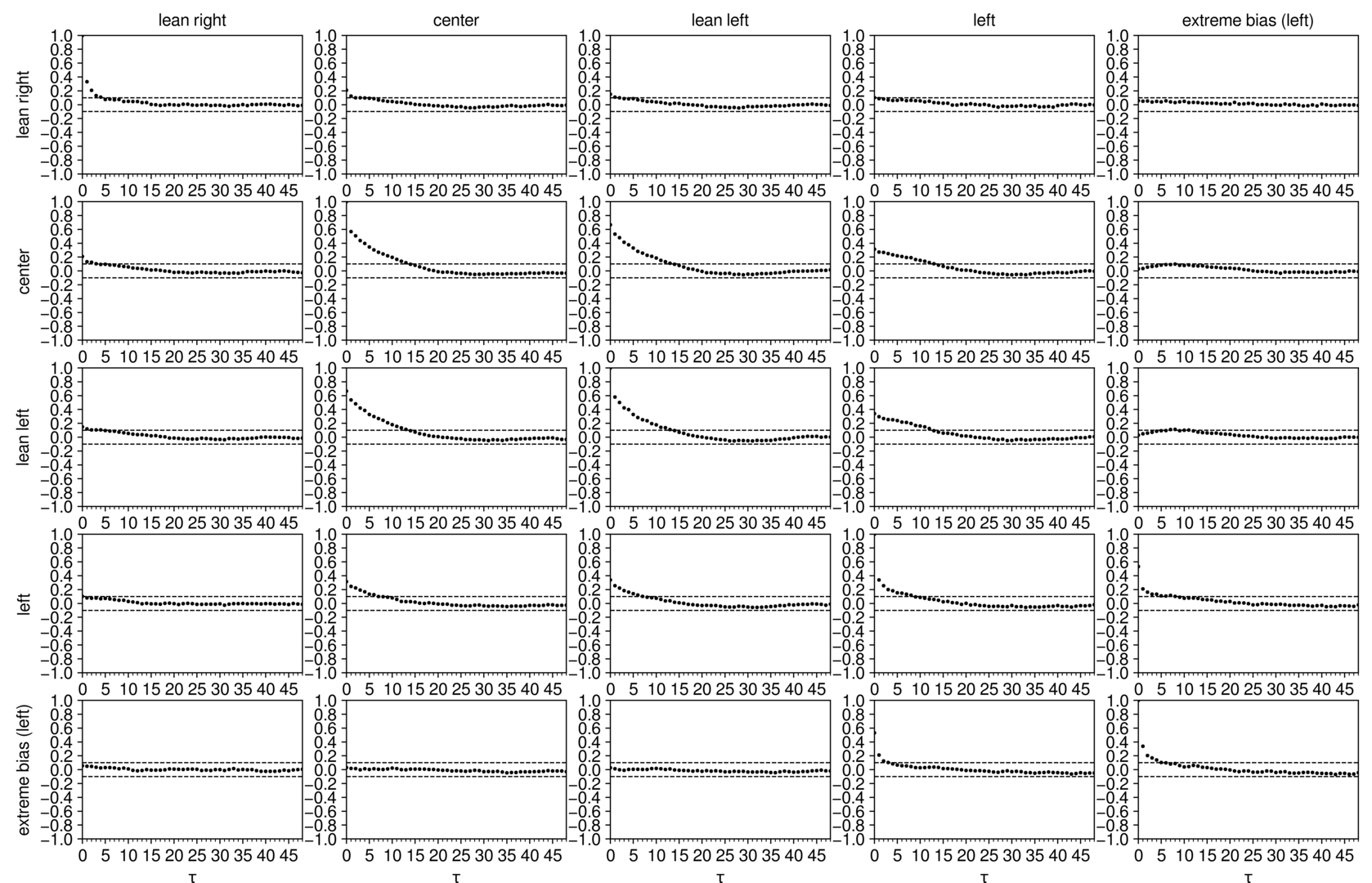}
\caption{\textbf{Pairwise lagged correlations between the activity time series of top 100 influencers 
of right leaning, center, left leaning, left and extreme bias (left) news}.
The time lag, $\tau$, is expressed in data time points corresponding to 15\,min interval.
The horizontal dashed lines represents a correlation value of 0.1 and -0.1.}
\label{fig:lagged_corr4}
\end{figure}


\begin{thebibliography}{10}
\expandafter\ifx\csname url\endcsname\relax
  \def\url#1{\texttt{#1}}\fi
\expandafter\ifx\csname urlprefix\endcsname\relax\def\urlprefix{}\fi
\providecommand{\bibinfo}[2]{#2}
\providecommand{\eprint}[2][]{\url{#2}}

\bibitem{Allcott2017}
\bibinfo{author}{Allcott, H.} \& \bibinfo{author}{Gentzkow, M.}
\newblock \bibinfo{title}{{Social Media and Fake News in the 2016 Election}}.
\newblock \bibinfo{type}{Tech. Rep.} \bibinfo{number}{2},
  \bibinfo{institution}{National Bureau of Economic Research},
  \bibinfo{address}{Cambridge, MA} (\bibinfo{year}{2017}).
\newblock \urlprefix\url{http://www.nber.org/papers/w23089.pdf}.

\bibitem{PoliticoFakeNews}
\bibinfo{author}{{Politico}}.
\newblock \bibinfo{title}{The long and brutal history of fake news}
  (\bibinfo{year}{2017}).
\newblock
  \urlprefix\url{https://www.politico.com/magazine/story/2016/12/fake-news-history-long-violent-214535}.
\newblock (\bibinfo{year}{2016}).

\bibitem{howell2013digital}
\bibinfo{author}{Howell, L.} \emph{et~al.}
\newblock \bibinfo{title}{Digital wildfires in a hyperconnected world}.
\newblock \emph{\bibinfo{journal}{WEF Report}} \textbf{\bibinfo{volume}{3}},
  \bibinfo{pages}{15--94} (\bibinfo{year}{2013}).
    \urlprefix\url{http://reports.weforum.org/global-risks-2013/risk-case-1/digital-wildfires-in-a-hyperconnected-world}.

  
  

\bibitem{Bessi2015PlosOne}
\bibinfo{author}{Bessi, A.} \emph{et~al.}
\newblock \bibinfo{title}{{Science vs Conspiracy: Collective Narratives in the
  Age of Misinformation}}.
\newblock \emph{\bibinfo{journal}{PloS one}} \textbf{\bibinfo{volume}{10}},
  \bibinfo{pages}{e0118093} (\bibinfo{year}{2015}).
\newblock \urlprefix\url{http://dx.plos.org/10.1371/journal.pone.0118093}.

\bibitem{Bessi2015}
\bibinfo{author}{Bessi, A.} \emph{et~al.}
\newblock \bibinfo{title}{{Viral Misinformation}}.
\newblock In \emph{\bibinfo{booktitle}{Proc. 24th International
  Conference on World Wide Web - WWW '15 Companion}}, \bibinfo{pages}{355--356}
  (\bibinfo{publisher}{ACM Press}, \bibinfo{address}{New York, New York, USA},
  \bibinfo{year}{2015}).
\newblock \urlprefix\url{http://dl.acm.org/citation.cfm?doid=2740908.2745939}.

\bibitem{Mocanu2015}
\bibinfo{author}{Mocanu, D.}, \bibinfo{author}{Rossi, L.},
  \bibinfo{author}{Zhang, Q.}, \bibinfo{author}{Karsai, M.} \&
  \bibinfo{author}{Quattrociocchi, W.}
\newblock \bibinfo{title}{{Collective attention in the age of
  (mis)information}}.
\newblock \emph{\bibinfo{journal}{Comput. Hum. Behav.}}
  \textbf{\bibinfo{volume}{51}}, \bibinfo{pages}{1198--1204}
  (\bibinfo{year}{2015}).
\newblock
  \urlprefix\url{http://linkinghub.elsevier.com/retrieve/pii/S0747563215000382}.


\bibitem{Bessi2015PlosTwo}
\bibinfo{author}{Bessi, A.} \emph{et~al.}
\newblock \bibinfo{title}{{Trend of Narratives in the Age of Misinformation}}.
\newblock \emph{\bibinfo{journal}{PloS one}} \textbf{\bibinfo{volume}{10}},
  \bibinfo{pages}{e0134641} (\bibinfo{year}{2015}).
\newblock \urlprefix\url{http://dx.plos.org/10.1371/journal.pone.0134641}.

\bibitem{Bessi2016}
\bibinfo{author}{Bessi, A.} \emph{et~al.}
\newblock \bibinfo{title}{{Homophily and polarization in the age of
  misinformation}}.
\newblock \emph{\bibinfo{journal}{Eur. Phys. J. Spec. Top.}} \textbf{\bibinfo{volume}{225}}, \bibinfo{pages}{2047--2059}
  (\bibinfo{year}{2016}).
\newblock \urlprefix\url{http://link.springer.com/10.1140/epjst/e2015-50319-0}.

\bibitem{DelVicario2016}
\bibinfo{author}{{Del Vicario}, M.} \emph{et~al.}
\newblock \bibinfo{title}{{The spreading of misinformation online}}.
\newblock \emph{\bibinfo{journal}{Proc. Natl. Acad. Sci.}} \textbf{\bibinfo{volume}{113}}, \bibinfo{pages}{554--559}
  (\bibinfo{year}{2016}).
\newblock
  \urlprefix\url{http://www.pnas.org/content/early/2016/01/02/1517441113}.

\bibitem{DelVicario2017}
\bibinfo{author}{{Del Vicario}, M.}, \bibinfo{author}{Gaito, S.},
  \bibinfo{author}{Quattrociocchi, W.}, \bibinfo{author}{Zignani, M.} \&
  \bibinfo{author}{Zollo, F.}
\newblock \bibinfo{title}{{Public discourse and news consumption on online
  social media: A quantitative, cross-platform analysis of the Italian
  Referendum}}.
\newblock {Preprint at \url{http://arxiv.org/abs/1702.06016}} (\bibinfo{year}{2017})

\bibitem{Shao2016}
\bibinfo{author}{Shao, C.}, \bibinfo{author}{Ciampaglia, G.~L.},
  \bibinfo{author}{Flammini, A.} \& \bibinfo{author}{Menczer, F.}
\newblock \bibinfo{title}{{Hoaxy}}.
\newblock In \emph{\bibinfo{booktitle}{Proc. 25th International
  Conference Companion on World Wide Web - WWW '16 Companion}},
  \bibinfo{pages}{745--750} (\bibinfo{publisher}{ACM Press},
  \bibinfo{address}{New York, New York, USA}, \bibinfo{year}{2016}).
\newblock
  \urlprefix\url{http://dx.doi.org/10.1145/2872518.2890098}.

\bibitem{Vosoughi2018}
\bibinfo{author}{Vosoughi, S.}, \bibinfo{author}{Roy, D.} \&
  \bibinfo{author}{Aral, S.}
\newblock \bibinfo{title}{{The spread of true and false news online}}.
\newblock \emph{\bibinfo{journal}{Science}} \textbf{\bibinfo{volume}{359}},
  \bibinfo{pages}{1146--1151} (\bibinfo{year}{2018}).
\newblock
  \urlprefix\url{http://www.sciencemag.org/lookup/doi/10.1126/science.aap9559}.

\bibitem{Shao2018}
\bibinfo{author}{Shao, C.} \emph{et~al.}
\newblock \bibinfo{title}{{Anatomy of an online misinformation network}}.
\newblock \emph{\bibinfo{journal}{PloS one}} \textbf{\bibinfo{volume}{13}},
  \bibinfo{pages}{1--23} (\bibinfo{year}{2018}).
\newblock \urlprefix\url{https://doi.org/10.1371/journal.pone.0196087}.


\bibitem{Bessi2016youtube}
\bibinfo{author}{Bessi, A.} \emph{et~al.}
\newblock \bibinfo{title}{{Users polarization on Facebook and Youtube}}.
\newblock \emph{\bibinfo{journal}{PloS one}} \textbf{\bibinfo{volume}{11}},
  \bibinfo{pages}{1--24} (\bibinfo{year}{2016}).
\newblock \urlprefix\url{https://doi.org/10.1371/journal.pone.0159641}.

\bibitem{Kumar2016}
\bibinfo{author}{Kumar, S.}, \bibinfo{author}{West, R.} \&
  \bibinfo{author}{Leskovec, J.}
\newblock \bibinfo{title}{{Disinformation on the Web}}.
\newblock In \emph{\bibinfo{booktitle}{Proc. 25th International
  Conference on World Wide Web - WWW '16}}, \bibinfo{pages}{591--602}
  (\bibinfo{publisher}{ACM Press}, \bibinfo{address}{New York, New York, USA},
  \bibinfo{year}{2016}).
\newblock \urlprefix\url{http://dl.acm.org/citation.cfm?doid=2872427.2883085}.

\bibitem{DelVicario2017SRep}
\bibinfo{author}{{Del Vicario}, M.}, \bibinfo{author}{Scala, A.},
  \bibinfo{author}{Caldarelli, G.}, \bibinfo{author}{Stanley, H.~E.} \&
  \bibinfo{author}{Quattrociocchi, W.}
\newblock \bibinfo{title}{{Modeling confirmation bias and polarization}}.
\newblock \emph{\bibinfo{journal}{Sci. Rep.}}
  \textbf{\bibinfo{volume}{7}}, \bibinfo{pages}{40391} (\bibinfo{year}{2017}).
\newblock \urlprefix\url{http://www.nature.com/articles/srep40391}.

\bibitem{Askitas2017}
\bibinfo{author}{Askitas, N.}
\newblock \bibinfo{title}{{Explaining opinion polarisation with opinion
  copulas}}.
\newblock \emph{\bibinfo{journal}{PloS one}} \textbf{\bibinfo{volume}{12}},
  \bibinfo{pages}{e0183277} (\bibinfo{year}{2017}).
\newblock \urlprefix\url{http://dx.plos.org/10.1371/journal.pone.0183277}.

\bibitem{Klayman1987}
\bibinfo{author}{Klayman, J.} \& \bibinfo{author}{Ha, Y.-W.}
\newblock \bibinfo{title}{{Confirmation, disconfirmation, and information in
  hypothesis testing.}}
\newblock \emph{\bibinfo{journal}{Psychol. Rev.}}
  \textbf{\bibinfo{volume}{94}}, \bibinfo{pages}{211--228}
  (\bibinfo{year}{1987}).
\newblock
  \urlprefix\url{http://doi.apa.org/getdoi.cfm?doi=10.1037/0033-295X.94.2.211}.

\bibitem{Qiu2017}
\bibinfo{author}{Qiu, X.}, \bibinfo{author}{{F. M. Oliveira}, D.},
  \bibinfo{author}{{Sahami Shirazi}, A.}, \bibinfo{author}{Flammini, A.} \&
  \bibinfo{author}{Menczer, F.}
\newblock \bibinfo{title}{{Limited individual attention and online virality of
  low-quality information}}.
\newblock \emph{\bibinfo{journal}{Nat. Hum. Behav.}}
  \textbf{\bibinfo{volume}{1}}, \bibinfo{pages}{0132} (\bibinfo{year}{2017}).
\newblock \urlprefix\url{http://www.nature.com/articles/s41562-017-0132}.

\bibitem{Schmidt2017}
\bibinfo{author}{Schmidt, A.~L.} \emph{et~al.}
\newblock \bibinfo{title}{{Anatomy of news consumption on Facebook}}.
\newblock \emph{\bibinfo{journal}{Proc. Natl. Acad. Sci.}} \textbf{\bibinfo{volume}{114}}, \bibinfo{pages}{3035--3039}
  (\bibinfo{year}{2017}).
\newblock
  \urlprefix\url{http://www.pnas.org/lookup/doi/10.1073/pnas.1617052114}.

\bibitem{DelVicario2017Brexit}
\bibinfo{author}{{Del Vicario}, M.}, \bibinfo{author}{Zollo, F.},
  \bibinfo{author}{Caldarelli, G.}, \bibinfo{author}{Scala, A.} \&
  \bibinfo{author}{Quattrociocchi, W.}
\newblock \bibinfo{title}{{Mapping social dynamics on Facebook: The Brexit
  debate}}.
\newblock \emph{\bibinfo{journal}{Soc. Networks}}
  \textbf{\bibinfo{volume}{50}}, \bibinfo{pages}{6--16} (\bibinfo{year}{2017}).
\newblock \urlprefix\url{http://dx.doi.org/10.1016/j.socnet.2017.02.002}.

\bibitem{Bakshy2015}
\bibinfo{author}{Bakshy, E.}, \bibinfo{author}{Messing, S.} \&
  \bibinfo{author}{Adamic, L.~A.}
\newblock \bibinfo{title}{{Exposure to ideologically diverse news and opinion
  on Facebook}}.
\newblock \emph{\bibinfo{journal}{Science}} \textbf{\bibinfo{volume}{348}},
  \bibinfo{pages}{1130--1132} (\bibinfo{year}{2015}).
\newblock
  \urlprefix\url{http://www.sciencemag.org/cgi/doi/10.1126/science.aaa1160}.

\bibitem{Lee2006}
\bibinfo{author}{Lee, K.}, \bibinfo{author}{Eoff, B.~D.} \&
  \bibinfo{author}{Caverlee, J.}
\newblock \bibinfo{title}{{Seven Months with the Devils: A Long-Term Study of
  Content Polluters on Twitter}}.
\newblock In \emph{\bibinfo{booktitle}{Proc. 5th Int. AAAI Conf. Weblogs Soc.
  Media}}, \bibinfo{pages}{185--192}
  (\bibinfo{year}{2006}).
  \urlprefix\url{https://www.aaai.org/ocs/index.php/ICWSM/ICWSM11/paper/viewFile/2780/3296}.


\bibitem{Bessi2016Bots}
\bibinfo{author}{Bessi, A.} \& \bibinfo{author}{Ferrara, E.}
\newblock \bibinfo{title}{{Social bots distort the 2016 U.S. Presidential
  election online discussion}}.
\newblock \emph{\bibinfo{journal}{First Monday}} \textbf{\bibinfo{volume}{21}}
  (\bibinfo{year}{2016}).
  \urlprefix\url{https://doi.org/10.5210/fm.v21i11.7090}.



\bibitem{Ferrara2016}
\bibinfo{author}{Ferrara, E.}, \bibinfo{author}{Varol, O.},
  \bibinfo{author}{Davis, C.}, \bibinfo{author}{Menczer, F.} \&
  \bibinfo{author}{Flammini, A.}
\newblock \bibinfo{title}{{The rise of social bots}}.
\newblock \emph{\bibinfo{journal}{Commun. ACM}}
  \textbf{\bibinfo{volume}{59}}, \bibinfo{pages}{96--104}
  (\bibinfo{year}{2016}).
\newblock \url{http://dl.acm.org/citation.cfm?doid=2963119.2818717}.

\bibitem{Shao2017}
\bibinfo{author}{Shao, C.}, \bibinfo{author}{Ciampaglia, G.~L.},
  \bibinfo{author}{Varol, O.}, \bibinfo{author}{Yang, K.}, 
  \bibinfo{author}{Flammini, A.} \&
  \bibinfo{author}{Menczer, F.}
\newblock \bibinfo{title}{{The spread of low-credibility content by social bots}}.
\newblock {Preprint at \url{http://arxiv.org/abs/1707.07592}} (\bibinfo{year}{2018}).

\bibitem{Bovet2017TwitterOpinion}
\bibinfo{author}{Bovet, A.}, \bibinfo{author}{Morone, F.} \&
  \bibinfo{author}{Makse, H.~A.}
\newblock \bibinfo{title}{{Validation of Twitter opinion trends with national
  polling aggregates: Hillary Clinton vs Donald Trump}}.
\newblock \emph{\bibinfo{journal}{Sci. Rep.}}
  \textbf{\bibinfo{volume}{8}}, \bibinfo{pages}{8673} (\bibinfo{year}{2018}).
\newblock \urlprefix\url{http://www.nature.com/articles/s41598-018-26951-y}.

\bibitem{Runge2015}
\bibinfo{author}{Runge, J.} \emph{et~al.}
\newblock \bibinfo{title}{{Identifying causal gateways and mediators in complex
  spatio-temporal systems}}.
\newblock \emph{\bibinfo{journal}{Nat. Commun.}}
  \textbf{\bibinfo{volume}{6}}, \bibinfo{pages}{8502} (\bibinfo{year}{2015}).
\newblock \urlprefix\url{http://www.nature.com/articles/ncomms9502}.

\bibitem{Barthelemy2004}
\bibinfo{author}{Barth{\'{e}}lemy, M.}, \bibinfo{author}{Barrat, A.},
  \bibinfo{author}{Pastor-Satorras, R.} \& \bibinfo{author}{Vespignani, A.}
\newblock \bibinfo{title}{{Velocity and Hierarchical Spread of Epidemic
  Outbreaks in Scale-Free Networks}}.
\newblock \emph{\bibinfo{journal}{Phys. Rev. Lett.}}
  \textbf{\bibinfo{volume}{92}}, \bibinfo{pages}{178701}
  (\bibinfo{year}{2004}).
\newblock
  \urlprefix\url{https://link.aps.org/doi/10.1103/PhysRevLett.92.178701}.

\bibitem{Vespignani2011}
\bibinfo{author}{Vespignani, A.}
\newblock \bibinfo{title}{{Modelling dynamical processes in complex
  socio-technical systems}}.
\newblock \emph{\bibinfo{journal}{Nat. Phys.}}
  \textbf{\bibinfo{volume}{8}}, \bibinfo{pages}{32--39} (\bibinfo{year}{2011}).
\newblock \urlprefix\url{http://www.nature.com/doifinder/10.1038/nphys2160}.



\bibitem{Goel2012}
\bibinfo{author}{Goel, S.}, \bibinfo{author}{Watts, D. J.},
  \& \bibinfo{author}{Goldstein, D. G.}
  \newblock \bibinfo{title}{{The Structure of Online Diffusion Networks}}.
\newblock \emph{\bibinfo{journal}{Proc. 13th ACM Conference on Electronic Commerce}}
  \textbf{\bibinfo{volume}{1}}, \bibinfo{pages}{623--638} (\bibinfo{year}{2012}).
\newblock \urlprefix\url{http://dl.acm.org/citation.cfm?id=2229058}.

\bibitem{Morone2015}
\bibinfo{author}{Morone, F.} \& \bibinfo{author}{Makse, H.~A.}
\newblock \bibinfo{title}{{Influence maximization in complex networks through
  optimal percolation}}.
\newblock \emph{\bibinfo{journal}{Nature}} 
 \textbf{\bibinfo{volume}{524}}, \bibinfo{pages}{65--68} (\bibinfo{year}{2015}).
\newblock \urlprefix\url{http://dx.doi.org/10.1038/nature14604}.


\bibitem{Cleveland1990}
\bibinfo{author}{Cleveland, R.~B.}, \bibinfo{author}{Cleveland, W.~S.},
  \bibinfo{author}{McRae, J.~E.} \& \bibinfo{author}{Terpenning, I.}
\newblock \bibinfo{title}{{STL: A seasonal-trend decomposition procedure based
  on loess}}.
\newblock \emph{\bibinfo{journal}{J. Off. Stat.}}
  \textbf{\bibinfo{volume}{6}}, \bibinfo{pages}{3--73} (\bibinfo{year}{1990}).
\newblock \urlprefix\url{http://www.jos.nu/Articles/abstract.asp?article=613}.


\bibitem{Margolin2018}
\bibinfo{author}{Margolin, D.~B.}, \bibinfo{author}{Hannak, A.} \&
  \bibinfo{author}{Weber, I.}
\newblock \bibinfo{title}{{Political Fact-Checking on Twitter: When Do
  Corrections Have an Effect?}}
\newblock \emph{\bibinfo{journal}{Political Commun.}}
  \textbf{\bibinfo{volume}{35}}, \bibinfo{pages}{196--219}
  (\bibinfo{year}{2018}).
\newblock \urlprefix\url{https://doi.org/10.1080/10584609.2017.1334018}.





\bibitem{Spirtes2000}
\bibinfo{author}{Spirtes, P.}, \bibinfo{author}{Glymour, C.} \&
  \bibinfo{author}{Scheines, R.}
\newblock \emph{\bibinfo{title}{{Causation, Prediction, and Search}}}
  (\bibinfo{publisher}{MIT Press}, \bibinfo{year}{2000}).

\bibitem{Runge2012}
\bibinfo{author}{Runge, J.}, \bibinfo{author}{Heitzig, J.},
  \bibinfo{author}{Petoukhov, V.} \& \bibinfo{author}{Kurths, J.}
\newblock \bibinfo{title}{{Escaping the Curse of Dimensionality in Estimating
  Multivariate Transfer Entropy}}.
\newblock \emph{\bibinfo{journal}{Phys. Rev. Lett.}}
  \textbf{\bibinfo{volume}{108}}, \bibinfo{pages}{258701}
  (\bibinfo{year}{2012}).
\newblock
  \urlprefix\url{https://link.aps.org/doi/10.1103/PhysRevLett.108.258701}.
  


\bibitem{Zhang2011}
\bibinfo{author}{Zhang, K.}, \bibinfo{author}{Peters, J.},
  \bibinfo{author}{Janzing, D.} \& \bibinfo{author}{Schoelkopf, B.}
\newblock \bibinfo{title}{{Kernel-based Conditional Independence Test and
  Application in Causal Discovery}}.
\newblock In \emph{\bibinfo{booktitle}{{UAI} 2011, Proc. 27th Conference on Uncertainty in Artificial Intelligence}},
  \bibinfo{pages}{804--813} (\bibinfo{address}{Barcelona},
  \bibinfo{year}{2011}).
\newblock \urlprefix\url{http://arxiv.org/abs/1202.3775}.

\bibitem{Strobl2017}
\bibinfo{author}{Strobl, E.~V.}, \bibinfo{author}{Zhang, K.} \&
  \bibinfo{author}{Visweswaran, S.}
\newblock \bibinfo{title}{{Approximate Kernel-based Conditional Independence
  Tests for Fast Non-Parametric Causal Discovery}}.
\newblock {Preprint at \url{http://arxiv.org/abs/1702.03877}} (\bibinfo{year}{2017}).

\bibitem{Varol2017}
\bibinfo{author}{Varol, O.}, \bibinfo{author}{Ferrara, E.},
  \bibinfo{author}{Davis, C.~A.}, \bibinfo{author}{Menczer, F.} \&
  \bibinfo{author}{Flammini, A.}
\newblock \bibinfo{title}{{Online human-bot interactions: detection,
  estimation, and characterization}}.
\newblock In \emph{\bibinfo{booktitle}{Proc. 11th Int. AAAI Conf. Weblogs Soc.
  Media}}, \bibinfo{pages}{280--289} (\bibinfo{year}{2017}).
\newblock \urlprefix\url{https://aaai.org/ocs/index.php/ICWSM/ICWSM17/paper/view/15587}.

  
  


\bibitem{Morone2016}
\bibinfo{author}{Morone, F.}, \bibinfo{author}{Min, B.}, \bibinfo{author}{Bo,
  L.}, \bibinfo{author}{Mari, R.} \& \bibinfo{author}{Makse, H.~A.}
\newblock \bibinfo{title}{{Collective Influence Algorithm to find influencers
  via optimal percolation in massively large social media}}.
\newblock \emph{\bibinfo{journal}{Sci. Rep.}}
  \textbf{\bibinfo{volume}{6}}, \bibinfo{pages}{30062} (\bibinfo{year}{2016}).
\newblock \urlprefix\url{http://www.nature.com/articles/srep30062}.

\bibitem{Teng2016}
\bibinfo{author}{Teng, X.}, \bibinfo{author}{Pei, S.}, \bibinfo{author}{Morone,
  F.} \& \bibinfo{author}{Makse, H.~A.}
\newblock \bibinfo{title}{{Collective Influence of Multiple Spreaders Evaluated
  by Tracing Real Information Flow in Large-Scale Social Networks}}.
\newblock \emph{\bibinfo{journal}{Sci. Rep.}}
  \textbf{\bibinfo{volume}{6}}, \bibinfo{pages}{36043} (\bibinfo{year}{2016}).
\newblock \urlprefix\url{http://www.nature.com/articles/srep36043}.

\bibitem{Katz1953}
\bibinfo{author}{Katz, L.}
\newblock \bibinfo{title}{{A new status index derived from sociometric
  analysis}}.
\newblock \emph{\bibinfo{journal}{Psychometrika}}
  \textbf{\bibinfo{volume}{18}}, \bibinfo{pages}{39--43}
  (\bibinfo{year}{1953}).
\newblock \urlprefix\url{http://link.springer.com/10.1007/BF02289026}.


  
  

\bibitem{MacKinnon1994}
\bibinfo{author}{MacKinnon, J.~G.}
\newblock \bibinfo{title}{{Approximate Asymptotic Distribution Functions for
  Unit-Root and Cointegration Tests}}.
\newblock \emph{\bibinfo{journal}{J. Bus. Econ. Stat.}} \textbf{\bibinfo{volume}{12}}, \bibinfo{pages}{167--176}
  (\bibinfo{year}{1994}).
\newblock
  \urlprefix\url{http://www.tandfonline.com/doi/abs/10.1080/07350015.1994.10510005}.

\bibitem{Runge2017}
\bibinfo{author}{Runge, J.}, \bibinfo{author}{Sejdinovic, D.} \&
  \bibinfo{author}{Flaxman, S.}
\newblock \bibinfo{title}{{Detecting causal associations in large nonlinear
  time series datasets}}.
\newblock {Preprint at \url{http://arxiv.org/abs/1702.07007}} (\bibinfo{year}{2017}).


\bibitem{Benjamini1995}
\bibinfo{author}{Benjamini, Y.} \& \bibinfo{author}{Hochberg, Y.}
\newblock \bibinfo{title}{{Controlling the False Discovery Rate : A Practical
  and Powerful Approach to Multiple Testing}}.
\newblock \emph{\bibinfo{journal}{J. Royal Stat. Soc. S. B}} 
\textbf{\bibinfo{volume}{57}}, \bibinfo{pages}{289--300}
  (\bibinfo{year}{1995}).
  \urlprefix\url{https://www.jstor.org/stable/2346101}.

  
   
\bibitem{Eichler2010}
\bibinfo{author}{Eichler, M.} \& \bibinfo{author}{Didelez, V.}
\newblock \bibinfo{title}{{On Granger causality and the effect of interventions
  in time series}}.
\newblock \emph{\bibinfo{journal}{Lifetime Data Anal.}}
  \textbf{\bibinfo{volume}{16}}, \bibinfo{pages}{3--32} (\bibinfo{year}{2010}).
  \urlprefix\url{https://doi.org/10.1007/s10985-009-9143-3}.

  
  
\end{thebibliography}

\clearpage

\begin{thebibliography}{1}
\expandafter\ifx\csname url\endcsname\relax
  \def\url#1{\texttt{#1}}\fi
\expandafter\ifx\csname urlprefix\endcsname\relax\def\urlprefix{}\fi
\providecommand{\bibinfo}[2]{#2}
\providecommand{\eprint}[2][]{\url{#2}}

\bibitem{Bakshy2015}
\bibinfo{author}{Bakshy, E.}, \bibinfo{author}{Messing, S.} \&
  \bibinfo{author}{Adamic, L.~A.}
\newblock \bibinfo{title}{{Exposure to ideologically diverse news and opinion
  on Facebook}}.
\newblock \emph{\bibinfo{journal}{Science}} \textbf{\bibinfo{volume}{348}},
  \bibinfo{pages}{1130--1132} (\bibinfo{year}{2015}).
\newblock
  \urlprefix\url{http://www.sciencemag.org/cgi/doi/10.1126/science.aaa1160}.

\end{thebibliography}
\end{document}